\documentclass[11pt,a4paper]{article}
\usepackage{longtable}
\usepackage{amsmath}
\usepackage{amssymb}
\usepackage{graphicx}
\usepackage{bm}% bold math
\usepackage{footmisc}
\usepackage{afterpage}
\usepackage{float}
\usepackage{lscape}
\usepackage{longtable}

\newcommand*{\La}{\cal{L}}
\newcommand*{\no}{\noindent}
\newcommand*{\bea}{\begin{eqnarray}}
\newcommand*{\eea}{\end{eqnarray}}
\newcommand*{\be}{\begin{equation}}
\newcommand*{\ee}{\end{equation}}
\newcommand*{\pd}{\partial}
\newcommand*{\pdm}{\pd_{\mu}}
\newcommand*{\pdn}{\pd_{\nu}}

\newcommand*{\pref}[1]{(\ref{#1})}

\newcommand*{\mn}{{\mu\nu}}

\newcommand*{\prefr}[2]{(\ref{#1}-\ref{#2})} 
\newcommand*{\nn}{\nonumber}

\newcommand*{\tr}{\mathrm{tr}}

\newcommand{\bma}{\begin{pmatrix}}
\newcommand{\ema}{\end{pmatrix}}

\usepackage[square,comma,sort&compress,numbers]{natbib}

\title{A spectroscopical analysis of the phase diagram of Yang-Mills-Higgs theory}

\author{Axel Maas$^{1}$, Tajdar Mufti$^{2}$\\
$^1$Institute of Physics, University of Graz,\\
Universit\"atsplatz 5, A-8010 Graz, Austria\\
$^2$Institute for Theoretical Physics, University of Jena,\\
Max-Wien-Platz 1, D-07743 Jena, Germany\\}

\begin{document}

\maketitle

\begin{abstract}
Yang-Mills-Higgs theory, being the standard-model Higgs sector for a suitable choice of gauge and custodial group, offers a rich set of physics. In particular, in some region of its parameter space it has QCD-like behavior, while in some other region it is Higgs-like. Therefore, it is possible to study a plethora of phenomena within a single theory. Here, the physics of the standard-model version is studied using lattice gauge theory at a qualitative level. To this end, the low-lying spectrum in several different channels is obtained for more than 160 different sets of bare parameters throughout the phase diagram. The theory shows quite different behaviors in the different regions, from almost Yang-Mills-like to the one of an essentially free gas of massive photons. Especially, not always is the behavior as naively expected.
\end{abstract}

\section{Introduction}

Yang-Mills-Higgs theory, the combination of a non-Abelian gauge theory with a fundamental scalar, called here for convenience the Higgs, yields a rich theory. Furthermore, for gauge group SU(2) with two Higgs flavors, i.\ e.\ SU(2) custodial symmetry, this is the Higgs sector of the standard model. The primary aim of the present study is to get a basic qualitative understanding of the physics of this theory by investigating its observable particle spectrum at a qualitative level. A secondary aim is to search for hints of reasonable stable states, which could eventually lead to unaccounted for background in searches for new physics \cite{Maas:2012tj,Maas:2014nya}. The expectation and motivation of this study will be discussed in more detail in section \ref{sphase}, but here already a brief outline will be given.

The theory exhibits two quite different behaviors: Besides the prominent Brout-Englert-Higgs (BEH) effect \cite{Bohm:2001yx} this theory can also exhibit a QCD-like behavior. This includes confinement in the same sense as in QCD, i.\ e.\ an intermediate distance linear-rising potential between fundamental charges, as well as a mass gap \cite{Knechtli:1998gf,Knechtli:1999qe}. Moreover, surprisingly both types of physics are not separated by a phase transition, at least when imposing a finite lattice-cutoff as regulator \cite{Osterwalder:1977pc,Fradkin:1978dv}. This has been confirmed by many numerical investigations \cite{Bonati:2009pf,Caudy:2007sf,Jersak:1985nf,Langguth:1985eu,Maas:2013aia}. However, it is still not finally settled if this theory is trivial \cite{Callaway:1988ya}. If it should be, all results become potentially regulator-dependent, including the statement about the phase structure. Especially, if no continuum limit exists, the choice of lattice action can influence the results. Though in the following this will not be mentioned again explicitly, this should always be kept in mind. But here it will be assumed that for the investigated low-energy physics this is only a minor quantitative effect.

It is then a long-standing question of how to characterize the two different domains \cite{Osterwalder:1977pc,Fradkin:1978dv,Kugo:1979gm}. It is clear that they show quite different behaviors in terms of the gauge-dependent correlation functions \cite{Bohm:2001yx,Caudy:2007sf,Maas:2013aia}, and how they relate to gauge-invariant degrees of freedom \cite{'tHooft:1979bj,Banks:1979fi,Frohlich:1980gj,Frohlich:1981yi,Maas:2012tj,Maas:2013aia}. But it is not so clear how this relates to gauge-invariant physics, though the transition seems to leave a trace in the spectrum \cite{Jersak:1985nf,Langguth:1985eu}.

Perturbatively, it is also expected that for weak gauge interactions the Higgs-like domain can be separated in four different regions \cite{Bohm:2001yx}, with a Higgs lighter than the $W$, heavier but stable, unstable but as a discernible resonances above the two $W$ threshold, and finally with a mass heavier than roughly 1 TeV in a region of strong Higgs-Higgs interaction.

The aim of the present paper is to investigate the physics of this theory, and hence also the previously mentioned expectations, qualitatively using spectroscopical methods: The low-lying spectrum in several different quantum number channels will be determined approximately at a multitude of points in the quantum phase diagram. The aim is to get a rough picture of the physics throughout the phase diagram, and to identify areas worthwhile for a more quantitative investigation in the future. As will be seen, the naive picture painted above is not fully reflected in our selection of points in the phase diagram, if it is correct at all.

The methods and strategies used for this investigation are detailed in section \ref{stech}. There also the different spectroscopical channels will be discussed. The phase diagram and its structure in the region relevant to this work is briefly detailed in section \ref{sphase}. The main results in the different regions are given in section \ref{sspectra}, substantially extending known spectroscopy results \cite{Jersak:1985nf,Langguth:1985eu,Evertz:1985fc,Langguth:1985dr,Karsch:1996aw,Philipsen:1996af,Knechtli:1998gf,Knechtli:1999qe,Iida:2007qp,Bonati:2009pf,Wurtz:2013ova,Gongyo:2014jfa} for this theory .

All of the present results are addressing bound states, which are inherently non-perturbative. The obvious question is, of course, whether they have relevance for the standard model. This question will be answered in section \ref{sphase}. Of course, the present theory is not the standard model, and especially dropping the fermions severely affects the running couplings \cite{Maas:2013aia}. Furthermore, without QED the 11 GeV mass splitting between the $W$ and the $Z$ vanishes. Hence, any result obtained here can at best at a semi-quantitative level comparable. Furthermore, this makes it especially complicated to define what a line-of-constant physics (LCP) would characterize when comparing it to the standard model \cite{Maas:2013aia}, and different choices, see e.\ g.\ \cite{Wurtz:2013ova}, can lead to quite different qualitative results.

This rather extensive study will then be wrapped up in section \ref{sconclusions}. Finally, remarks on systematic errors are given in appendix \ref{a:sys}, and tables with the numerical values of the results are compiled in appendix \ref{numval}.

This work extends the previous results presented in \cite{Maas:2012tj,Maas:2012zf,Mufti:2013uxa}\footnote{Note that in the proceedings \cite{Maas:2012zf,Mufti:2013uxa} the states $1^-_1$ and $0^+_3$ have been wrongly identified, and the corresponding results should be ignored.}. There is a companion paper, which addresses the gauge-dependent aspects of the physics of this theory \cite{Maas:2013aia}, which motivates several of the investigations in section \ref{sspectra}. In particular, there we discussed and checked in detail the relation between the masses of the gauge-dependent $W$ and Higgs particles and the corresponding gauge-invariant states in terms of the Fr\"ohlich-Morchio-Strocchi (FMS) mechanism \cite{Frohlich:1981yi,Frohlich:1980gj}, as detailed in section \ref{sphase}, and for which evidence in numerical lattice simulations was first obtained in \cite{Maas:2012tj}.

We would like to emphasize again that our interest is the whole phase diagram, not only those regions most relevant to the standard model physics. We would like to understand which type of physics is encountered where in the phase diagram, and what kind of trajectories in parameter/theory space are present. The latter is particularly interesting for renormalization group studies \cite{Gies:2014xha,Gies:2013pma,Degrassi:2012ry}. It is therefore necessary to have results on as many points as possible, including those with small lattice cutoff, rather than to have (yet) every point of the phase diagram as precise as possible. We therefore perform only a qualitative study and accept larger systematic errors, especially by choice of the lattice volume, in favor of are more comprehensive scanning of the phase diagram. This choice of small volumes offers at the same time the advantage of a better separation of scattering states \cite{Luscher:1990ux}, which improves the background situation in the search for genuine resonances.

\section{Technical details}\label{stech}

\subsection{Continuum setup}\label{ssetup}

The theory to be investigated here consists of two flavors of scalar particles $\phi$ coupled to a non-Abelian gauge field $W$, with the (Euclidean) action
\bea
{\La}&=&-\frac{1}{4}W_\mn^aW^\mn_a+(D_\mu\phi)^\dagger D^\mu\phi-\gamma(\phi\phi^\dagger)^2-\frac{m_0^2}{2}\phi\phi^\dagger\label{action}\\
W_\mn^a&=&\pdm W_\nu^a-\pdn W_\mu^a-g f^{abc} W_\mu^b W_\nu^c\nn\\
D_\mu^{ij}&=&\pd_\mu\delta^{ij}-igW_\mu^a\tau^{ij}_a\nn,
\eea
\no where $g$ is the gauge-coupling, $\gamma$ and $m_0$ the parameters of the Higgs potential, and $f^{abc}$ and $\tau^a$ are the structure constants and generators of the gauge group, respectively. In presence of the BEH effect the bare mass can also be exchanged for a Higgs expectation value $v$ in a suitable gauge, $m_0^2=2\gamma v^2$ \cite{Bohm:2001yx}. The gauge group is chosen to be the weak isospin gauge group SU(2). Hence, the complex doublet $\phi$ contains four real scalar degrees of freedom, and the flavor symmetry acts as a SU(2) custodial symmetry. This flavor symmetry appears to be intact throughout the phase diagram. It is therefore convenient to make it explicit using the notation \cite{Shifman:2012zz}
\be
X=\bma \phi_1 & -\phi_2^* \cr \phi_2 & \phi_1^* \ema=(\phi^\dagger_i\phi_i)^\frac{1}{2}\varphi\label{x}.
\ee
\no Gauge transformations act on this matrix as a left multiplication, while flavor transformations act as a right multiplication. As given by the second equality, this can be written as the length of the Higgs field multiplied by an SU(2)-valued matrix $\varphi$, except at special points where the Higgs field vanishes.

The parameters in the Lagrangian \pref{action} are the bare ones, i.\ e.\ at the ultraviolet cutoff, if one is introduced. This will be the (inverse) lattice spacing $a^{-1}$ below.

It is important to make a remark here concerning the naming conventions. In this work, we will adhere strictly to the above prescribed naming scheme of calling the (gauge-dependent) elementary fields Higgs and $W$, in accordance with the PDG \cite{pdg}, and the phenomenological language \cite{Bohm:2001yx}. In contrast, based on the works \cite{Banks:1979fi,'tHooft:1979bj,Frohlich:1980gj,Frohlich:1981yi}, certain gauge-invariant composite operators to be introduced below in section \ref{s:bs} have in the lattice literature been denoted as Higgs and $W$ boson, for reasons discussed in \cite{Maas:2012tj} and in section \ref{sphase}. Thus, one should be wary when comparing these different sources.

\subsection{Creation of configurations}\label{s:configs}

To perform lattice simulations of \pref{action}, the techniques described in \cite{Cucchieri:2006tf,Maas:2010nc,Maas:2012tj,Maas:2013aia} will be used. For the sake of completeness, the details will be repeated here.

Under the assumption that a naive lattice discretization of \pref{action} captures all the pertinent qualitative features of the theory, the starting point is the unimproved lattice action \cite{Montvay:1994cy},
\bea
S&=&\beta\sum_x\Big(1-\frac{1}{2}\sum_{\mu<\nu}\Re\tr U_\mn(x)\Big)+\phi^\dagger(x)\phi(x)+\lambda\left(\phi(x)^\dagger\phi(x)-1\right)^2\nonumber\\
&&-\kappa\sum_\mu\Big(\phi(x)^\dagger U_\mu(x)\phi(x+e_\mu)+\phi(x+e_\mu)^\dagger U_\mu(x)^\dagger\phi(x)\Big)\label{laction}\\
U_\mn(x)&=&U_\mu(x)U_\nu(x+e_\mu)U_\mu(x+e_\nu)^{+}U_\nu(x)^{+}\label{plaq}\\
W_\mu&=&\frac{1}{2agi}(U_\mu(x)-U_\mu(x)^\dagger)+{\cal O}(a^2)\label{wdef}\\
\beta&=&\frac{4}{g^2}\nn\\
a^2m_0^2&=&\frac{(1-2\lambda)}{\kappa}-8\nn\\
\lambda&=&\kappa^2\gamma\nn.
\eea
\no In this expression $a$ is the lattice spacing, $U_\mu$ the link variable $\exp(iga W_\mu)$, $\phi$ again the Higgs field, the bare lattice couplings depend on the bare continuum couplings in the described way, and $e_\mu$ is the unit vector in $\mu$ direction. The parameters $\beta$, $\kappa$, and $\lambda$ are therefore the couplings at the lattice cut-off, which is essentially given by $1/a$, with the largest energy accessible being $4/a$, corresponding to a momentum across the body-diagonal of the cubic lattice of extension $N$ in each direction.

Choosing a physical scale is not an entirely trivial issue, especially when a consistent scale throughout the phase diagram should be implemented. The simplest choice is to give the lightest state in the spectrum a fixed mass. For the regions of the phase diagram investigated here, this is either the $0^{+}$ flavor singlet or the $1^{-}$ flavor triplet, obtained with the methods described below in section \ref{s:bs}. Due to the relation between the $1^-$ triplet and the $W$ boson mass at the physical point \cite{Maas:2013aia}, the most convenient choice is, however, to always set the mass of the $1^-_3$ ground state to 80.375 GeV. Nonetheless, in general, all results will be given in lattice units or in the form of dimensionless ratios, so this matters little. The set of lattice parameters and the obtained lattice spacing used for the spectra is noted in each case below, and also reported in the appendix \ref{numval}. Errors on these lattice spacings are usually small, below 1\%, and therefore not explicitly included in the following.

The generation of configurations follows \cite{Maas:2010nc,Maas:2013aia}. For the gauge fields a combination of one heat-bath and five-overrelaxation sweeps have been used \cite{Cucchieri:2006tf}. In between each of these 6 gauge fields updates one Metropolis sweep for the Higgs field has been performed using a Gaussian proposal. The width of the proposal is adaptively tuned to achieve a 50\% acceptance probability. This should balance the movement through configuration space compared to the finding of relevant configurations. These updates have been performed lexicographically. These 12 sweeps together constitute a single update for the field configuration. The auto-correlation time of the plaquette is of the order of 1 or less such updates. Thus, $N$ such updates separate a measurement of a gauge-invariant observable, to reduce the auto-correlation time. For the thermalization, $2(10N+300)$ such updates have been performed. Furthermore, all calculations involved many independent runs, at least 25, to further reduce correlations.

All errors have been calculated using bootstrap with 1000 re-samplings and give a, possibly asymmetric, 67.5\% interval, i.\ e.\ approximately a 1$\sigma$ interval.

The code, including the one to determine the bound states in section \ref{s:bs}, has been checked, where available, by comparing to the results in \cite{Langguth:1985dr,Jersak:1985nf}.

Throughout, symmetric lattices\footnote{To prevent any problems with polarization effects for the vector states on the rather small lattices \cite{Cucchieri:2006za}, no asymmetric lattices have been used.} of size $N^4=24^4$ have been used. Since the main question here was to identify the ground-state spectrum, and possibly some low-lying further states, this is sufficient. A deeper analysis of possible resonances will require more lattice volumes, and is under way for the various regions in the phase diagram identified in the present paper.

We should note that, based upon the arguments of \cite{Frohlich:1980gj,Frohlich:1981yi}, it is possible to perform a variable transformation into a theory described by gauge-invariant variables \cite{Langguth:1985dr,Jersak:1985nf}, though this incurs the danger of topological defects \cite{Maas:2013aia}. Based upon these variables it is possible to develop a perturbative expansion \cite{Buchmuller:1995sf,Buchmuller:1994vy,Fodor:1994sj}, which in principle could also be used to estimate several parameters of the theory. However, this reformulation includes a Jacobian, which is neglected in the perturbative expansion, but creates in general an infinite series of additional tree-level terms. Its neglect is a good approximation when the Higgs fluctuations are small compared to its expectation value, and therefore well suited when addressing standard-model-like physics. Here, with the focus widened to the full phase diagram, this may no longer be a good approximation in general, and would have to be checked for every point separately, therefore requiring the calculations anyway. It is hence not done.

\subsection{Bound states}\label{s:bs}

The central observables in the following will be the spectrum in channels of different quantum numbers. The channels investigated are $0^+_1$, $0^-_1$, $1^-_3$, and $2^+_1$, where the $J^P_f$ quantum numbers give the continuum notation, and $f$ indicates the flavor structure, being either singlets or triplets. No attempt was made to resolve the different hypercubic representations, see \cite{Wurtz:2013ova} for a detailed investigation of this question. The aim here was primarily a qualitative one. Most interesting are the $0^+_1$ and $1^-_3$ channels, as they correspond to the observable Higgs and $W$ states, as discussed in section \ref{sphase}.

As one interesting question below will be the structure above the ground state a variational analysis \cite{Gattringer:2010zz} with a base of 5-10 operators per channel has been performed. Always a preconditioning using the correlation matrix at time zero turned out to be most useful. To create the operator basis, APE smearing \cite{DeGrand:1997ss,Philipsen:1996af} has been used, which yields the links and Higgs fields after $n\ge 1$ checker-board smearing sweeps as 
\bea
U_\mu^{(n)}(x)&=&\frac{1}{\sqrt{\det R_\mu^{(n)}(x)}}R_\mu^{(n)}(x)\nn\\
R_\mu^{(n)}(x)&=&\alpha U_\mu^{(n-1)}(x)+\frac{1-\alpha}{2(d-1)}\nn\\
&&\times\sum_{\nu\neq\mu}\left(U^{(n-1)}_\nu(x+e_\mu)U^{(n-1)\dagger}_\mu(x+e_\nu)U^{(n-1)\dagger}_\nu(x)\right.\nn\\
&&\left.+U_\nu^{(n-1)\dagger}(x+e_\mu-e_\nu)U_\mu^{(n-1)\dagger}(x-e_\nu)U_\nu^{(n-1)}(x-e_\nu)\right)\nn\\
\phi^{(n)}(x)&=&\frac{1}{1+2(d-1)}\left(\phi^{(n-1)}(x)\right.\nn\\
&&+\left.\sum_\mu(U^{(n-1)}_\mu(x)\phi^{(n-1)}(x+e_\mu)+U^{(n-1)}_\mu(x-e_\mu)\phi^{(n-1)}(x-e_\mu))\right)\nn,
\eea
\no with $\alpha=0.55$ and $d=4$. The maximum number of iterations was $n=4$. The impact of this smearing on the correlators will be discussed below. Note that the smearing also helped in terms of the noise, as in the relevant domain of the phase diagram all correlation functions are quite noisy, even at the ${\cal O}(10000-100000)$ configurations used here for each lattice setup.

A number of basic operators have been used to construct the correlation matrix. The local operators are \cite{Jersak:1985nf,Langguth:1985eu,DeGrand:2006zz}
\bea
{\cal O}_{\rho}(x)&=&\phi_i^\dagger(x)\phi^i(x)=\rho(x)\label{higgsonium}\\
{\cal O}_{W}(x)&=&\tr U_\mu(x)U_\nu(x+e_\mu)U_\mu(x+e_\nu)^\dagger U_\nu(x)^\dagger\nn\\
{\cal O}_{1^-\mu}^a(x)&=&\tr \tau_a \det(-X(x))^{-\frac{1}{2}} X^\dagger(x) U_\mu(x)\det(-X(x+e_\mu))^{-\frac{1}{2}} X(x+e_\mu)\label{wl}\\
{\cal O}_-&=&\sum_{\mu\neq\nu\neq\rho\neq\sigma}\tr U_\mu(x) U_\nu(x+e_\mu)U_\mu^\dagger(x+e_\nu)U_\nu^\dagger(x)U_\rho(x)U_\sigma(x+e_\rho)U_\rho^\dagger(x+e_\sigma)U_\sigma^\dagger(x)\nn\\
{\cal O}_2&=&\Re\tr(U_{xy}(x)+U_{yz}(x)-2U_{xz}(x))\nn,
\eea
\no which are operators in the $0^+_1$, $0^+_1$, $1^-_3$, $0^-_1$, and $2^+_1$ channels, respectively. In the language of a naive constituent interpretation, the operators can be viewed as bound states of the Higgs and the $W$:
\begin{itemize}
 \item ${\cal O}_{\rho}$ describes a two-Higgs bound-state, similar to a QCD meson
 \item ${\cal O}_{W}$ is a $W$-ball, the weak version of a glueball
 \item ${\cal O}_{1^-\mu}^a$ is a $W$ dressed by two Higgs particles in a non-trivial flavor multiplet, similar to a isospin triplet vector meson in QCD
 \item ${\cal O}_-$ is a pseudoscalar $W$-ball, or glueball
 \item ${\cal O}_2$ is a tensor $W$-ball, or glueball
\end{itemize}
All of these operators have been integrated over the spatial volume to obtain the zero momentum component. The final correlation functions $C(t)\equiv C(t_1-t_2)=\langle O(t_1) O(t_2)\rangle$ have been averaged over time slices, and, where appropriate, flavor and Lorentz indices. Note that any disconnected parts have been implicitly removed. For many of the present lattice spacings, non-zero momentum states have been too noisy, and hence have not been included at all, see \cite{Wurtz:2013ova} for such an investigation in a different region of the parameter space. Note that also different operators can, and have been, used in the literature, see \cite{Jersak:1985nf,Langguth:1985eu,Evertz:1985fc,Langguth:1985dr,Karsch:1996aw,Philipsen:1996af,Knechtli:1998gf,Knechtli:1999qe,Iida:2007qp,Bonati:2009pf,Wurtz:2013ova}.

The operators in the channels $0^-_1$, and $2^+_1$ turn out to be never reliably lighter than the ones in the $0^+_1$ and/or $1_3^-$ channels. Due to this, and the significant noise associated consequently with them, only the ground states have been investigated here. These operators are especially interesting, as Yang-Mills theory has an inverted hierarchy with the tensor being lighter than the pseudoscalar. Hence, the ordering of these states can be seen as an indication of  how Yang-Mills-like the theory is. Hence, for them the correlation matrix was given by the 5$\times$5 matrix obtained from the operator $O^n_i(t_1)O^n_i(t_2)$, where $n$ counts the number of APE sweeps .

For the more interesting operators in the $0_1^+$ and $1_3^-$ channels, a 10 and 8 operator basis has been used, respectively. For the $0^+_1$, these are
\bea
O^{0^+_1}_1&=&O_\rho^4\nn\\
O^{0^+_1}_2&=&O_\rho^3\nn\\
O^{0^+_1}_3&=&O_W^4\nn\\
O^{0^+_1}_4&=&O_W^3\nn\\
O^{0^+_1}_5&=&O_{1^-}^{4} O_{1^-}^{4}\nn\\
O^{0^+_1}_6&=&O_{1^-}^{3} O_{1^-}^{3}\nn\\
O^{0^+_1}_7&=&O_\rho^4O_\rho^4\nn\\
O^{0^+_1}_8&=&O_\rho^3O_\rho^3\nn\\
O^{0^+_1}_9&=&O_W^4O_W^4\nn\\
O^{0^+_1}_{10}&=&O_W^3O_W^3\nn,
\eea
\no where the numbers 3 and 4 give the number of APE smearing levels. The first four operators are single particle ones, while the remainder are $s$-wave 2-particle operators.

Similarly, for the $1^-_3$ the operators
\bea
O^{1^-_3}_1&=&O_{1^-}^4\nn\\
O^{1^-_3}_2&=&O_{1^-}^3\nn\\
O^{1^-_3}_3&=&O_{1^-}^4O_\rho^4\nn\\
O^{1^-_3}_4&=&O_{1^-}^3O_\rho^3\nn\\
O^{1^-_3}_5&=&O_{1^-}^4O_W^4\nn\\
O^{1^-_3}_6&=&O_{1^-}^3O_W^3\nn\\
O^{1^-_3}_7&=&O_{1^-}^4(O_{1^-}^4O_{1^-}^4)\nn\\
O^{1^-_3}_8&=&O_{1^-}^3(O_{1^-}^3O_{1^-}^3)\nn
\eea
\no have been used. Besides the one and two particle operators, the latter also again in zero momentum $s$ wave, also three-particle operators with zero angular momentum have been used. In this case, the flavor and Lorentz indices of two of the operators, indicated by the parentheses, are fully contracted, while the third operator carries the total spin and flavor.

For all channels, however, in most cases the full correlation matrix was too noisy to extract reliable eigenvalues at larger times, and hence only a subset was used. Furthermore, the eigenvectors were in many components far too noisy to be useful. This entails the possibility of missorting the eigenvalues when associating them with a state. However, a detailed analysis shows that this almost never occurred, as the effective masses are sufficiently smooth functions of time, see section \ref{s:fitex} for an example. Even if intersections occurred, the eigenvalue assignment did mostly not create any kinks, and thus provided a reliable assignment of eigenvalues to states.

Furthermore, in many cases the number of operators were reduced, starting with the ones with the highest number, i.\ e.\ the 'heaviest' and least smeared ones, until a statistical sufficiently reliable extraction was possible. In most cases, this reduced to the first seven operators for the $0^+_1$ and to the first four operators for the $1^-_3$. Furthermore, it was observed that even for the ground state usually no perfect unmixing was possible, but higher state contaminations prevailed. For the higher states, usually contamination by both heavier and lighter states has been observed. This was also evident when evaluating the eigenvectors. An example of this will be shown in the next section.

\subsection{Mass determination}

To determine the masses, a fit just using plateaus was often not possible. Rather than that, two-state fits of the type
\be
C(t)=A\cosh\left(am\left(t-\frac{N}{2}\right)\right)+B\cosh\left(an\left(t-\frac{N}{2}\right)\right)\label{stdfit},
\ee
\no have been performed for the correlators. If possible, information about adjacent states or the ground state were used to provide one of the masses, to improve the fit results, but no simultaneous fit of the whole spectrum was attempted. Errors on the fitted masses were obtained by varying the correlation functions coherently, i.\ e.\ at all times either up or down, within the 67.5\% error margin from the bootstrap analysis. Since as a consequence of the contamination with different states there were often no good plateaus, a fit was deemed to be acceptable if it described the data points of the effective mass for $t>1$ for the ground state and for $t>0$ for higher states within twice the determined error band, up to times where the correlation function error became comparable to the value of the correlation functions. Of course, this is rather optimistic. The second masses in \pref{stdfit} were also included in the spectra, provided that within errors they did not coincide with the leading mass of any other level.

\begin{figure}
\centering
\includegraphics[width=0.5\linewidth]{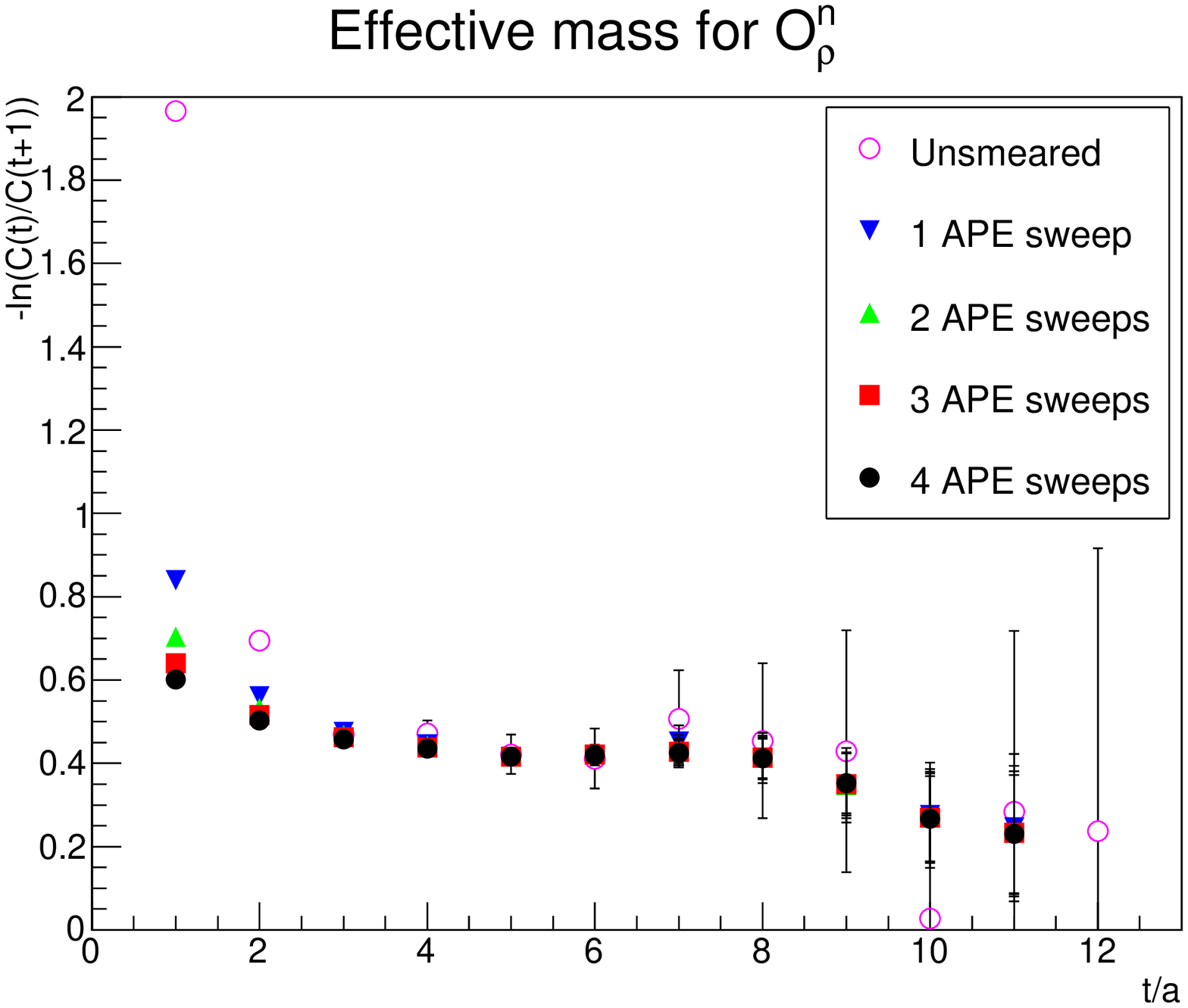}\includegraphics[width=0.5\linewidth]{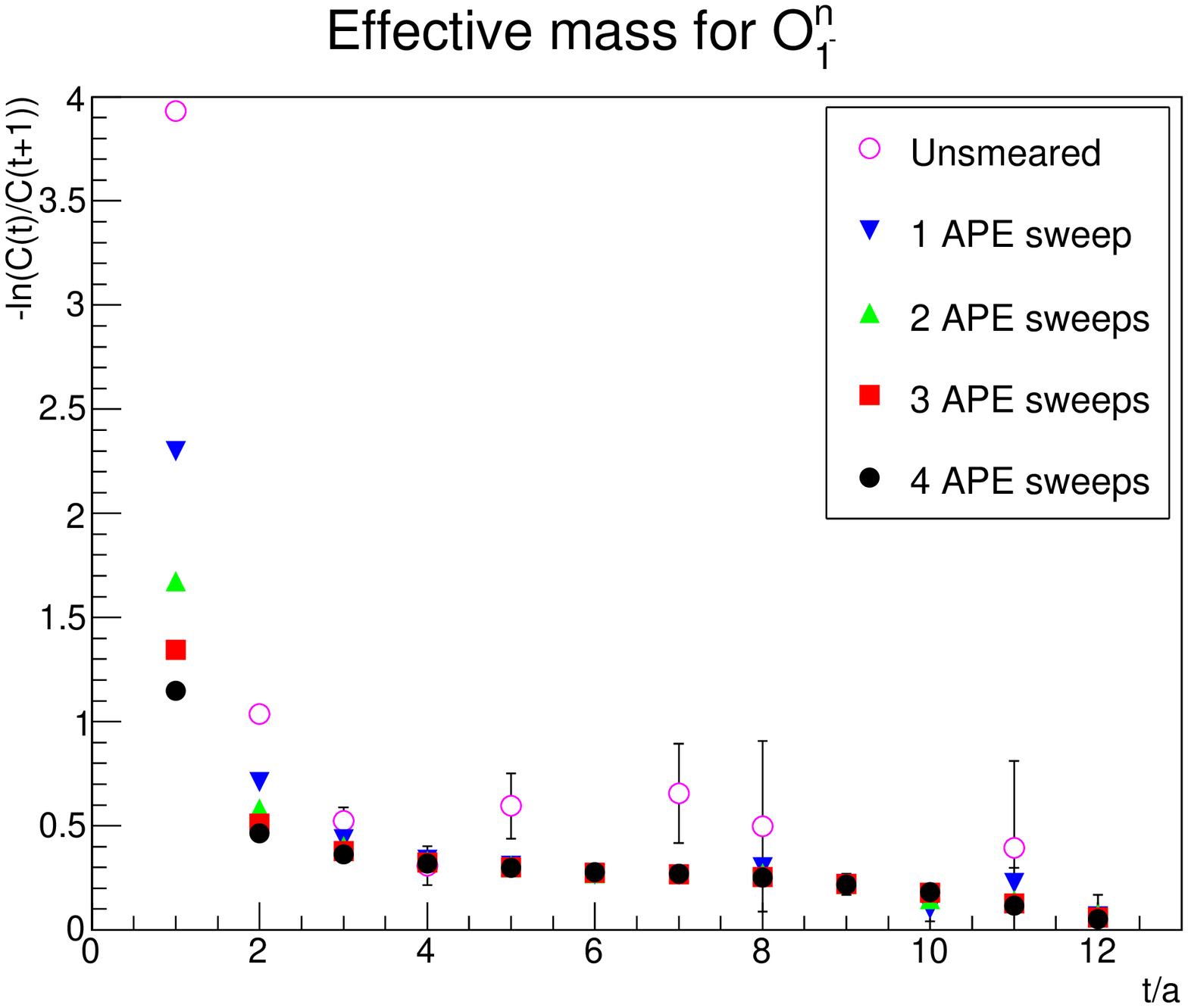}
\caption{\label{smearing}The effective mass of the operators $O_\rho^n$ (left panel) and $O_{1^-}^n$ as a function of time for different numbers $n$ of APE sweeps. Results are at $\beta=2.4728$, $\kappa=0.2939$, and $\lambda=1.036$, which is in the Higgs-like domain with a physical $0^+_1$ to $1^-_3$ ratio and 88192 configurations.}
\end{figure}

Since one of the most interesting questions with respect to physical consequences will be the nature of the heavier states in the $0^+_1$ and $1^-_3$ channels, an important question is how much the APE smearing affects the results. This is studied in figure \ref{smearing} for the operators $O_\rho$ and $O_{1^-}^a$ for an example, but representative, lattice setting. As is visible, the $1^-_3$ state is much stronger affected by the smearing than the $0^+_1$ state. Without smearing, the statistical noise, especially for the $1^-_3$ state is so large as to make the correlators effectively useless. Even 1 and 2 level of smearing still show significant statistical fluctuations at large times. At the same time, the difference between 3 and 4 level of smearing is already rather small, so that more smearing would be eliminating too much of the higher state contribution to make it still extractable. Hence, the choice for 3 and 4 levels of APE smearing for the operators in the $1^-_3$ and $0^+_1$ channels. Including also the lower level of APE smearing for the remaining operators is considered just as a possibility, and usually only up to and including 2-3 level of APE smearing yielded a statistically useful result.

Finally, at several points throughout the higher states in a given channel will be compared to scattering states. These are obtained in the most naive way without lattice spacing corrections as
\be
W_j=\sum_i\sqrt{m_i^2+\left(\frac{2\pi j_i}{N}\right)^2}\nn,
\ee
\no where the masses of the constituents $m_i$ are taken at face value, and the lattice momenta $j_i$ are included as required, see section \ref{s:qld} for details.

Note that for the lattice volume employed the smallest momentum in lattice unit is about 0.26, which implies that scattering states with non-zero momentum will be substantially far spread. At the same time, this permits to have around four units of momenta, and therefore up to two non-zero-momentum scattering states at the finest lattice discretization, until reaching an energy of about 1 in lattice units, where discretization errors become dominant. Since no more than three scattering states (including the one at rest) will be analyzed below, this implies that the lattice volume is nonetheless large enough to make meaningful statements.

\subsection{An example spectrum extraction}\label{s:fitex}

To exemplify the fitting procedure and mass extraction, as it is not possible to provide plots for the more than 160 sets of lattice parameters later, a single case will be treated here. This will be the case of $\beta=2.7984$, $\kappa=0.2954$, and $\lambda=1.317$. This case exhibits all the kinds of ambiguous or noteworthy situations which have been encountered, and is therefore suitable to show how the fitting has been performed.

\begin{figure}
\centering
\includegraphics[width=\linewidth]{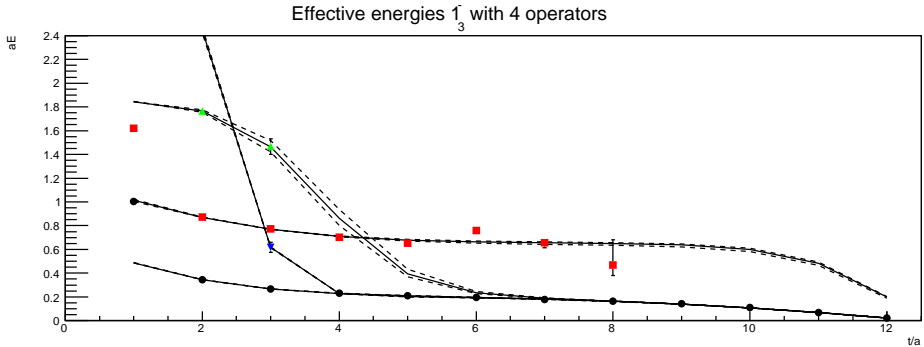}
\caption{\label{exp-1-3}The effective energy levels in the $1^-_3$ channel, together with fits of type \pref{stdfit}. Different styles correspond to different eigenvalues. Points with too large error bars have been omitted for visibility. See text for details. Here and hereafter, the full line is the fit to the average values, while the dashed one is the fit to the $1\sigma$-shifted data, and thus gives an error band.}
\end{figure}

The first channel is the $1^-_3$ channel, which is also used to set the scale. In figure \ref{exp-1-3} the effective energy levels
\be
E_i\left(\frac{t}{a}\right)=\ln\frac{\lambda_i\left(\frac{t}{a}-1\right)}{\lambda_i\left(\frac{t}{a}\right)}\nn
\ee
\no obtained from the eigenvalues $\lambda_i$ after the eigenvalue analysis are shown. Only when performing the eigenvalue analysis using the operators 1-4, with preconditioning at time $t=0$, the results were statistically sufficiently good to proceed further.

The lowest energy level is well-fitted for all times $t/a>1$, but this fit requires both components of \pref{stdfit}. At $t/a=1$, even this lowest level is too strongly affected by contributions from higher levels, as the agreement with the fit for the second state shows. The second state itself is well described with the two parameter fit. The point at $t/a=6$ indicates that the statistical errors are probably an underestimation of the actual errors.

The two further levels have only three statistical significant points available, and none agree with a plateau, though this is not generally so for such short data sequences. They show both a strong contamination with lighter states. The fits in figure \ref{exp-1-3} are performed under the assumption that the contributing state is the ground state, which performs quite well. However, fitting with such few points is highly unreliable. The energy of these two states are both above 1.5, and such states will be omitted in the discussion below.

However, the sub-leading contributions in the lower two levels have energies that are smaller than 1.5, and are therefore distinctively different from the leading contribution of the two highest states, and thus signify two further levels. Such levels have hence been included in the results below, if their energies were found to be sufficiently small.

\begin{figure}
\centering
\includegraphics[width=\linewidth]{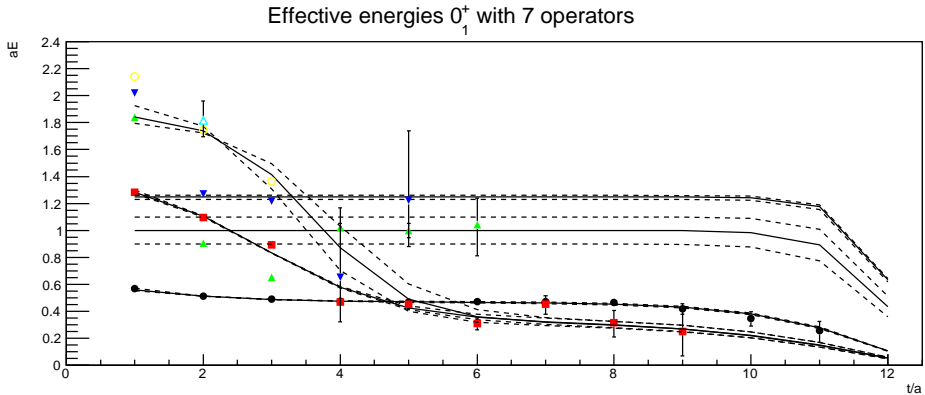}
\caption{\label{exp-0+1}The effective energy levels in the $0^+_1$ channel, together with fits of type \pref{stdfit}. Different styles correspond to different eigenvalues. Points with too large error bars have been omitted for visibility. See text for details.}
\end{figure}

The situation in the $0^+_1$ channel is, and this is almost generally true, more involved. This can be seen in figure \ref{exp-0+1}. In this case, 7 operators could be included in the eigenvalue analysis. Still, for two levels not enough points are statistically relevant to perform any kind of fit. However, the lower levels have not been affected by these uncertainties, as would occur when including further operators.

The first interesting observation are the two lowest states, signified by a black, filled circle and the red squares, to be called level one and two, respectively, as these have been the corresponding eigenvalues. Level one is very stable, could be well-fitted with a suitable plateau, and the higher-state contamination could be fitted as well. State two is more troublesome. It has a substantially stronger excited state contribution, which can be fitted using a two-level fit of type \pref{stdfit} still in an acceptable manner. Due to this excited state contamination, the lowest state needs much longer to isolate itself, and does so with a rather small correlator. Hence, it has large statistical errors. The fits yield for the lighter state in state one 0.466(7) and for state two 0.33(2). They are therefore separated by more than 4$\sigma$. Hence, they should be identified as two separate states, with state two being lighter, even though its correlation function for the time interval available is larger than that of state one, and the individual effective energy points have larger statistical errors. This situation appears actually rather often, especially in the Higgs-like domain of the phase diagram close to the threshold of a $0^+_1$ decaying into two $1^-_3$. So far, in almost all cases an increase of statistics has corroborated such an interpretation of the data, and will be hence followed throughout.

The next level, signified by green triangles, is very complicated to interpret. It cannot be excluded that the point of level two and three at $t/a=3$ has been wrongly sorted. However, the eigenvector components are too noisy to unambiguously identify this. Hence, in this case a single component fit was made, which included all but this point, with a rather conservative error estimate. Such situations are neither exceptionally rare, nor common. In doubt, always the error margin, as in the present case, was substantially increased compared to the one obtained from a direct fit.

Level four, signified by the blue upside-down triangles is a typical representative of the case where substantial errors are present, though not so large as to be completely useless. In such cases, usually a single-level fit was performed, demanding that all points are compatible with the fit result within 2$\sigma$.

The last level fitted, the fifth level, signified by yellow open circles, has again only data at three points, and decays quickly with time. Though a fit is possible, under the assumption of mixing with the ground-state, this state is again unreliable. Since it has a mass above 1.5, however, it would anyhow not be included below. Also, below only the three lowest energy levels will be discussed, so that this state would also be excluded from discussion anyway.

\begin{figure}
\centering
\includegraphics[width=\linewidth]{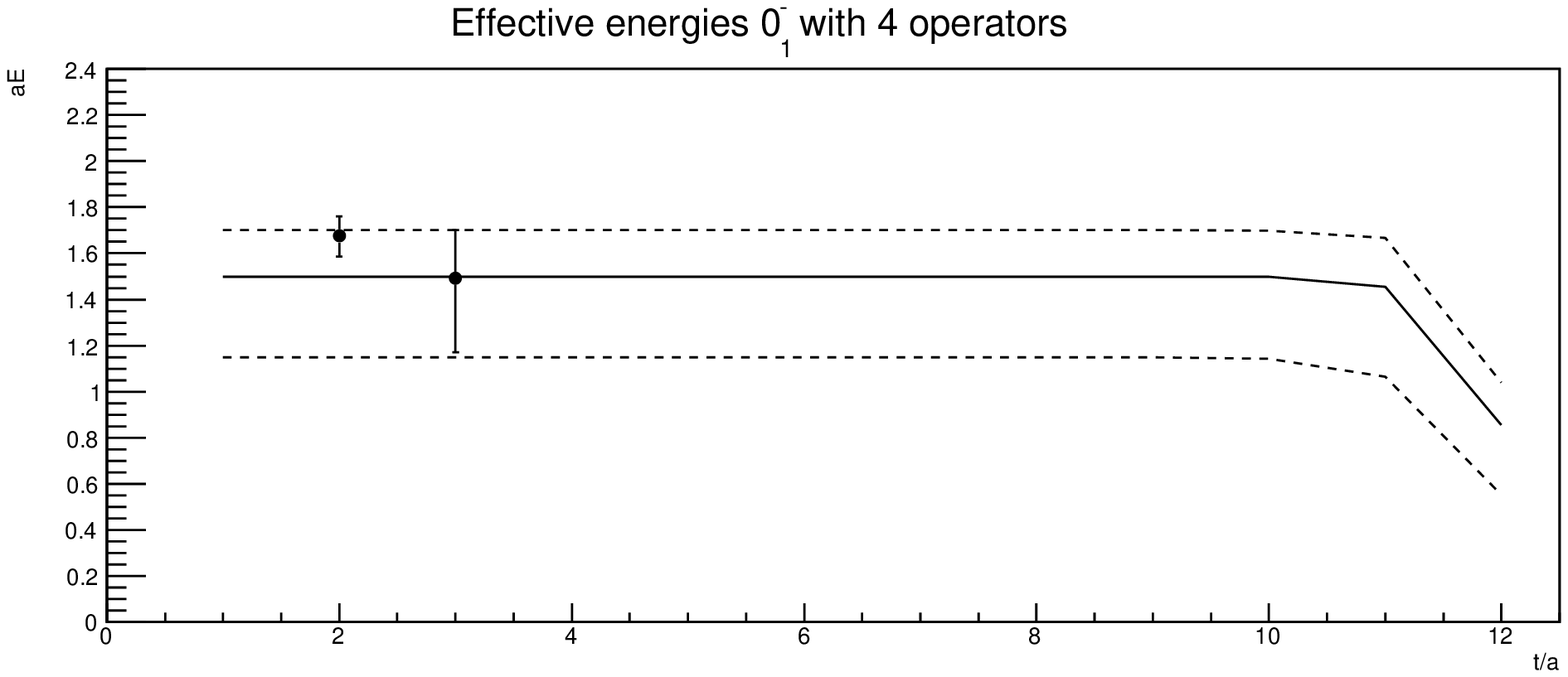}\\
\includegraphics[width=\linewidth]{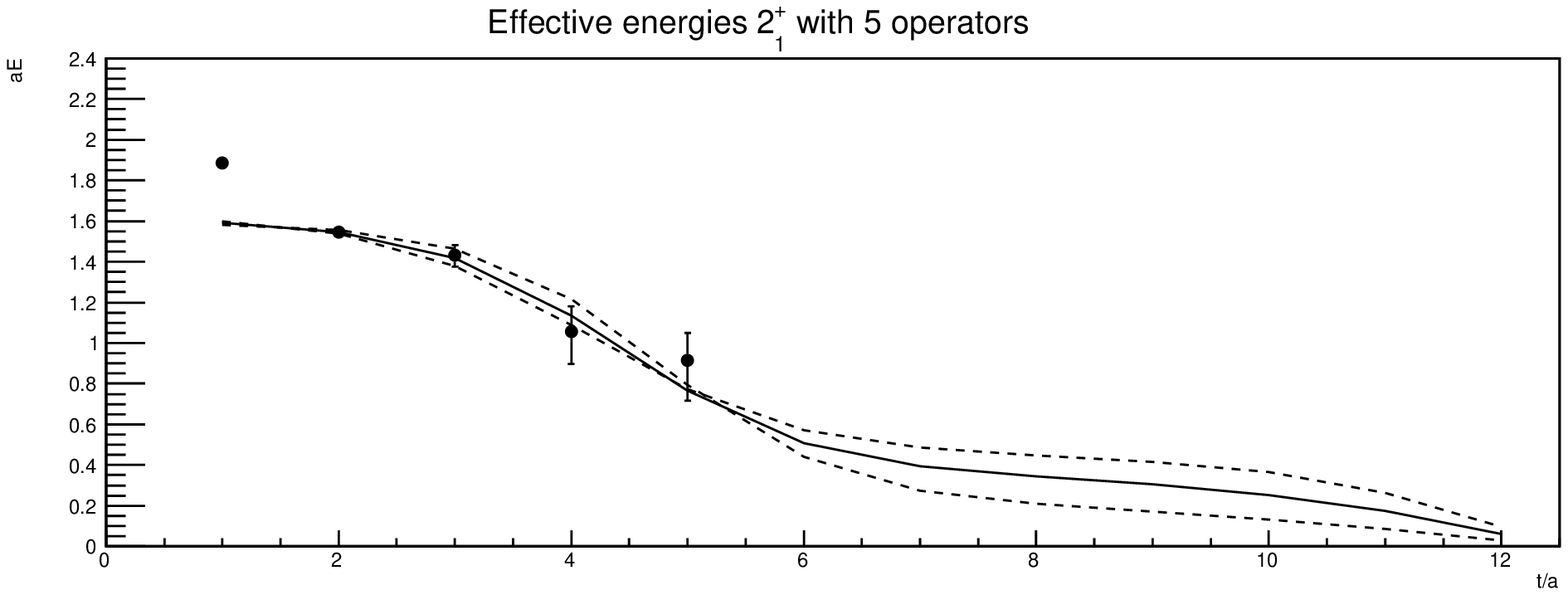}\\
\caption{\label{exp-rest}The effective energy levels in the $0^-_1$ channel (top panel) and $2^+_1$ channel (bottom panel), together with fits of type \pref{stdfit}. Different styles correspond to different eigenvalues. Points with too large error bars have been omitted for visibility. See text for details.}
\end{figure}

The remaining channels are shown in figure\footnote{Note that since energy relations are not enforced when fitting the data or the $1\sigma$-shifted data, it can happen that the fit for the average value lies outside the fit from the error bands.} \ref{exp-rest}.

The $0^-_1$ channel shows a behavior which is found to be rather typical, if the lowest energy level is substantially above 1: In this case, there are very few points at best. Extracting an energy is therefore highly unreliable, and anyhow only possible under the assumption that the variational analysis has projected close to the ground state at short times. As has been seen above, this is only approximately true. As a consequence, in all graphs below, levels with an energy above one will be signified, and should be rather considered to be upper limits.

The situation in the $2^+_1$ channel is quite similar: A number of points, of which the latest show a decline, and a reasonably accurate fit can be found which indicates a (significantly) lower energy than inferred from the points alone. At the same time, the error bands of these points increase, but even taking these into account, no much larger energy can really fit the decline. In such a situation, it is not entirely clear to which extent the result can be trusted. In all such cases, however, the obtained energy levels do not push it below the masses of the lightest state, and therefore such results are reluctantly included in the below presentation of results, for the sake of completeness. However, it should be kept in mind that this will require much more detailed investigations to provide a reasonably reliable final estimate of the smallest energy in these channels.

These categories cover all of the observed ones. Hence in all cases, all channels show one of the present behavior, albeit especially the channels $0^-_1$ and $2^+_1$ permute their behavior constantly throughout the phase diagram.

As a note, it turned out that, if the ground-state is stable, inclusion of further operators never altered the ground-state level beyond a few percent. If it is not stable, however, the lowest level was often substantially above the elastic threshold, indicating the absence of a state. Including the corresponding two-particle state for the decay product usually fixed this, except for some cases to be discussed below in section \ref{s:a}. Note that since the basis in \cite{Maas:2013aia,Maas:2012tj} did not include scattering states the masses used there should be regarded rather as upper limits to the actual masses.

A detailed discussion of the systematic errors can be found in appendix \ref{a:sys}. This assessment shows that the qualitative conclusions of the following appear solid. However, there are certainly quantitative errors which are larger the higher the energy of a state. Also, in some cases the states are substantially above an expected scattering state, but essentially never below, and thus systematic errors seem rather to mess up the identification of the correct scattering state rather than make a scattering state appear as a non-trivial state.

\section{QCD-like vs.\ Higgs-like}\label{sphase}

To interpret the findings here in comparison to both perturbative calculations and, eventually, experiment, it is important to recall a number of features of Yang-Mills-Higgs theory. Foremost, the (lattice) theory has a continuously connected phase diagram \cite{Osterwalder:1977pc,Fradkin:1978dv}. Hence, the physical states are the gauge-invariant composite states, and necessarily bound states or collections of bound states, throughout the phase diagram, irrespective of the couplings. Of course, it may happen that there is only one stable bound state in the system, and all asymptotic states are consisting only out of collections of this state. But there is at least this one state. Nonetheless, there can be, of course, some phase transition in the phase diagram in principle, as long as it can be circumvented. But older findings indicating the presence of such a transition \cite{Langguth:1985eu,Evertz:1985fc} have been shown to be a finite-volume lattice artifact \cite{Bonati:2009pf}. In turn, only at the ultimate limit of employed resources signals of a phase transition have been seen in  \cite{Bonati:2009pf}. But it is not clear, whether a similar increase in lattice parameters as from [17,18] to [8] would again alter this result, and this is at the present a somewhat daunting challenge. Finding such a phase transition would be of tremendous importance: As long as it is not an isolated point of a first order phase transition, it is expected to be either a second-order phase transition or a first-order transition line ending in a second-order critical end point. But if either would exist, any second-order point would locate a continuum limit, which may be potentially non-trivial.

\begin{figure}
\centering
\includegraphics[width=0.8\linewidth]{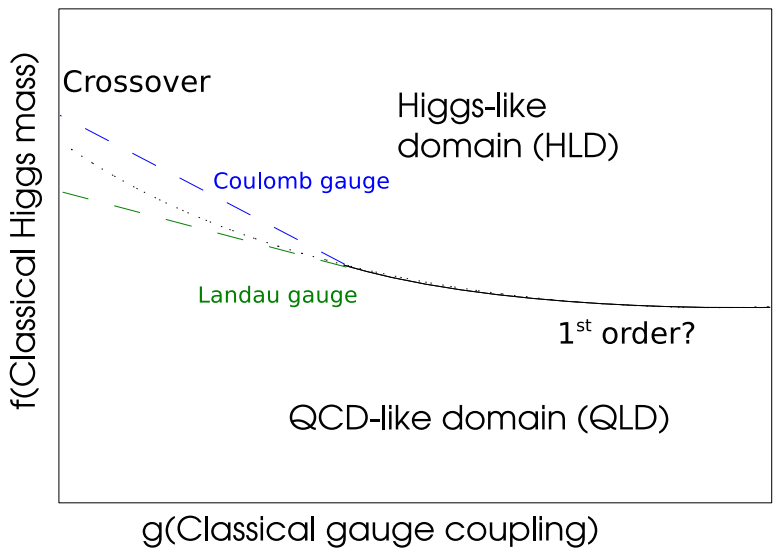}
\caption{\label{pds}A sketch of a two-dimensional hyper-plane of the phase diagram. The axis are some functions of the parameters. For a quantitative version, see \cite{Bonati:2009pf,Caudy:2007sf}.}
\end{figure}

The connectedness of the phase diagram implies that, in contradistinction to perturbation theory \cite{Bohm:2001yx}, a distinction between the BEH effect and the confinement mechanism is a purely gauge-dependent statement, and therefore physically irrelevant \cite{Osterwalder:1977pc,Fradkin:1978dv,Caudy:2007sf,Maas:2012ct}. Especially, there is no gauge-invariant, local order parameter distinguishing both regions\footnote{Note that the non-Abelian nature is here relevant, and Abelian gauge theories may be different \cite{Kennedy:1986ut}.}. But if a gauge is chosen, then the BEH effect affects the gauge-dependent correlation functions in a well-defined region of the phase diagram \cite{Caudy:2007sf,Kugo:1979gm,Alkofer:2000wg}. In particular, it is possible to define gauge-dependent order parameters signaling this transition, though the location depends on the chosen gauge \cite{Caudy:2007sf}. This situation is sketched in figure \ref{pds}.

Still, it has been found that there exist two regions of the phase diagram, in which the physics shows quantitatively a distinctively different behavior. The most marked difference is the ordering of the ground states of the $0^+_1$ and $1^-_3$ channels, which changes between them \cite{Langguth:1985eu,Evertz:1985fc}. This will also be confirmed here. In addition, in the region where the $0^+_1$ is lighter also an intermediate-distance string tension has been observed \cite{Knechtli:1998gf,Knechtli:1999qe}. Sufficiently far away from the cross-over region, defined to be the region where both masses are roughly equal, also in all investigated gauges the BEH effect is absent. This has been checked here explicitly as well, using the order parameter introduced for Landau gauge in \cite{Caudy:2007sf} and the behavior of gauge-dependent correlation functions \cite{Maas:2013aia}. This allows to speak of a QCD-like domain (QLD). On the other hand, the other region is characterized by the absence of an intermediate-distance string tension, and sufficiently far away from the transition shows clear signals for BEH-type physics, e.\ g.\ a non-vanishing vacuum-expectation value of the Higgs field in suitable gauges, is observed. Thus, this region can be characterized as a Higgs-like domain (HLD). This is also noted in the sketch in figure \ref{pds}.

If the physics is so different from the perturbative picture, this begs the question, why does perturbation theory work so well in phenomenology. In particular, as therefore the gauge-dependent $W$ boson and the Higgs boson are not asymptotically observable states as assumed in perturbation theory, this requires to clarify the connection between the gauge-invariant states and the elementary fields states. This relation has been established in form of the FMS mechanism \cite{Frohlich:1980gj,Frohlich:1981yi}, which has been confirmed on the lattice \cite{Maas:2012tj,Maas:2013aia}. In the HLD, for not too large masses of the $0^+_1$, there is a relation between the gauge-invariant $0^+_1$ and $1^-_3$ states' masses with the masses of the gauge-dependent Higgs and $W$ particles, respectively. It is obtained from an expansion in the quantum fluctuations of the Higgs.

These relations require an aligned gauge, i.\ e.\ one with non-vanishing Higgs expectation value \cite{Lee:1974zg,Maas:2012ct}. Taking then the correlators \pref{higgsonium} and \pref{wl} in the continuum and expanding the Higgs field around its expectation value $vn^i$, $\phi^i(x)=\eta^i(x)+n^iv$, with $n^i$ some constant isospin vector, yields \cite{Frohlich:1980gj,Frohlich:1981yi}
\bea
&&\langle\phi_i^\dagger(x)\phi^i(x)\phi_j^\dagger(y)\phi^j(y)\rangle\nn\\
&\approx& v^4+4v^2(c+\langle\eta^\dagger_i(x) n^i n_j^\dagger\eta_j(y)\rangle)+{\cal O}(\eta^3)\label{correl},
\eea
\no and
\bea
&&\langle(\tau^a\varphi^\dagger D_\mu\varphi)(x)(\tau^a\varphi^\dagger D_\mu\varphi)(y)\rangle\nn\\
&\approx& \tilde{c}\tr(\tau^a\tilde{n}\tau^b\tilde{n}\tau^a\tilde{n}\tau^c\tilde{n})\langle W^b_\mu W^c_\mu\rangle+{\cal O}(\eta W)\label{correl2},
\eea
\no with $c$ and $\tilde{c}$ some constants, and $\tilde{n}$ an SU(2)-valued representation of $n^i$ like \pref{x}. Thus, the masses determined by the poles of the correlators on both sides have to coincide to this order. That this holds true beyond this expansion is supported by lattice simulations \cite{Maas:2012tj,Maas:2013aia}. Hence, in the part of the HLD relevant to the standard model, a description in terms of the gauge-invariant and gauge-dependent degrees of freedom give an equally good picture of the physics of the ground states, explaining the great success of perturbation theory\footnote{Similar relations hold for the standard-model fermions, except for the gauge-singlet right-handed neutrinos, where it is unnecessary \cite{Frohlich:1980gj,Frohlich:1981yi}.}. However, this relation does not hold throughout the phase diagram \cite{Maas:2013aia}, and therefore here the name of the Higgs and $W$ will be reserved for the gauge-dependent elementary degrees of freedom. Still, this motivates the scale-setting procedure of section \ref{stech}. Note that possible excited states or resonances\footnote{Here, excited states will be states which are not the ground state, but stable, while resonances are unstable.} are not covered by this expansion, and can therefore not be determined using a perturbative analysis. The same applies to bound states, excited states or resonances in different quantum number channels. In these cases, the expansion has no leading term, and the first non-vanishing one is not a propagator, but a scattering state.

The FMS mechanism is hence both the justification of the success of perturbation theory in electroweak physics, and the reconciliation of field theory with the experimental observation of Higgs and $W$/$Z$ bosons. It shows how the electroweak physics is, in a sense, as confining as QCD, just with the exception that it is just a minor dressing effect. However, the FMS mechanism in its original form is based on an expansion  \cite{Frohlich:1980gj,Frohlich:1981yi}, as can be seen in \prefr{correl}{correl2}. It may therefore be of limited validity, as in fact both this work and \cite{Maas:2013aia} explicitly show. Fortunately, it applies to the standard model. However, this may no longer be true in extensions of the standard model \cite{Maas:2014nya,Maas:2015gma}, and therefore understanding its domain of validity is mandatory. Furthermore, the mechanism does not preclude the existence of internal excitations of the $0^+_1$ and the $1^-_3$, and thus effectively from an experimental point of view of the Higgs and the $W$, nor the presence of bound states in other quantum number channels \cite{Maas:2012tj}. Such states would mimic new physics. E.\ g., an excited $0^+_1$ has the same signature as a second Higgs, an excited $1^-_3$ as a $W'$. The other channels investigated, $0^-_1$, and $2^+_1$, correspond, e.\ g., to a pseudoscalar Higgs and a Kaluza-Klein graviton. By virtue of the expansion they will be much weaker coupled, and would therefore likely be compatible with experiment. Excluding them is hence a necessity to avoid false positive observations of new physics.

That said, a remaining problem is that there are three independent parameters in the theory. Thus, after fixing the $0^+_1$ and $1^-_3$ masses, e.\ g.\ to the experimental results, a third external input is necessary. At the current time, no quantity is both experimentally and theoretically in lattice terms good enough under control to serve as this input parameter. However, the present theory is anyway not quantitatively comparable to the standard model, as the $W$-$Z$ mass splitting is missing, and the gauge coupling runs much faster than in the standard model \cite{Maas:2013aia}. Hence a quantitative comparison to the standard model is of limited reliability. Here, therefore, the aim is to gain a qualitative understanding of the phase diagram and the bound state dynamics. For an alternative approach, see e.\ g.\ \cite{Wurtz:2013ova}.

\begin{figure}
\centering
\includegraphics[width=\linewidth]{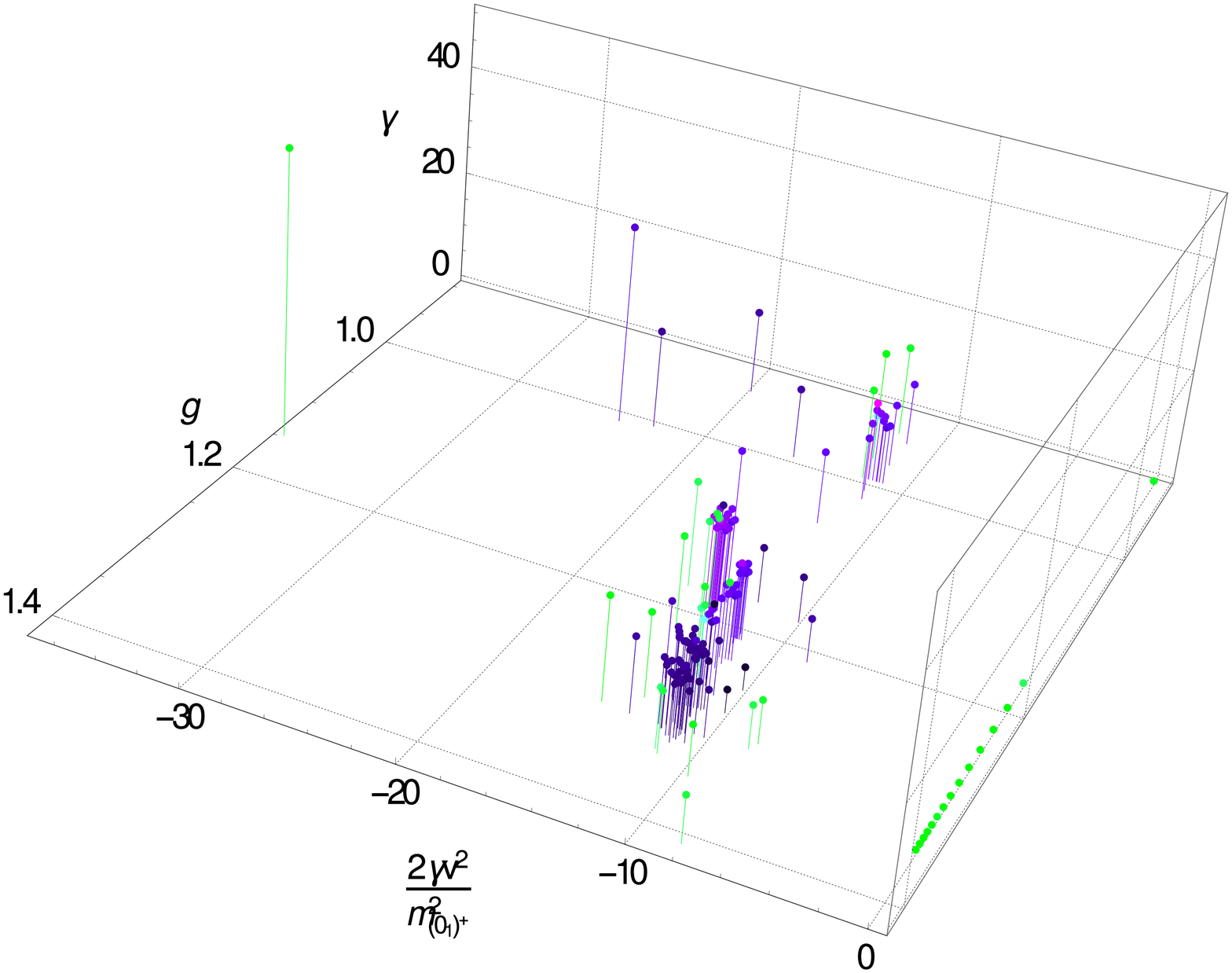}\\
\includegraphics[width=\linewidth]{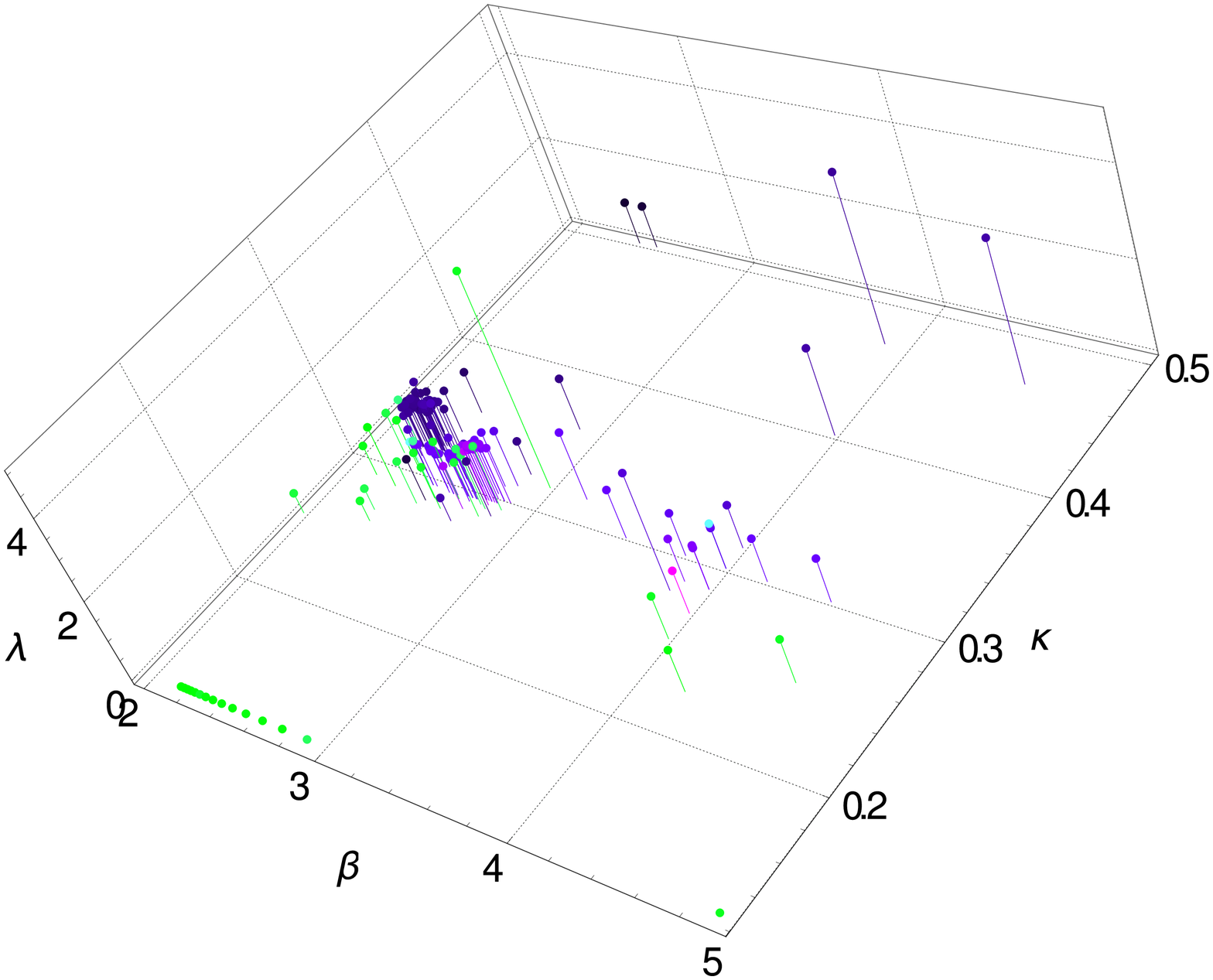}
\caption{\label{lcp}The top plot shows the phase diagram of Yang-Mills-Higgs theory as a function of bare gauge coupling, Higgs 4-point coupling, and the bare Higgs mass in units of the $0^+_1$ mass. Green points are confinement-like, and purple points are Higgs-like. The lighter the points, the smaller is the lattice spacing. The bottom plot shows the same in terms of the lattice bare parameters of inverse gauge coupling, hopping parameter, and four-Higgs coupling, see \pref{action} for their relation to the continuum parameters.}
\end{figure}

The part of the phase diagram covered here is shown in figure \ref{lcp}. The phase diagram disconnects into two parts, the HLD and the QLD. Due to the additive mass renormalization, the QLD region persists even deep into the negative $m_0^2$ region, where classically already the Higgs effect would be operative. In the following the spectrum will be studied in the various regions in more detail.

As it is not entirely trivial to follow the LCPs, due to the fine-tuning problem especially in $\kappa$, at the current time only a very limited amount of different lattice spacing effects can be studied. This is further complicated by the fact that most ground states are unstable or even scattering states, and hence there are often not enough observables to parameterize the trajectories. Hence, below classes of LCPs, rather than individual LCPs, will be discussed.

\section{Results}\label{sspectra}

Though the phase diagram is not disjunct in different phases, it can still be worthwhile to classify different regions, which exhibit quantitatively a similar behavior. This behavior is expected to be best characterized by the lightest excitation in the spectrum. In all cases investigated here, this excitation will be either in the $0_1^+$ or in the $1^-_3$ channel. Inspired by the standard model situation, the mass of the $1^-_3$ will always be set to 80.375 GeV, and the different domains will then be characterized by the mass of the $0^+_1$ mass relative to this scale. In ascending order, this will be 
\begin{itemize}
 \item the QLD with $m_{0^+_1}<75$ GeV
 \item the cross-over region with 75 GeV$<m_{0^+_1}<85$ GeV
 \item the light-Higgs region with 85 GeV$<m_{0^+_1}<115$ GeV
 \item the physical-Higgs region with 115 GeV$<m_{0^+_1}<135$ GeV
 \item the heavy-Higgs region with 135 GeV$<m_{0^+_1}<155$ GeV
 \item the threshold region with 155 GeV$<m_{0^+_1}<170$ GeV
 \item an anomalous region. This is a set of points in the phase diagram where with the operator basis employed the lowest level in the $0^+_1$ channel is above 170 GeV. This implies that for some reason the lowest scattering state could not be identified.
\end{itemize}
As already the selection shows, the regions cannot be distinguished by where perturbation theory works. Perturbation theory should work for not too large gauge and 4-Higgs couplings, but irrespective of the relative masses of the $0^+_1$ and $1^-_3$, viz.\ the Higgs and the $W$. This is not true, and rather the regions are characterized where the FMS mechanism works, which turns out to be essentially the region with roughly 80 GeV $<m_{0^+_1}<160$ GeV. Why this region is still splitted into three regions will be discussed below in section \ref{s:lh}. The situation at and above threshold is special, and will be addressed in sections \ref{s:th} and \ref{s:a}.

\subsection{The QCD-like domain}\label{s:qld}

\begin{figure}
\centering
\textsf{The QCD-like region}
\includegraphics[width=0.5\linewidth]{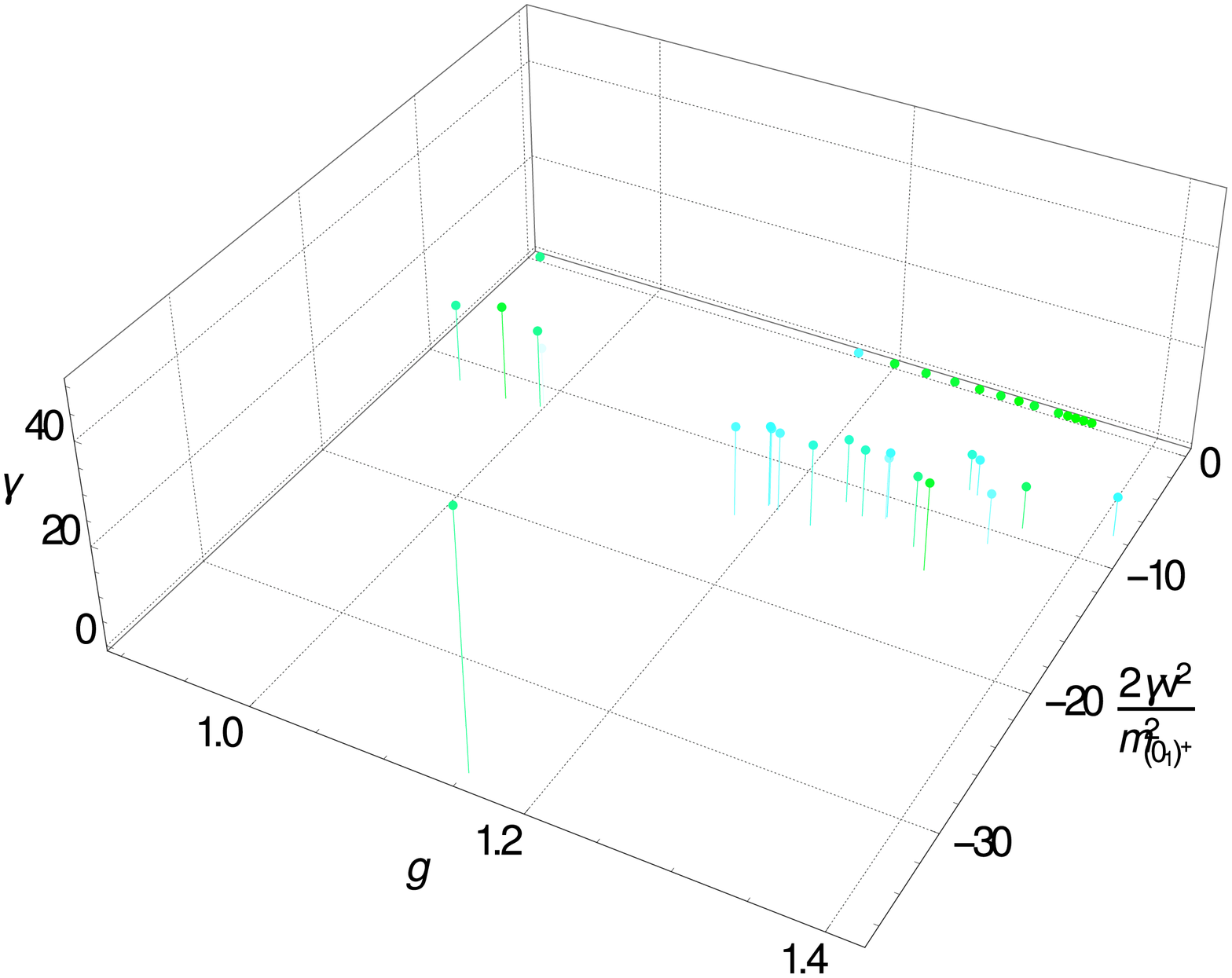}\includegraphics[width=0.5\linewidth]{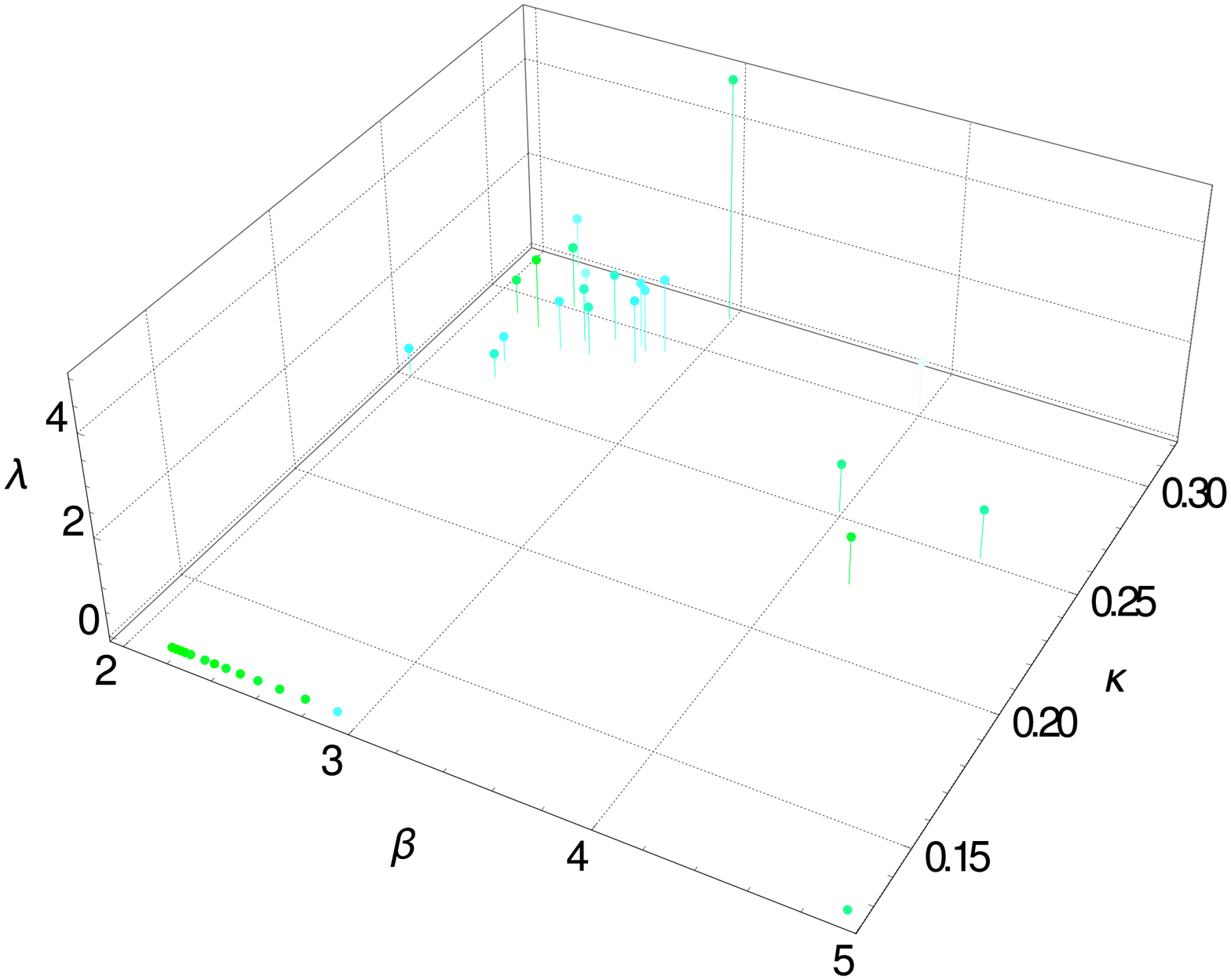}
\caption{\label{lcp-qld}The left-hand plot shows the bare continuum parameters qualifying as to be in the QLD, while the right-hand plot shows the same in terms of the lattice bare parameters.}
\end{figure}

The points which could be classified as to be in the QLD are shown in figure \ref{lcp-qld}. Especially in the continuum representation it is visible that for strong coupling these points are essentially only at not too negative bare Higgs mass squared, where more negative masses are possible at larger Higgs self-coupling. With a flat potential at the ultraviolet cutoff the system always ended up in the QLD, as naively expected.

\begin{figure}
\centering
\textsf{The spectrum in the QLD, normalized to the lightest mass}
\includegraphics[width=\linewidth]{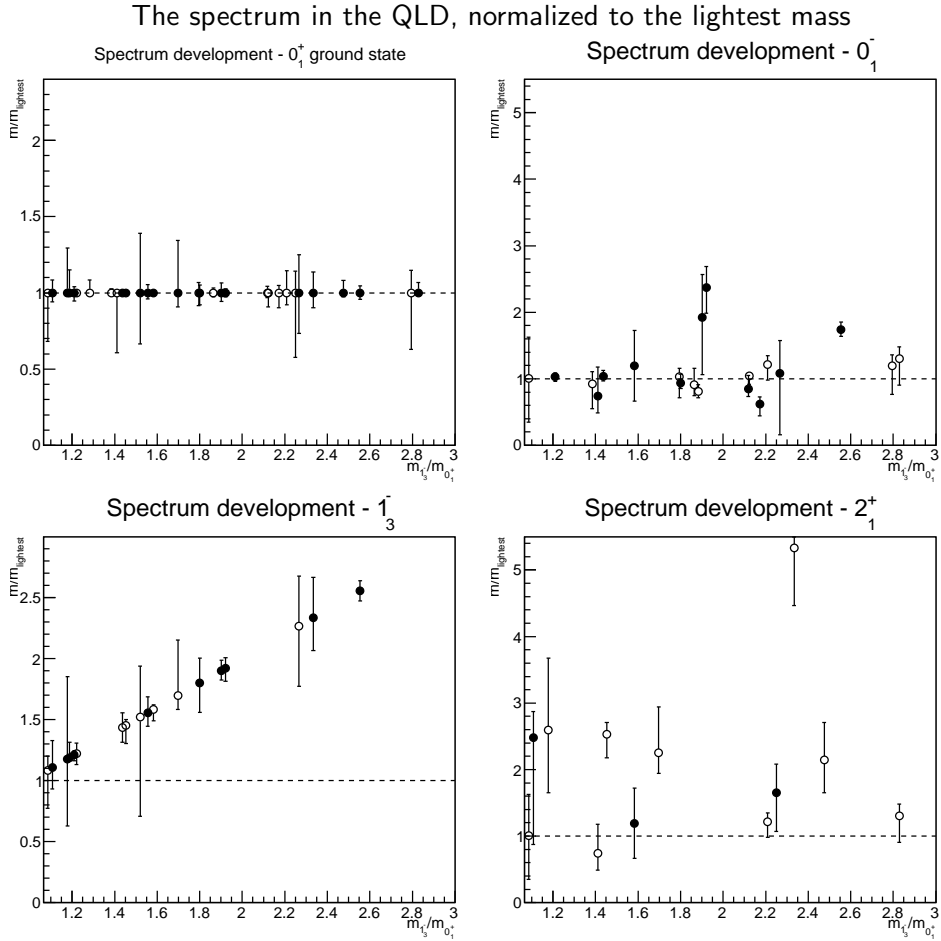}
\caption{\label{qld-gs}The ground states in the different quantum number channels in the QLD. Open symbols have an energy greater than one, but below 1.5, in lattice units, while closed symbols are below 1. All levels are normalized to the lightest mass.}
\end{figure}

\begin{figure}
\centering
\textsf{The spectrum in the QLD, normalized to the decay channels}
\includegraphics[width=\linewidth]{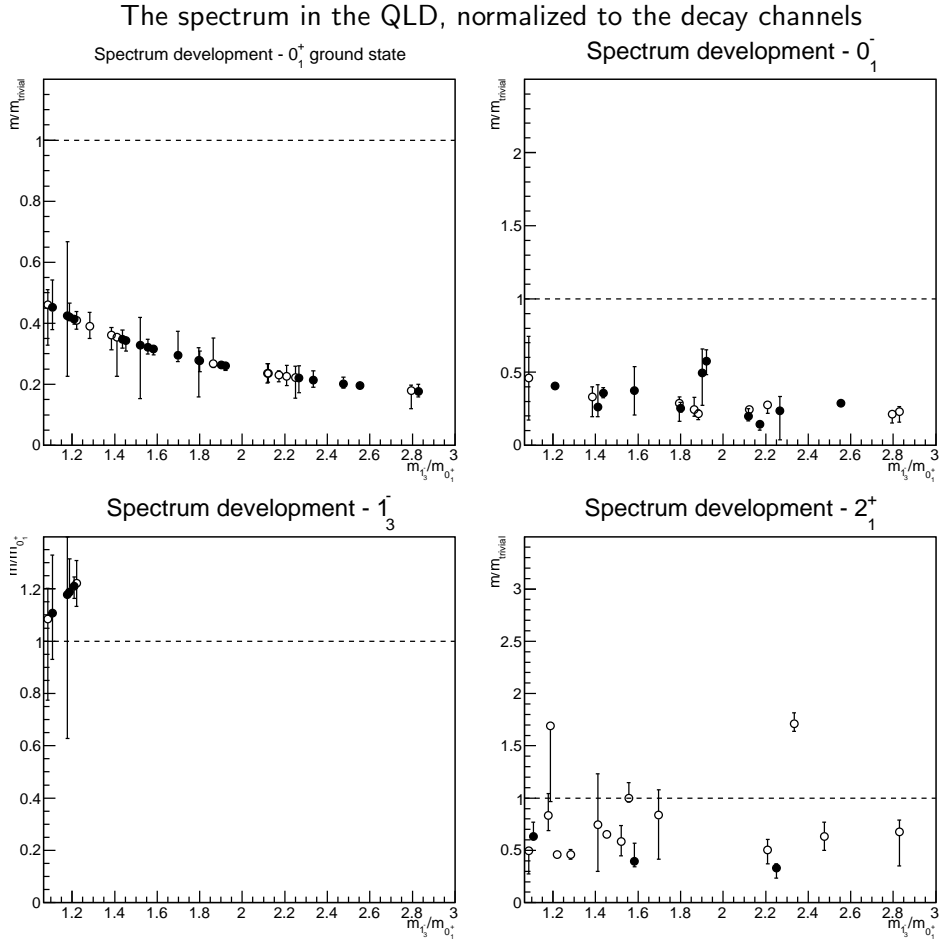}
\caption{\label{qld-gs-dc}The ground states in the different quantum number channels in the QLD. Open symbols have an energy greater than one, but below 1.5, in lattice units, while closed symbols are below 1. For the normalization, see text.}
\end{figure}

To study the physics further, it is useful to characterize the physics in terms of the energy levels. The ground states, normalized to the lightest mass, i.\ e.\ $\min(m_{0^+_1},m_{1^-_3})$, are shown\footnote{Note that due to the normalization displaying the $0^+_1$ and $1^-_3$ channels is to some extent redundant. However, the normalization makes no statement about the size of the errors, and therefore showing these is the main reason to include both channels in the plots.} in figure \ref{qld-gs}, as a function of the mass ratio $m_{0_1^+}/m_{1_3^-}$. In addition, in figure\footnote{The error in the $x$-axis have been suppressed for clarity, but could be reinstated using the raw data in appendix \ref{numval}. In most cases, this error does not affect the results, but in some, especially on coarse lattices, the classification of the parameter sets is not beyond the statistical error.} \ref{qld-gs-dc}, the same levels are shown, but now normalized by a trivial mass $m_\text{trivial}$ in a particular way to characterize them:
\begin{itemize}
 \item The energy of the $0^+_1$ channel was divided by twice the lowest energy in the $1^-_3$ channel. It therefore becomes a monotonous increasing function describing the approach towards the threshold of decay into two $1^-_3$ particles. In the QLD, by definition, it is therefore always below 1/2.
 \item For the other channels, since they are at most degenerate, but never found lighter than the $0^+_1$ and the $1^-_3$ channels, the normalization is to the lowest of the possible two-body decay channels. The possible decay channels with the available channels are:
 \begin{itemize}
  \item For the $0^-_1$ channel: two $1^-_3$ in a $p$-wave
  \item For the $2^+_1$ channel: two $0^+_1$ in a $d$-wave; two $1^-_3$ in a $s$-wave; two $0^-_1$ in a $d$-wave
 \end{itemize}
\end{itemize}
\no Since no further operator in the triplet channel was considered, there was no possibility to construct decays of the ground-state in the $1^-_3$ channel. Rather, the relative mass to the $0^+_1$ channel, as the other important channel, will be given.

In no case the stability of the target levels is taken into account, since if any of these would be unstable, a different channel containing the decay products of this channel would anyhow be lighter. Hence, if the normalized values are below one, the ground state in the respective channel is stable against the considered decays.

The results from these two figures show that the low-energy physics is quite rich. In almost all cases in all channels the ground state is found significantly below one, and therefore the ground state is stable. In fact, in many cases the states have a similar mass as the lightest state, as can be seen in figure \ref{qld-gs}. Thus, the low energy physics appears as rich as expected from the analogy to QCD. Unfortunately, the results for the $2^+_1$ state are too noisy to really quantify the Yang-Mills-likeness of the theory in the QLD. Beyond the QLD the lightness of the $1^-_3$ will anyhow make the theory very different from Yang-Mills theory.

\begin{figure}
\centering
\textsf{2$^\text{nd}$ and 3$^\text{rd}$ levels in the QLD}
\includegraphics[width=\linewidth]{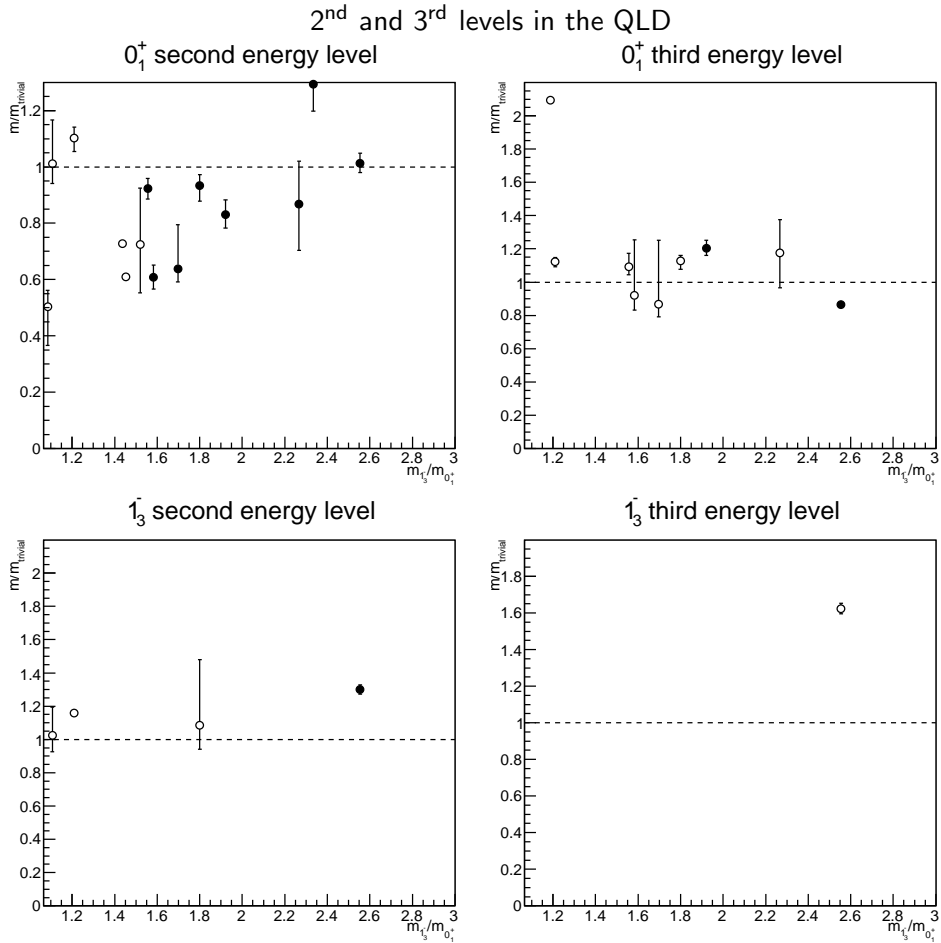}
\caption{\label{qld-higher}The two next states after the ground state in the $0^+_1$ channel (upper panels) and the $1^-_3$ channel (bottom panels) in the QLD. Open symbols have an energy greater than one, but below 1.5, in lattice units, while closed symbols are below 1. All levels are normalized to scattering states, as detailed in the text.}
\end{figure}

It is, however, also instructive to have a look at the higher-lying levels. These are shown in figure \ref{qld-higher}. The results are normalized to decays in the $0^+_1$ and $1^-_3$ channels, since the FMS mechanism suggests that they should be, at least later in the HLD, the dominant ones. Also, since the other channels are all not lighter than these, kinematical arguments support this. The normalizations, which for completeness includes also the ones used in the remainder of this section, are thus such that
\begin{itemize}
\item for the $0^+_1$, the higher levels are normalized to the lighter of either two $1^-_3$ or two $0^+_1$ in an $s$-wave, possibly with relative momenta, provided the $0^+_1$ ground state is stable against decays into two $1^-_3$. If not, only decays into two $1^-_3$ in an $s$-wave with an increasing amount of momenta in back-to-back kinematics are considered.
\item for the $1^-_3$ channel, in case of a stable $0^+_1$ ground state the decay channels of one $0^+_1$ and one $1^-_3$ in an $s$-wave and into three $1^-_3$ with two of them forming a $0^+_1$ state are considered. Otherwise, only the three-body decay is considered. For the higher levels, corresponding momenta in back-to-back kinematics are considered, for the three-body decay between two of the decay products with the third one at rest.
\end{itemize}
Note that, due to the inherent uncertainties of mass determinations, it was assumed that the decay channel was open, if 51\%, instead of 50\%, of the mass of the ground state was larger than the ground state mass of the $1^-_3$ state.

The results in figure \ref{qld-higher} show that most of the second states are close to, but below, one, and thus could still be just scattering states. There are some states with masses significantly below one in the $0^+_1$ channel, though none in the $1^-_3$ channel. These are potentially stable, but still much heavier than the ground state, at least 60\% heavier. That is not what is observed in QCD, here the excited states of, e.\ g.\, the nucleon are densely packed already below 50\% mass excess \cite{pdg}. This contrast becomes even more pronounced when looking at the, significantly more unreliable, third level. Here, essentially all states cluster close to one, and are therefore likely scattering states. Thus, even if there should be (stable) excited states in the $0^+_1$ channel, they are few. For the $1^-_3$ channel, there is so far only one candidate, and this one is unreliable and much above the expected trivial level, so little can be said.

In total, the physics in the QLD is not entirely what is expected from the comparison to QCD. Though there are stable ground states in most channels, there appear to be few, if at all, excited states. This would be contrary to the naive expectations, and requires further investigations, especially using multiple volumes.

\subsection{The crossover region}\label{s:cor}

\begin{figure}
\centering
\textsf{The crossover region}
\includegraphics[width=0.5\linewidth]{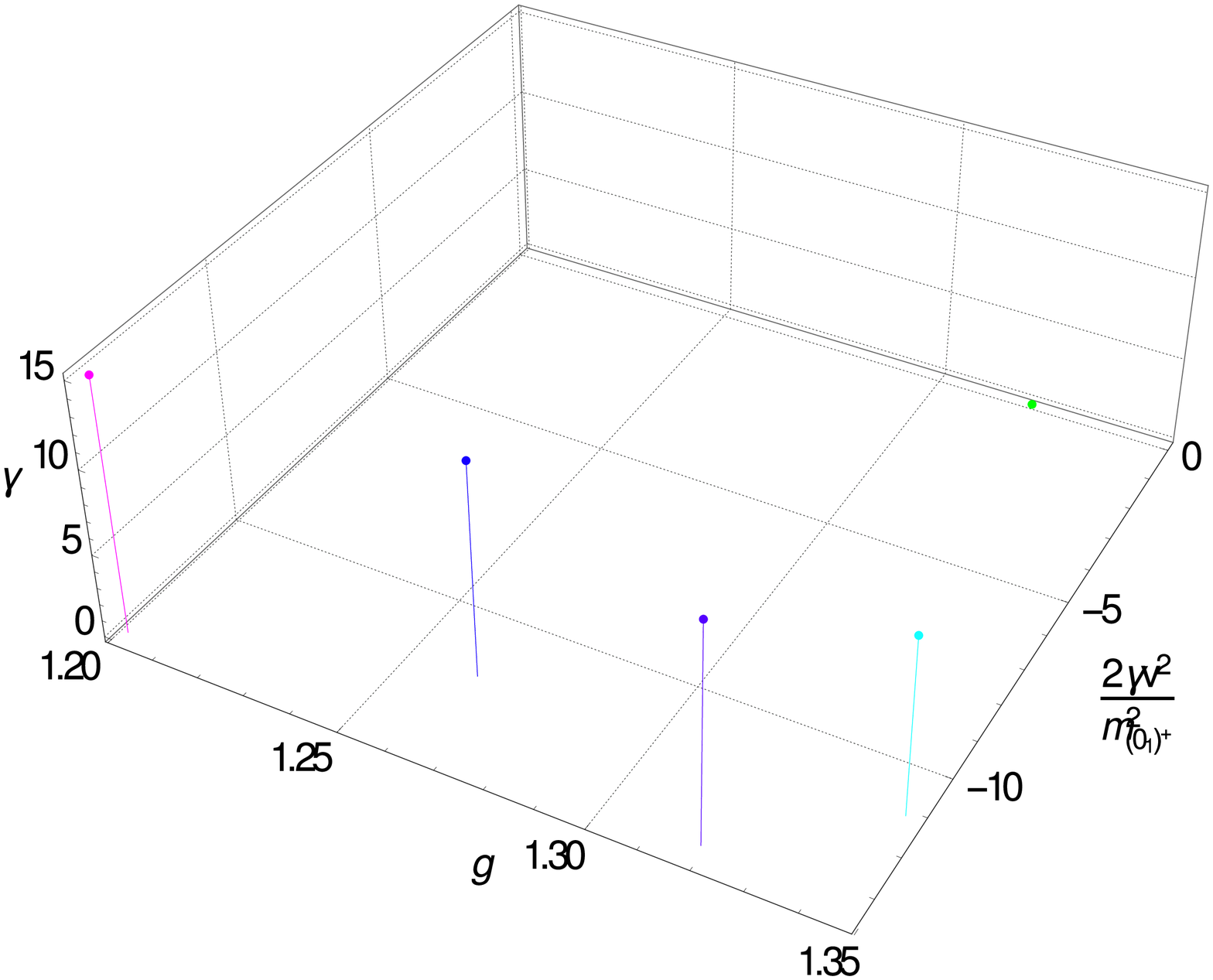}\includegraphics[width=0.5\linewidth]{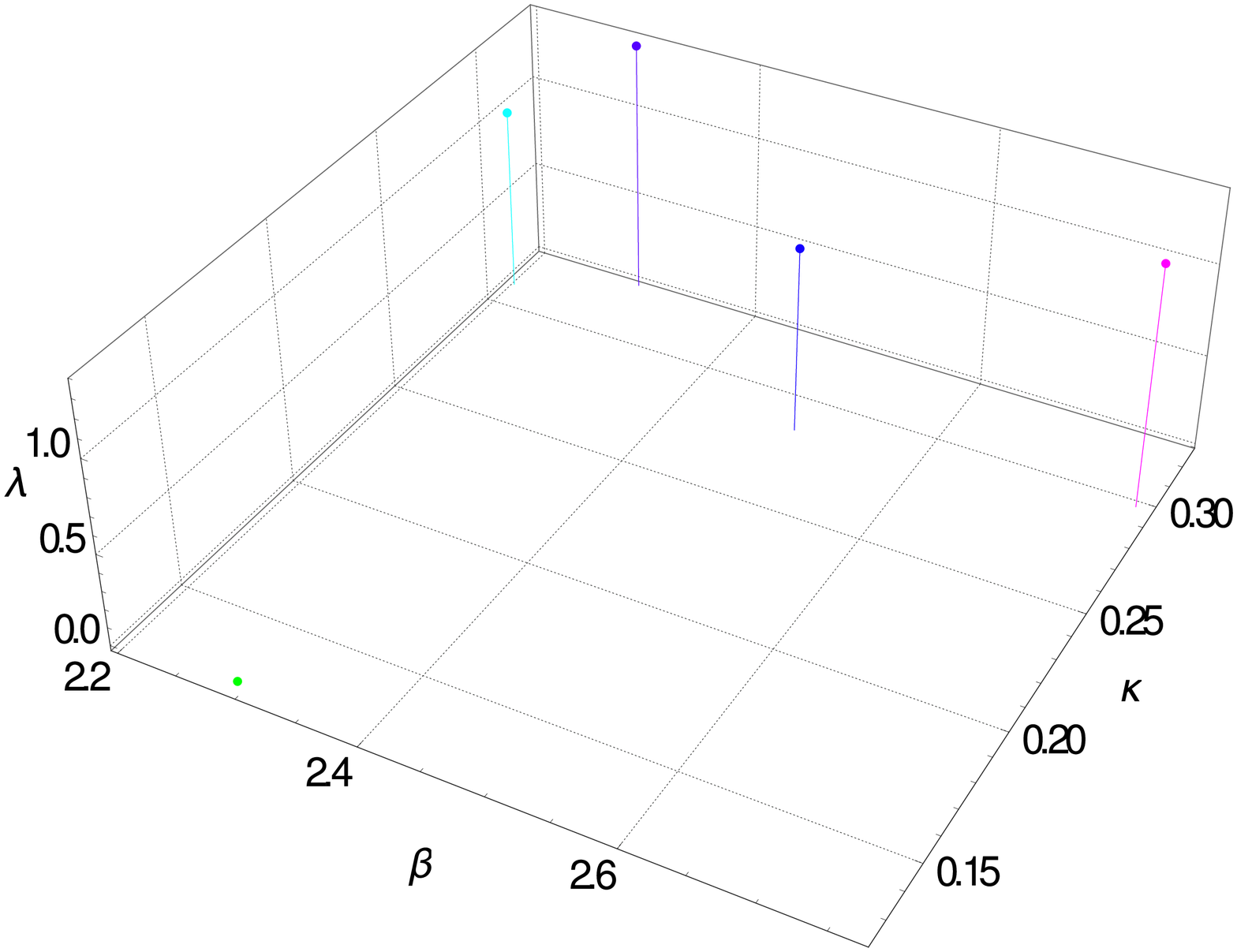}
\caption{\label{lcp-cor}The left-hand plot shows the bare continuum parameters qualifying as to be in the cross-over region, while the right-hand plot shows the same in terms of the lattice bare parameters.}
\end{figure}

The second interesting region is the one where the ground states in the $0^+_1$ and the $1^-_3$ channels are roughly of equal size. Here, this will be taken to be the range 75 GeV$<m_{0^+_1}<85$ GeV, if the $1^-_3$ ground state is assigned the $W$ mass of 80.375 GeV. The corresponding points in the phase diagram are shown in figure \ref{lcp-cor}. Only a few points are currently available, and they are located in a rather narrow region of the phase diagram. If the cross-over indeed becomes sharper with increasing $\beta$, as various studies suggests \cite{Caudy:2007sf,Bonati:2009pf}, it should not be too surprising that none are found here: It just becomes very hard to precisely find these points.

\begin{figure}
\centering
\textsf{The spectrum in the crossover region, normalized to the lightest mass}
\includegraphics[width=\linewidth]{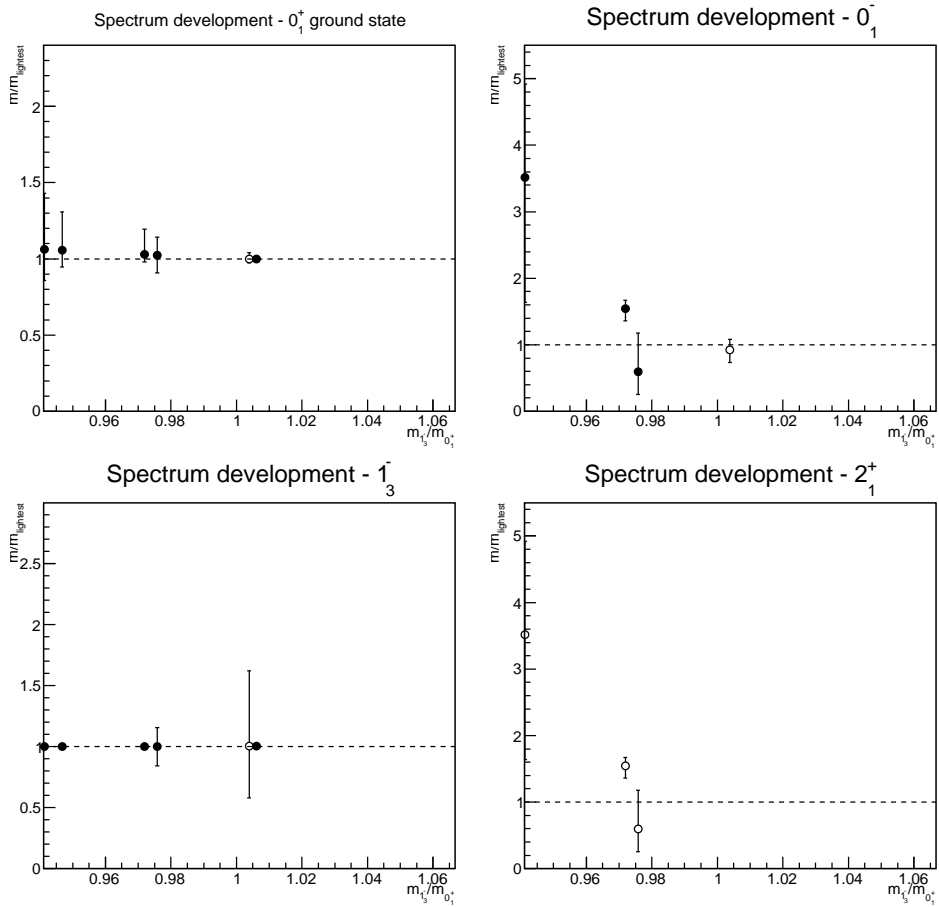}
\caption{\label{cor-gs}The ground states in the different quantum number channels in the crossover region. Open symbols have an energy greater than one, but below 1.5, in lattice units, while closed symbols are below 1. All levels are normalized to the lightest mass.}
\end{figure}

\begin{figure}
\centering
\textsf{The spectrum in the crossover region, normalized to the decay channels}
\includegraphics[width=\linewidth]{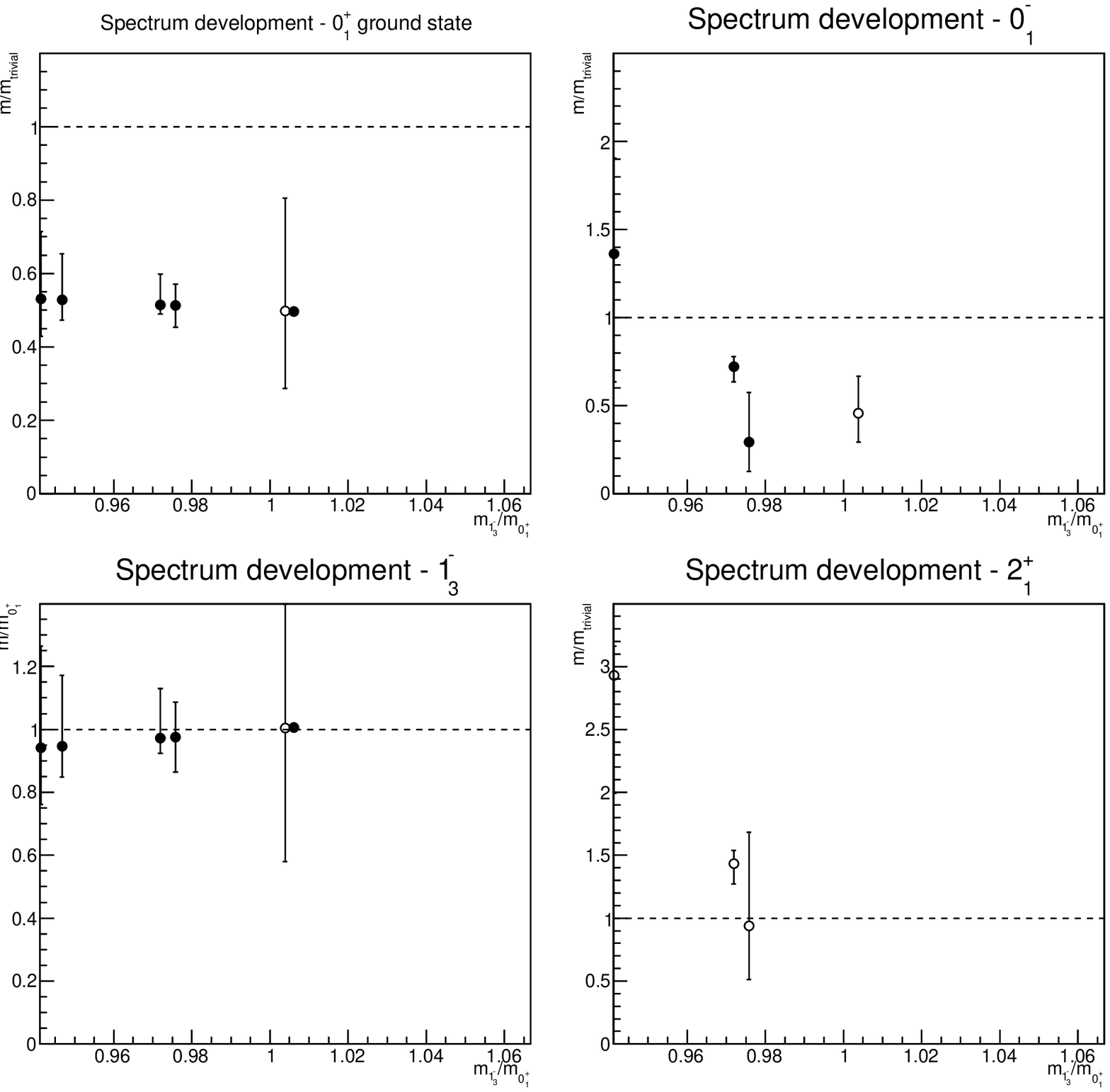}
\caption{\label{cor-gs-dc}The ground states in the different quantum number channels in the crossover region. Open symbols have an energy greater than one, but below 1.5, in lattice units, while closed symbols are below 1. For the normalization, see text.}
\end{figure}

\begin{figure}
\centering
\textsf{2$^\text{nd}$ and 3$^\text{rd}$ levels in the crossover region}
\includegraphics[width=\linewidth]{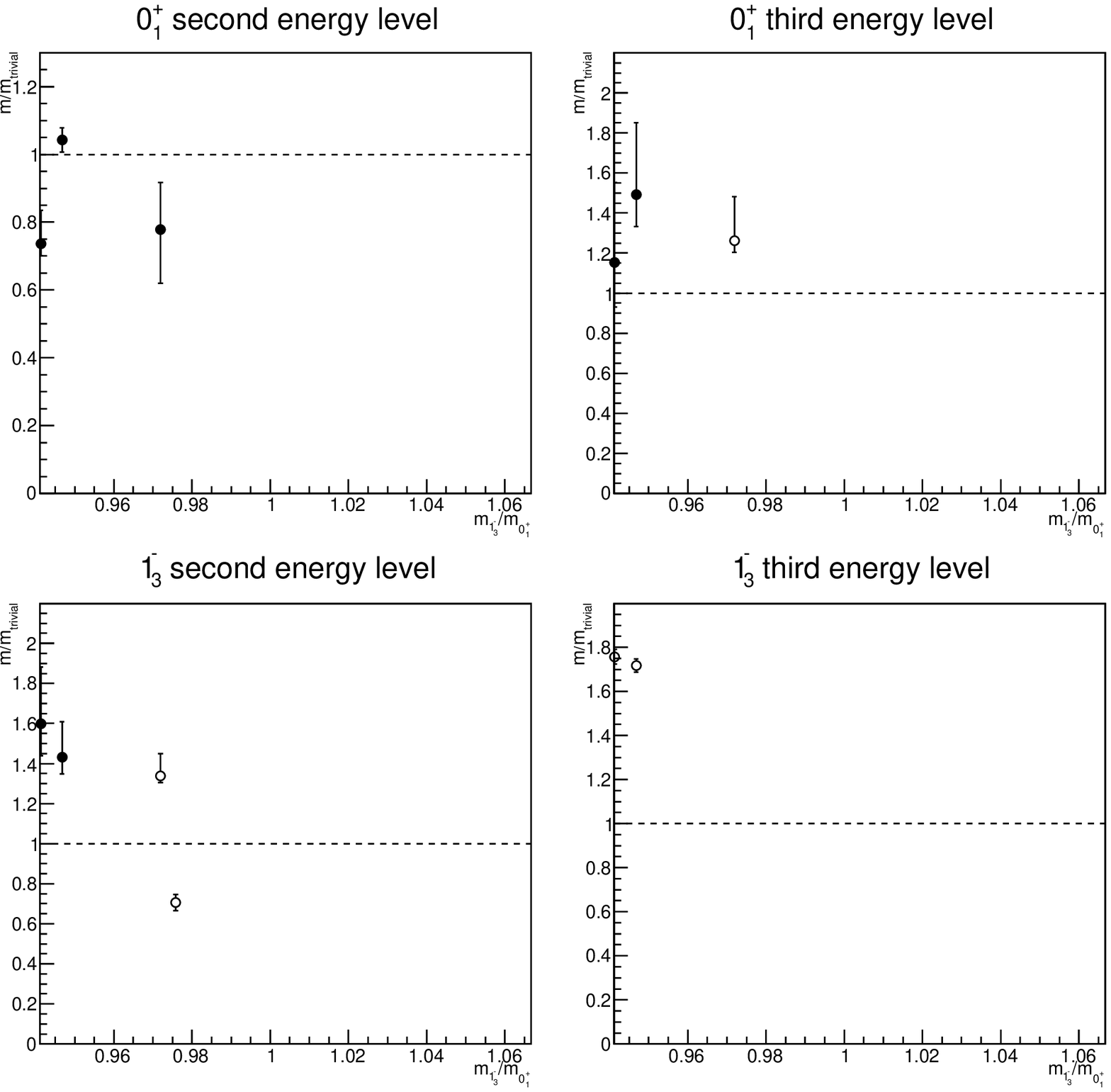}
\caption{\label{cor-higher}The two next states after the ground state in the $0^+_1$ channel (upper panels) and the $1^-_3$ channel (bottom panels) in the crossover region. Open symbols have an energy greater than one, but below 1.5, in lattice units, while closed symbols are below 1. All levels are normalized to scattering states, as detailed in the text.}
\end{figure}

The results for the spectrum are shown in figures \ref{cor-gs}-\ref{cor-higher}. The ground state masses are now almost always reliable. In the $0^-_1$ and $2^+_1$ channels there appears to be an abrupt rise in the mass of the ground states when moving into the HLD. While the $2^+_1$ channel is always above threshold, this development shifts the $0^-_1$ channel above threshold. The situation for the higher levels in figure \ref{cor-higher} is that the higher states in the $0^+_1$ are essentially compatible within 1-3$\sigma$ with scattering states. In the $1^-_3$ channel, they are substantially above the threshold, and therefore here either the determination is not good enough, or other decay channels still play a role.

At any rate, the cross-over seems to leave a strong imprint on the other channels, and there the states become quickly unbound. That is certainly what is naively expected when moving from the QLD into the HLD, if the latter is weakly interacting.

\subsection{Light Higgs}\label{s:lh}

In this and the next two subsections the situation will be investigated for a stable $0^+_1$ ground state heavier than the $1^-_3$, and thus inside the HLD, but separately in three different mass regions. One is below the physical mass region, one above, and the last one is the physical mass region. A priori, it seems to be odd to distinguish these three regions, as there is no obvious reason why the physical Higgs mass should be distinct. However, there are several arguments that the physical Higgs mass is actually distinguished due to the ultraviolet properties of the theory \cite{Shaposhnikov:2009pv,Gies:2014xha,Gerhold:2010bh,Gies:2013pma,Degrassi:2012ry}. Hence, the three situations will be studied here separately, starting with the case of a light Higgs, i.\ e.\ for 85 GeV$<m_{0^+_1}<115$ GeV. The upper bound comes from the fact that due to the absence of the $W$-$Z$ mass splitting the masses can be expected to be off by (at least) 10 GeV, and thus the physical region is taken to be a $\pm$10 GeV interval around the physical Higgs mass of about 126 GeV \cite{pdg}.

\begin{figure}
\centering
\textsf{The light-Higgs region}
\includegraphics[width=0.5\linewidth]{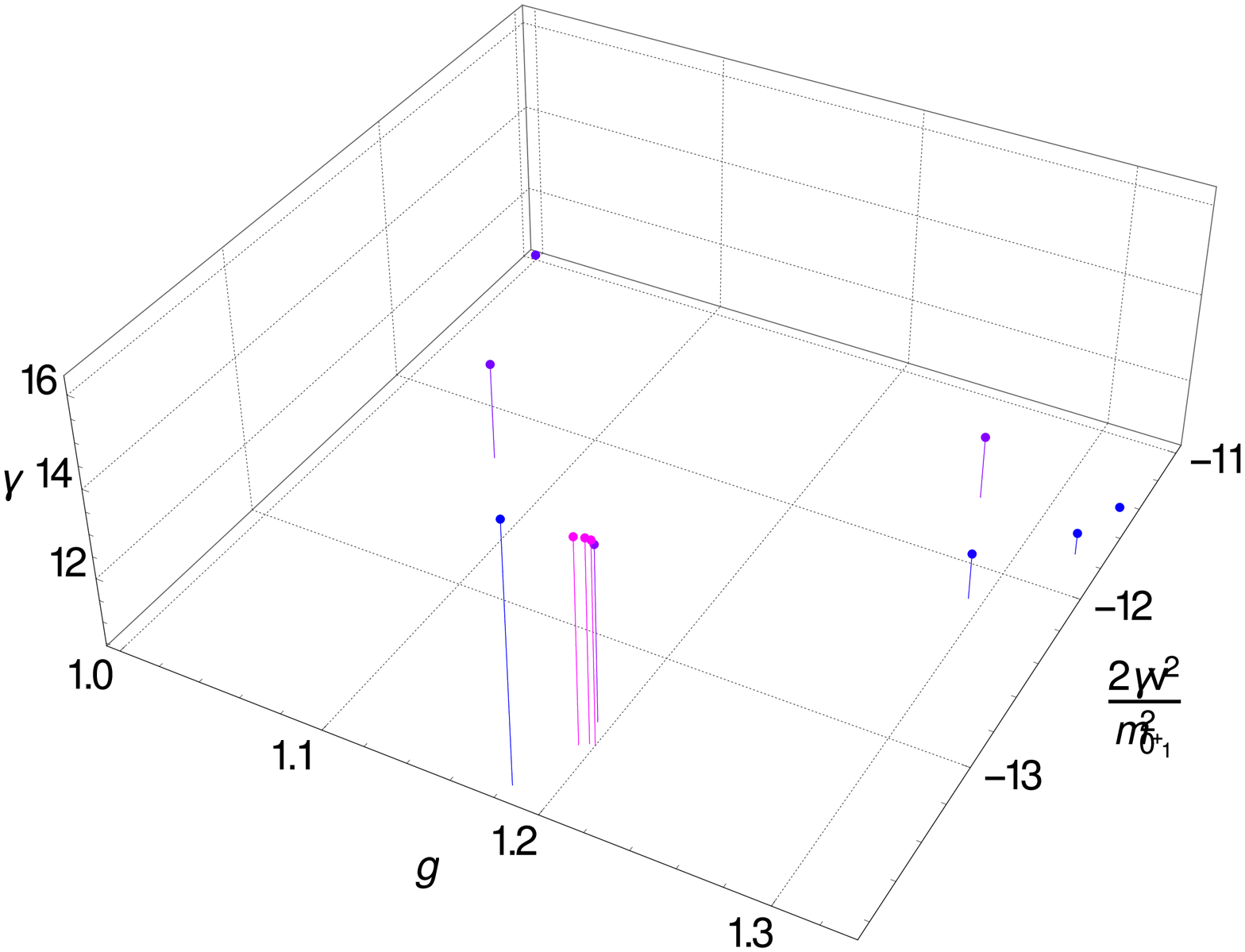}\includegraphics[width=0.5\linewidth]{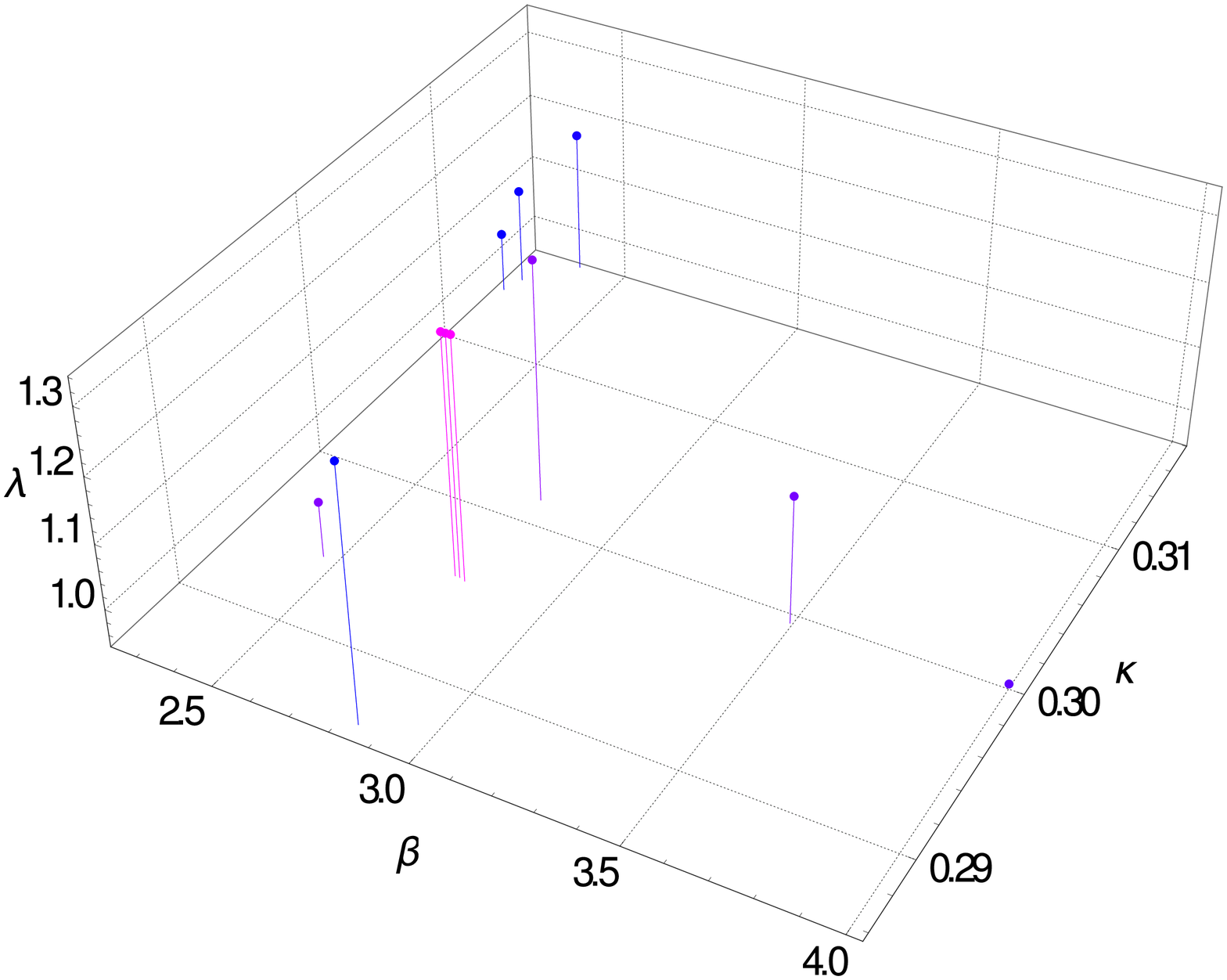}
\caption{\label{lcp-lh}The left-hand plot shows the bare continuum parameters exhibiting a light $0^+_1$ in the HLD, while the right-hand plot shows the same in terms of the lattice bare parameters.}
\end{figure}

The parameters of the phase diagram where a light $0^+_1$ in the HLD has been found are shown in figure \ref{lcp-lh}. In contrast to the cross-over region, this region extends to much smaller, but not too small, gauge couplings, though the majority still clusters around intermediate values of the coupling. Also, the regions populate only a rather narrow strip in both the hopping parameter and the self-coupling.

\begin{figure}
\centering
\large{\textsf{The spectrum in the light-Higgs region, normalized to the lightest mass}}
\includegraphics[width=\linewidth]{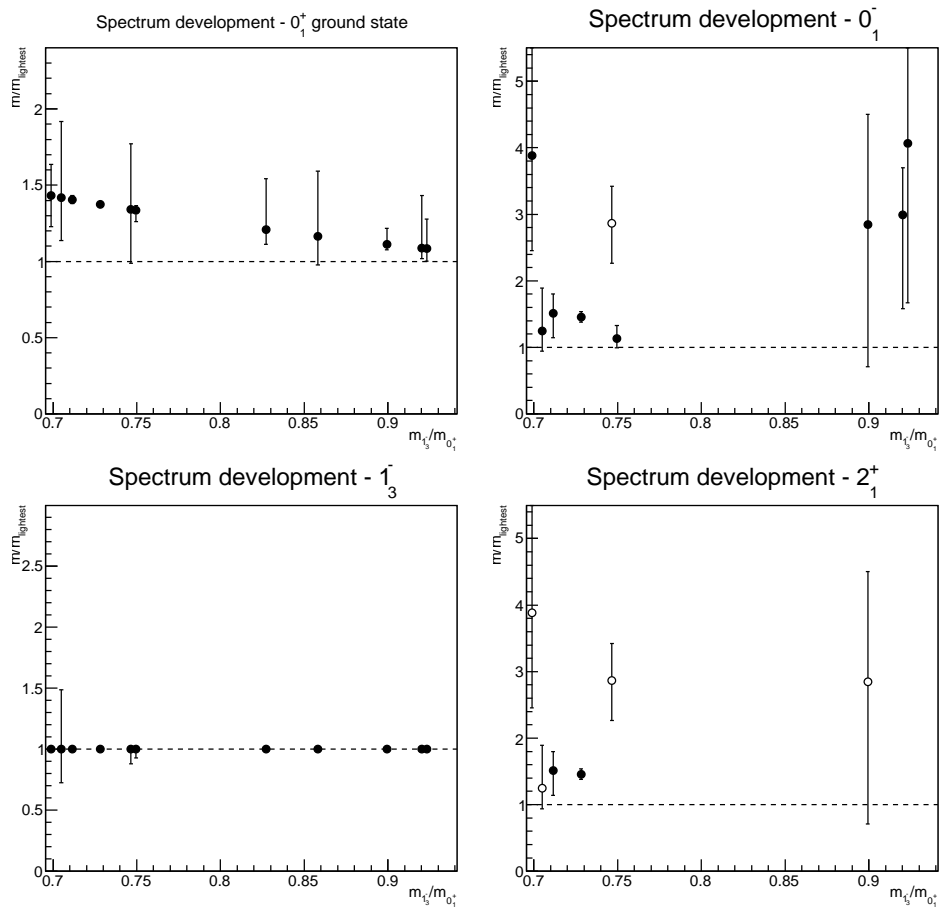}
\caption{\label{lh-gs}The ground states in the different quantum number channels. Open symbols have an energy greater than one, but below 1.5, in lattice units, while closed symbols are below 1. All levels are normalized to the lightest mass.}
\end{figure}

\begin{figure}
\centering
\textsf{The spectrum in the light-Higgs region, normalized to the decay channels}
\includegraphics[width=\linewidth]{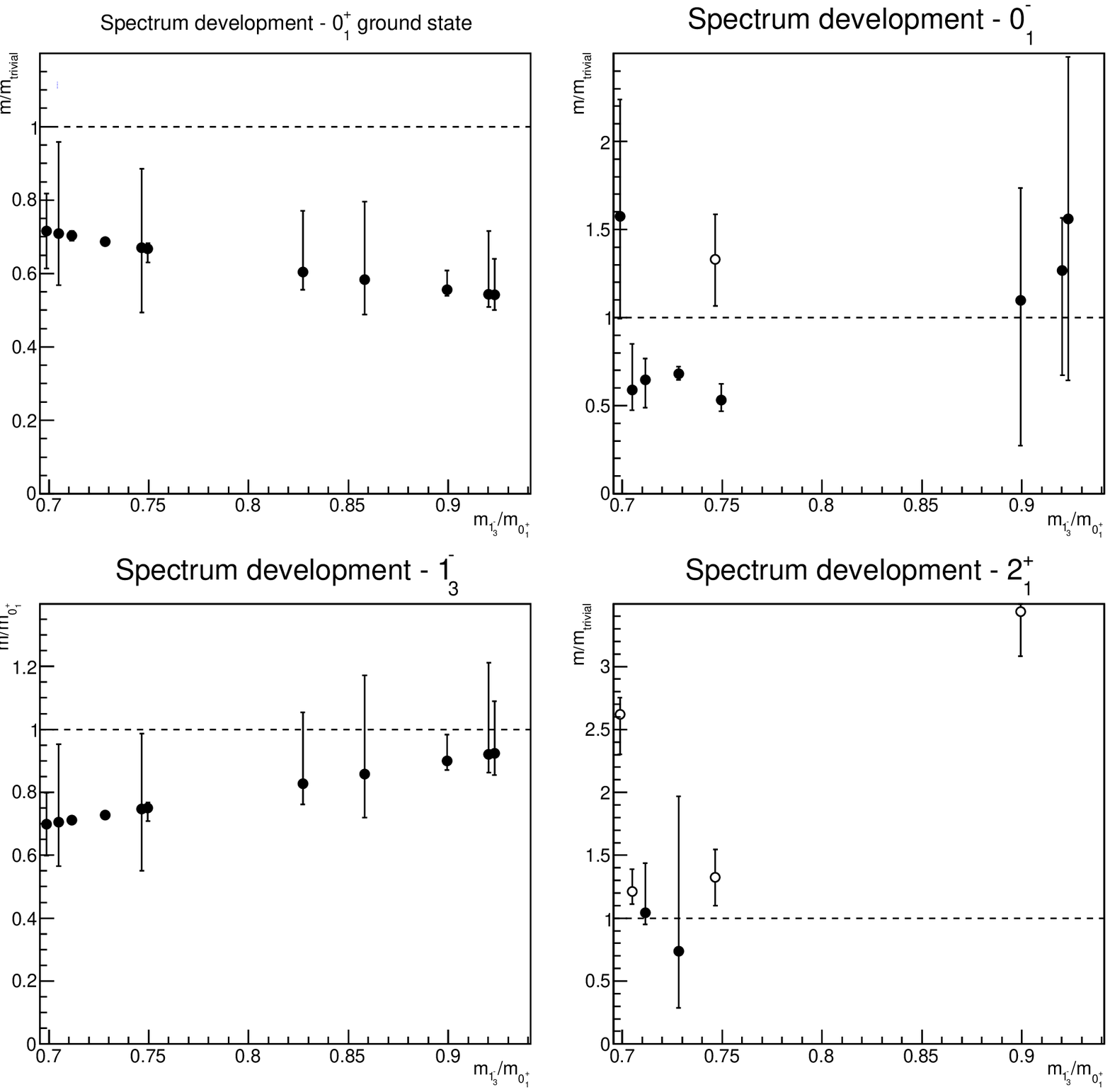}
\caption{\label{lh-gs-dc}The ground states in the different quantum number channels. Open symbols have an energy greater than one, but below 1.5, in lattice units, while closed symbols are below 1. For the normalization, see text.}
\end{figure}

The ground-state spectrum, normalized to the lightest mass and the elastic threshold, is shown in figures \ref{lh-gs} and \ref{lh-gs-dc}, respectively. Interestingly, the states seem to become more stable with increasing $0^+_1$ mass, especially the pseudoscalar appears, though within large errors, to become lighter. In the $2^+_1$ channel, the state is generally around or above the elastic threshold, and thus within errors compatible with scattering states.

\begin{figure}
\centering
\textsf{2$^\text{nd}$ and 3$^\text{rd}$ levels in the light-Higgs region}
\includegraphics[width=\linewidth]{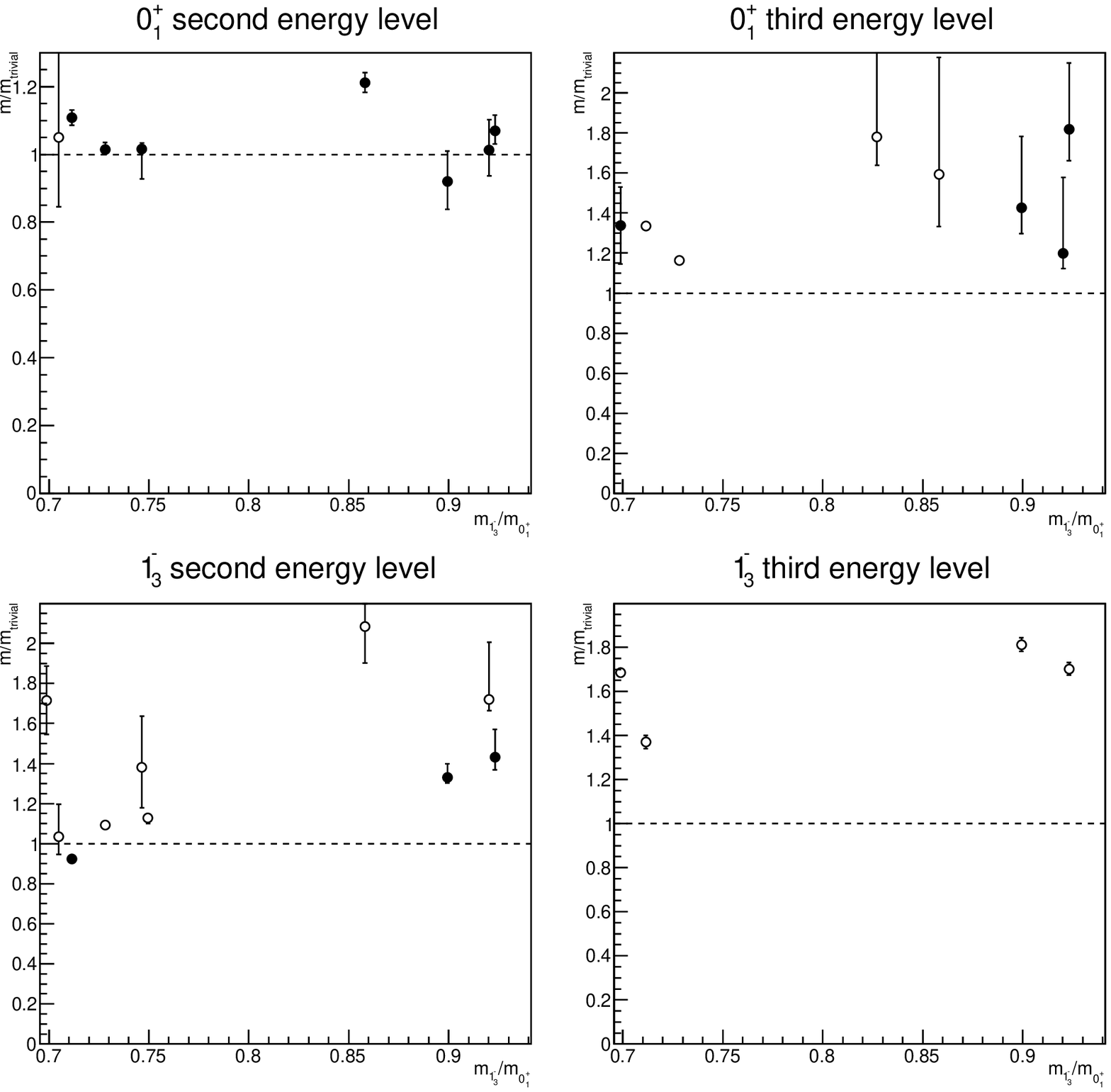}
\caption{\label{lh-higher}The two next states after the ground state in the $0^+_1$ channel (upper panels) and the $1^-_3$ channel (bottom panels). Open symbols have an energy greater than one, but below 1.5, in lattice units, while closed symbols are below 1. All levels are normalized to scattering states, as detailed in the text.}
\end{figure}

This feature is also shared by the higher levels in the $0^+_1$ and $1^-_3$ channels, shown in figure \ref{lh-higher}. The second level of the $0^+_1$ is either at threshold, or somewhat heavier. The third state is also always at most heavier than the inelastic threshold. Hence, there does not appear to be any obvious low-lying excited state. The situation for the second state in the $1^-_3$ channel is similar. Though generally decreasing, the results cluster around the threshold, and are therefore likely scattering states. The third level is above the next scattering state, indicating still strong systematic uncertainties.

\subsection{Physical Higgs}\label{s:ph}

As noted in the previous section \ref{s:lh}, here the physical mass region of 115 GeV$<m_{0^+_1}<135$ GeV will be investigated. After realization that the FMS mechanism implies the presence of bound states, the possibility of finding an internal excitation of the $0^+_1$ state dual to the Higgs was, of course, an interesting option, as any such state could mimic new physics and therefore form a genuinely new background to new physics searches \cite{Maas:2012tj}. Though an investigation of only Yang-Mills-Higgs theory can hardly provide any quantitative results, the observation of such internal excitations in this case would significantly strengthen this case, given that this theory is weaker interacting than the standard model \cite{Maas:2013aia}. After all, since bound states are genuinely not part of perturbation theory, at some point deviations between the FMS scenario and the perturbative results should always surface, and this would be an ideal candidate as it is a rather clean signature.

\begin{figure}
\centering
\textsf{The physical-Higgs region}
\includegraphics[width=0.5\linewidth]{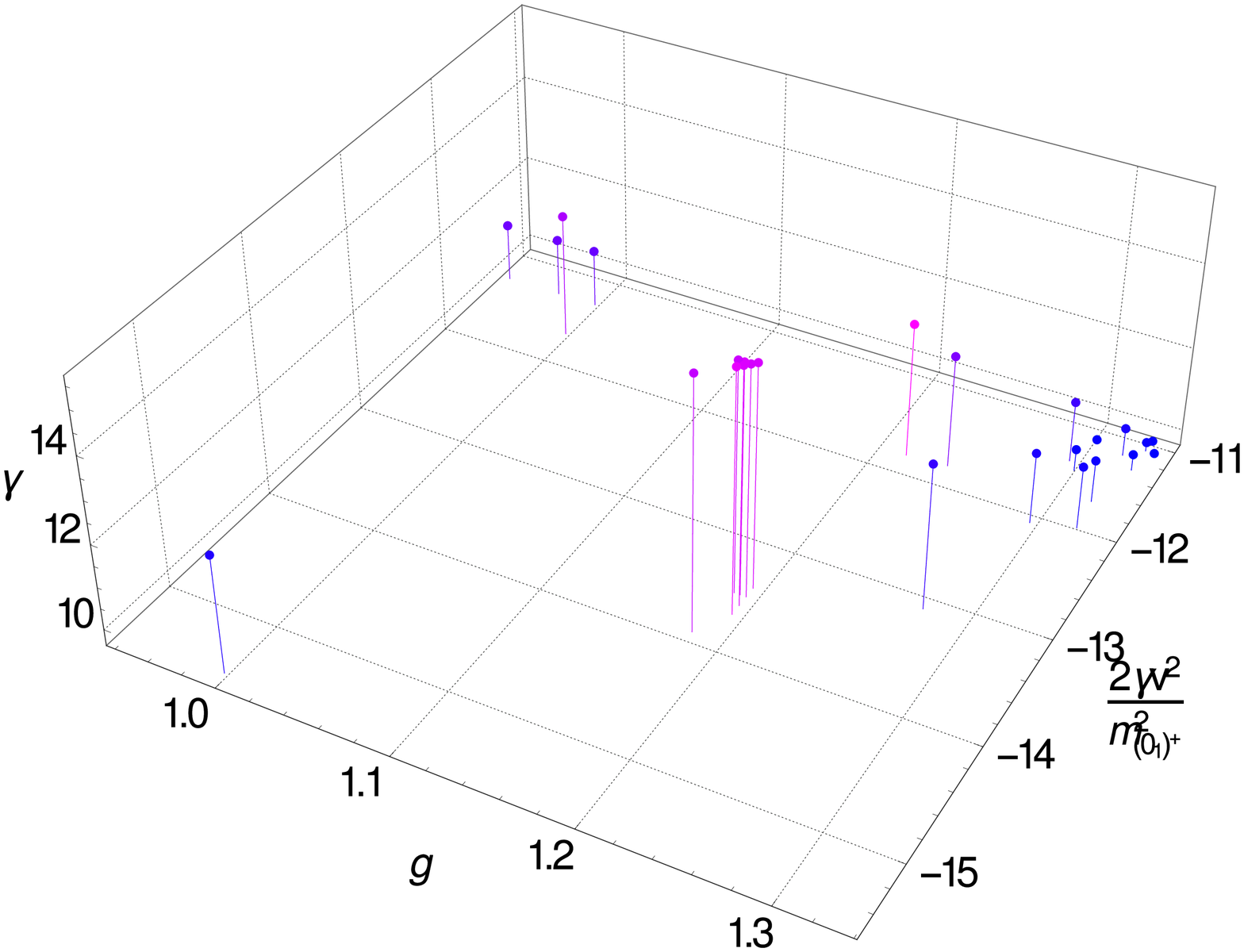}\includegraphics[width=0.5\linewidth]{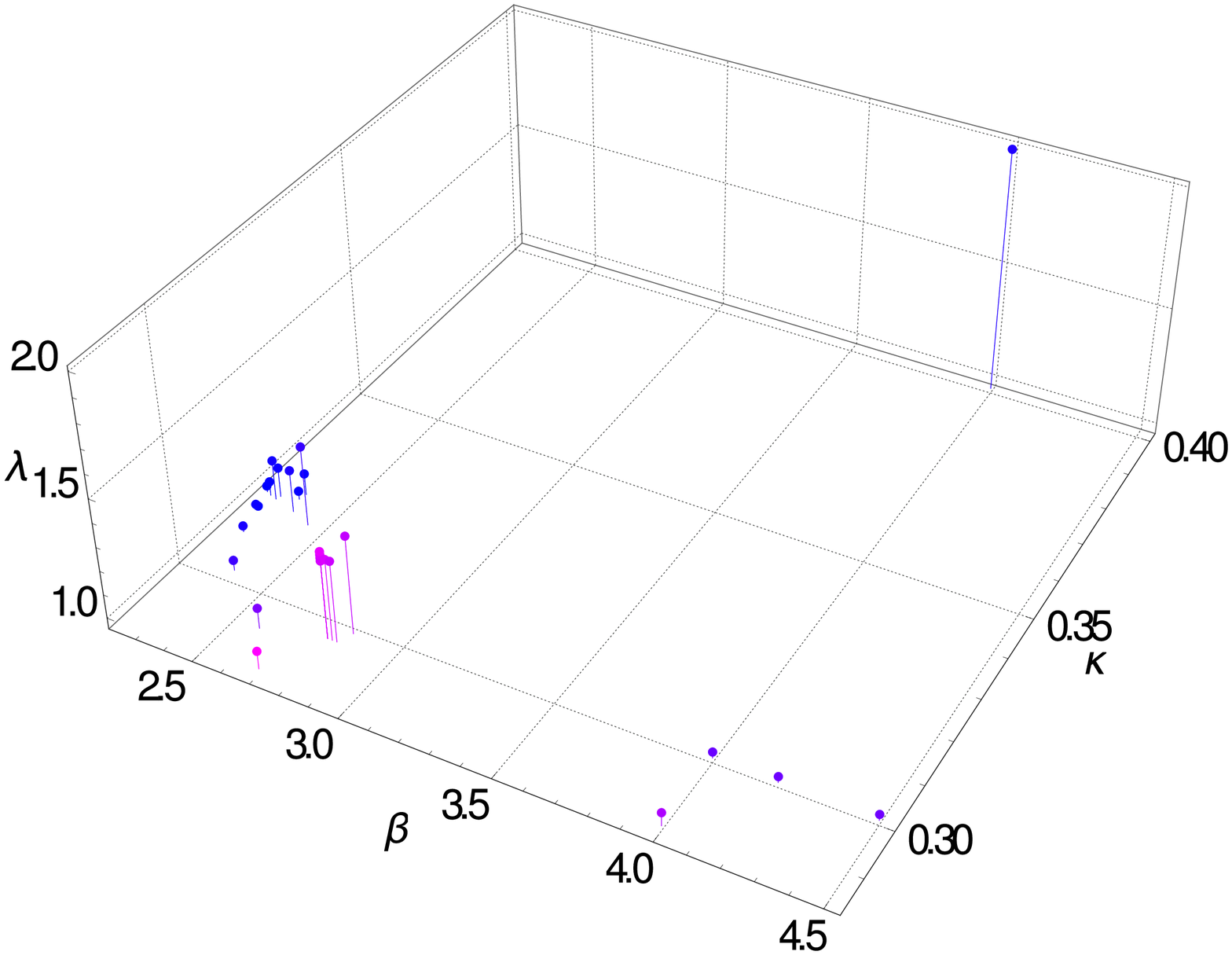}
\caption{\label{lcp-ph}The left-hand plot shows the bare continuum parameters exhibiting a $0^+_1$ ground state in the HLD within the physical mass window, while the right-hand plot shows the same in terms of the lattice bare parameters.}
\end{figure}

The points in the phase diagram satisfying the selection criterion are shown in figure \ref{lcp-ph}. They spread in the phase diagram, and have also been found at substantially lower bare gauge and 4-Higgs coupling than investigated here \cite{Wurtz:2013ova}.

\begin{figure}
\centering
\textsf{The spectrum in the physical-Higgs region, normalized to the lightest mass}
\includegraphics[width=\linewidth]{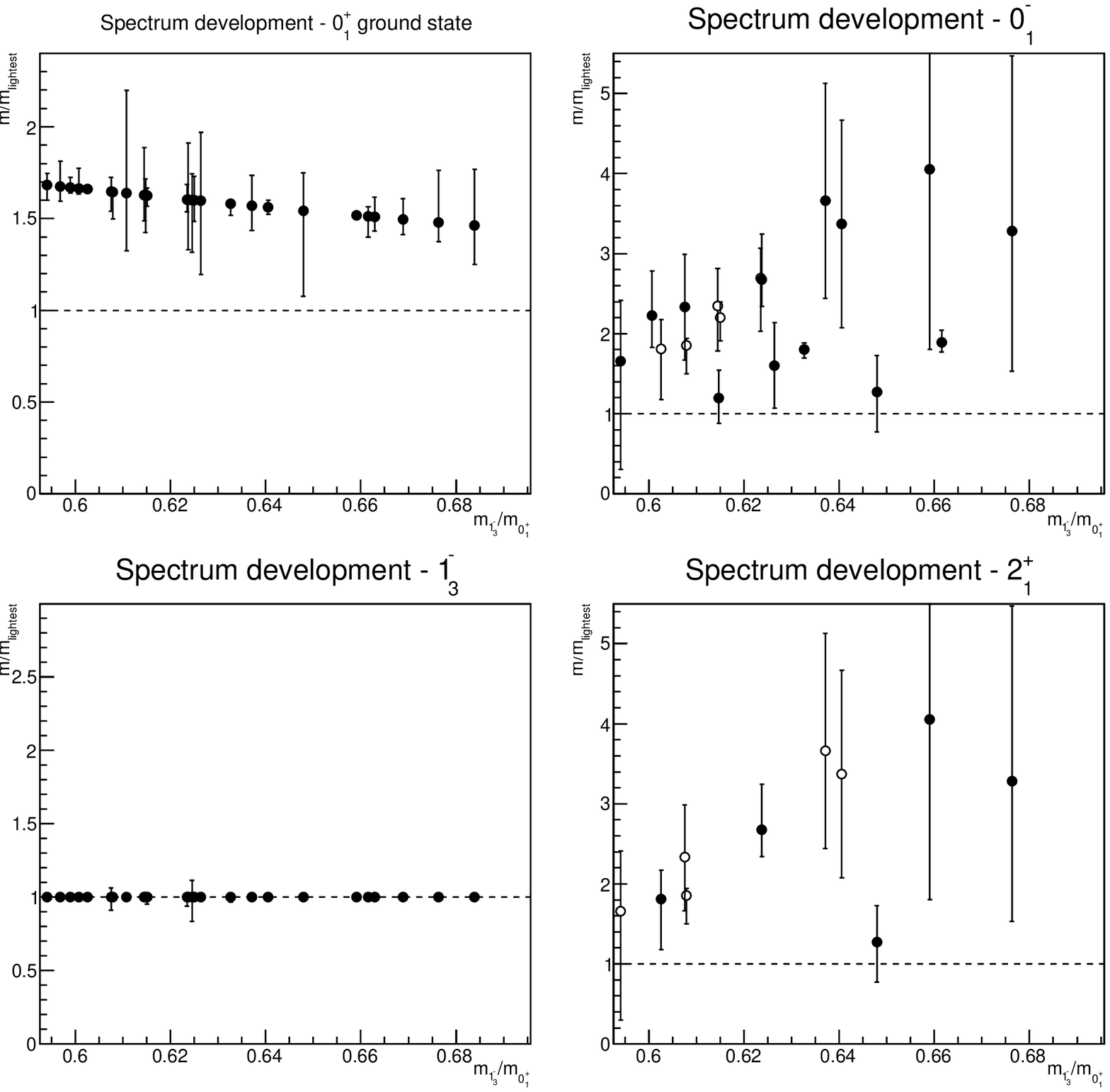}
\caption{\label{ph-gs}The ground states in the different quantum number channels. Open symbols have an energy greater than one, but below 1.5, in lattice units, while closed symbols are below 1. All levels are normalized to the lightest mass.}
\end{figure}

\begin{figure}
\centering
\textsf{The spectrum in the physical-Higgs region, normalized to the decay channels}
\includegraphics[width=\linewidth]{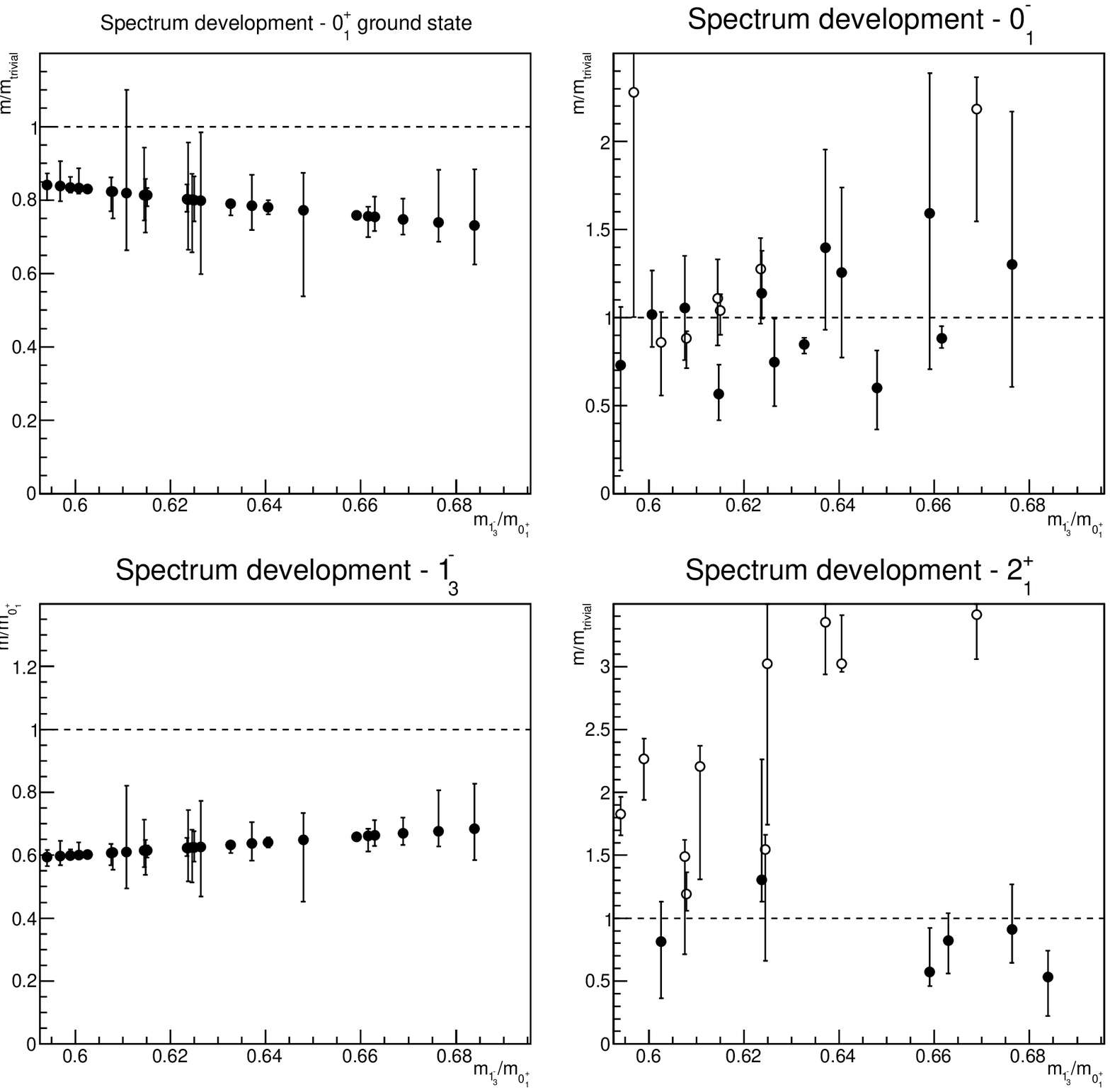}
\caption{\label{ph-gs-dc}The ground states in the different quantum number channels. Open symbols have an energy greater than one, but below 1.5, in lattice units, while closed symbols are below 1. For the normalization, see text.}
\end{figure}

The spectrum in this region, shown in figure \ref{ph-gs}, and especially the decay pattern shown in figure \ref{ph-gs-dc}, are quite interesting. Generically, all quantum numbers cluster mostly around or above the elastic decay threshold, especially if the mass is large in lattice units. There are also some points in the $2^+_1$ channel, where the state appears stable, but here the errors still include the threshold. At any rate, the states are so close to the decay threshold that a more careful investigation should be performed, before any conclusions are drawn. Especially, explicit scattering states must be included.

\begin{figure}
\centering
\textsf{2$^\text{nd}$ and 3$^\text{rd}$ levels in the physical-Higgs region}
\includegraphics[width=\linewidth]{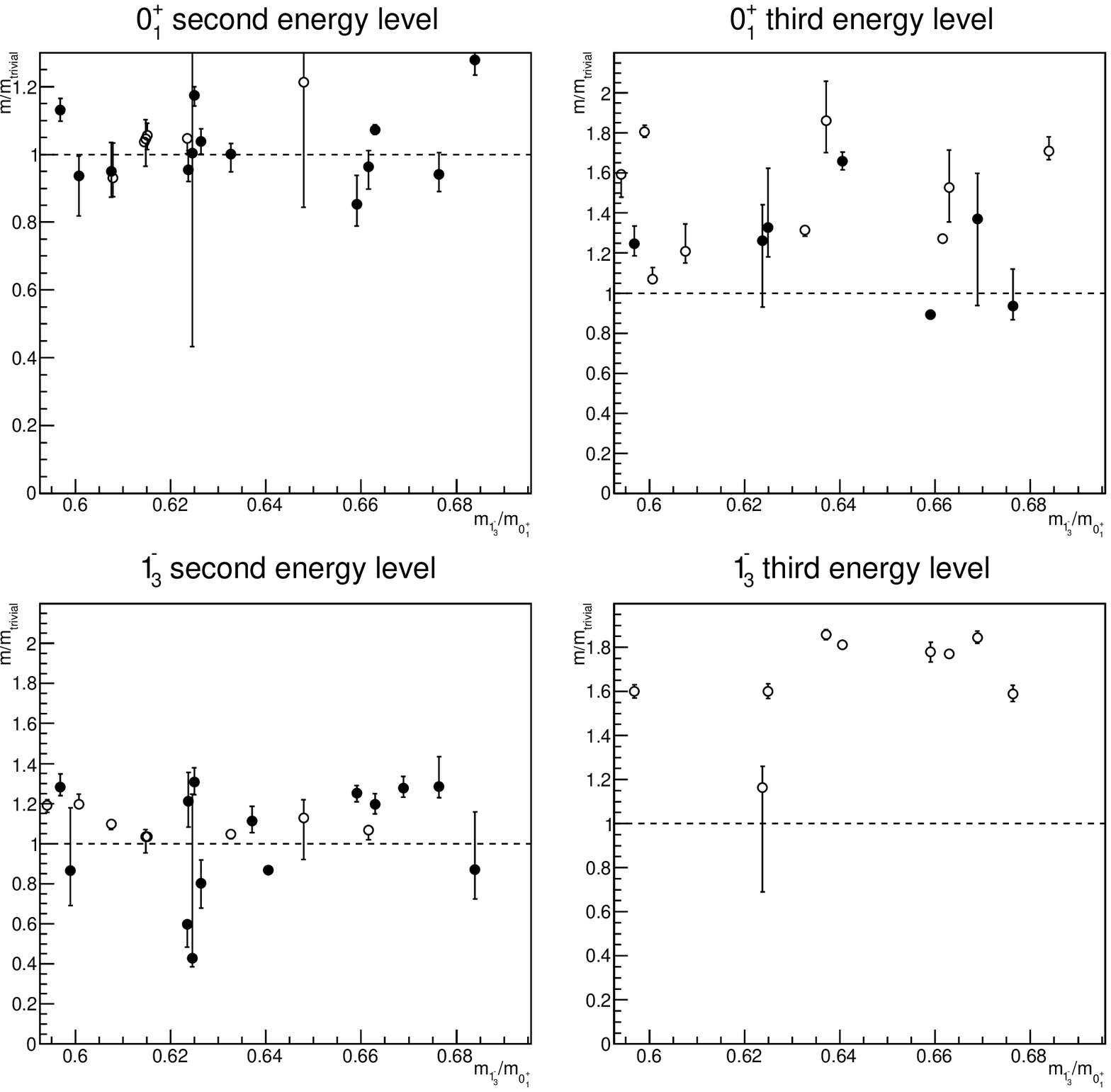}
\caption{\label{ph-higher}The two next states after the ground state in the $0^+_1$ channel (upper panels) and the $1^-_3$ channel (bottom panels). Open symbols have an energy greater than one, but below 1.5, in lattice units, while closed symbols are below 1. All levels are normalized to scattering states, as detailed in the text.}
\end{figure}

The situation for the higher levels, shown in figure \ref{ph-higher}, is much simpler. All states in the second level cluster essentially around the elastic threshold, and all in the third around the inelastic decay threshold or above. The latter is particularly true if the mass in lattice units is large, indicating large lattice artifacts. The only exception are a few states for the second level in the $1^-_3$ channel around $m_{1^-_3}/m_{0^+_1}\approx0.625$, where some seem to form a systematic dip. This may be an interesting signal, but must also be confirmed with more systematic investigations.

Still, the bottom line is that no trace of internal excitations are observed which would provide a good signature of the FMS mechanism.

\subsection{Heavy Higgs}\label{s:hh}

The next part is the one with a stable, but heavy, $0^+_1$ ground state, 135 GeV$<m_{0^+_1}<155$ GeV. Perturbatively, such a Higgs is not very different from a lighter Higgs \cite{Bohm:2001yx}, and thus little changes are expected compared to the previous sections \ref{s:lh} and \ref{s:ph}.

\begin{figure}
\centering
\textsf{The heavy-Higgs region}
\includegraphics[width=0.5\linewidth]{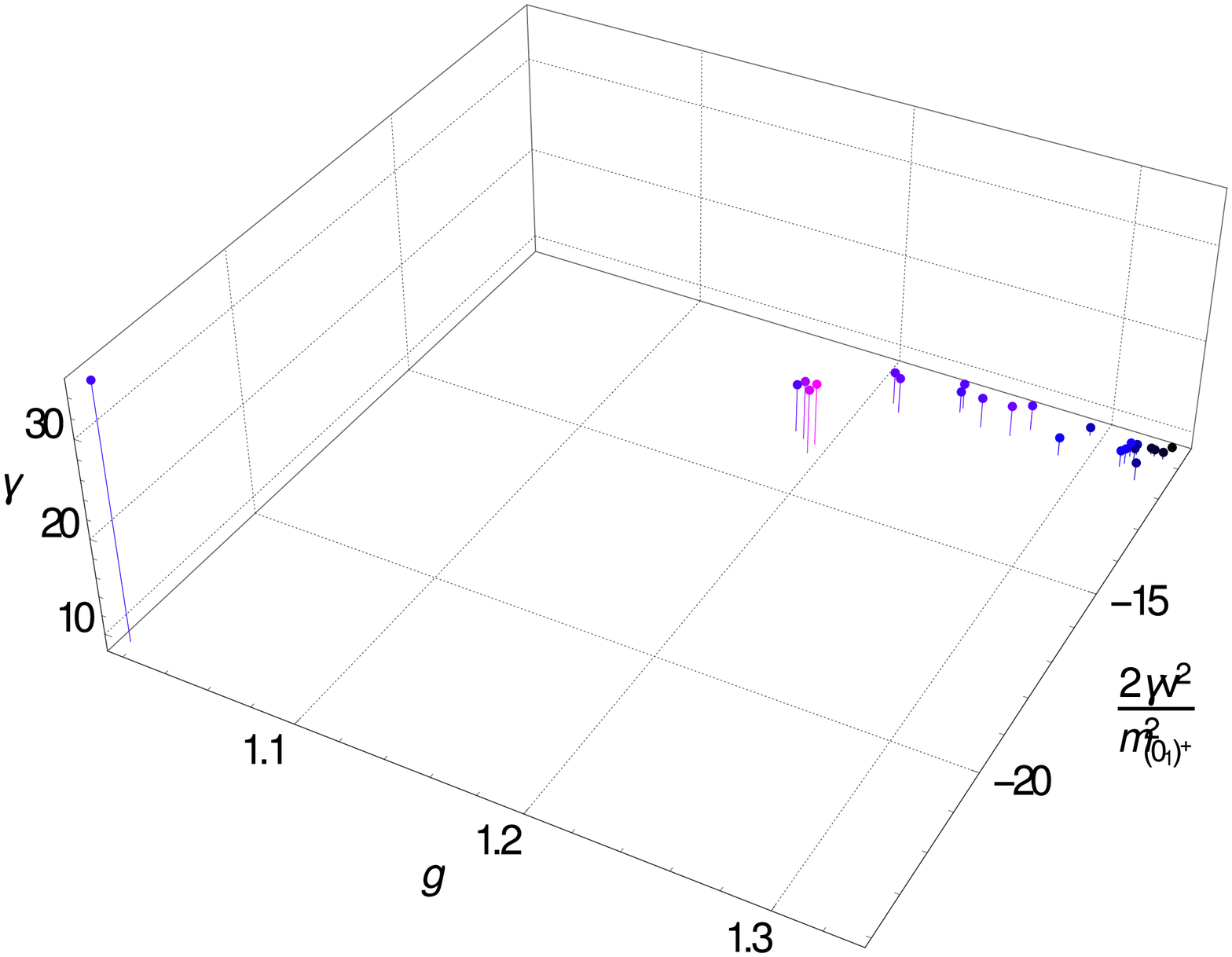}\includegraphics[width=0.5\linewidth]{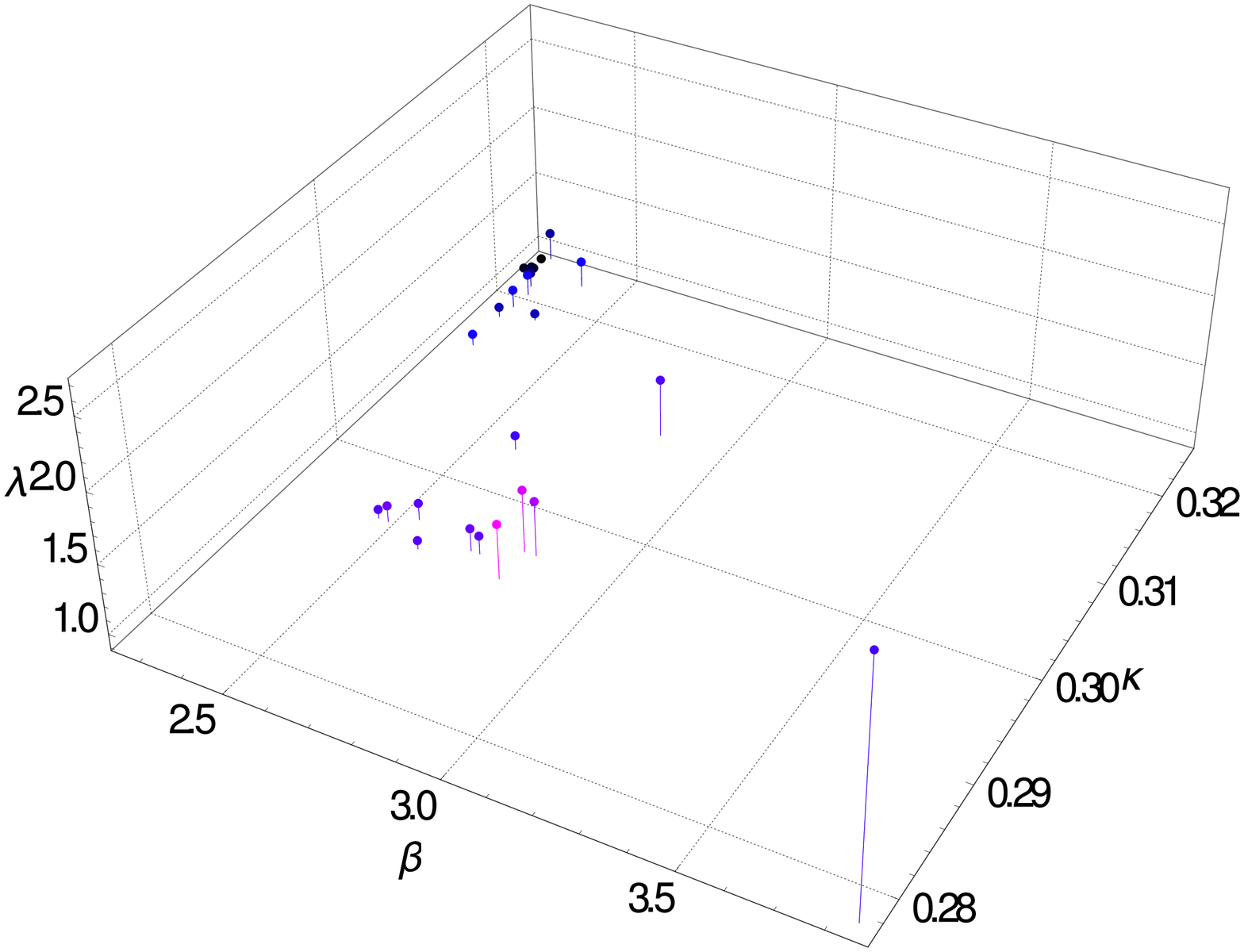}
\caption{\label{lcp-hh}The left-hand plot shows the bare continuum parameters exhibiting a $0^+_1$ ground state in the HLD above the physical mass window, but still stable, while the right-hand plot shows the same in terms of the lattice bare parameters.}
\end{figure}

The parameters exhibiting this behavior are shown in figure \ref{lcp-hh}. Interestingly, when comparing the situation with stable $0^+_1$ as a function of the mass from \ref{lcp-lh} over \ref{lcp-ph} to \ref{lcp-hh}, it appears that the heavier the ground state, the smaller tends the hopping parameter to become, though this is not an exclusive process, but seems to be especially correct the finer the lattice. Interestingly, in terms of the gauge coupling, also small gauge couplings, i.\ e.\ larger values of $\beta$, seem to favor smaller lattice spacings.

\begin{figure}
\centering
\textsf{The spectrum in the heavy-Higgs region, normalized to the lightest mass}
\includegraphics[width=\linewidth]{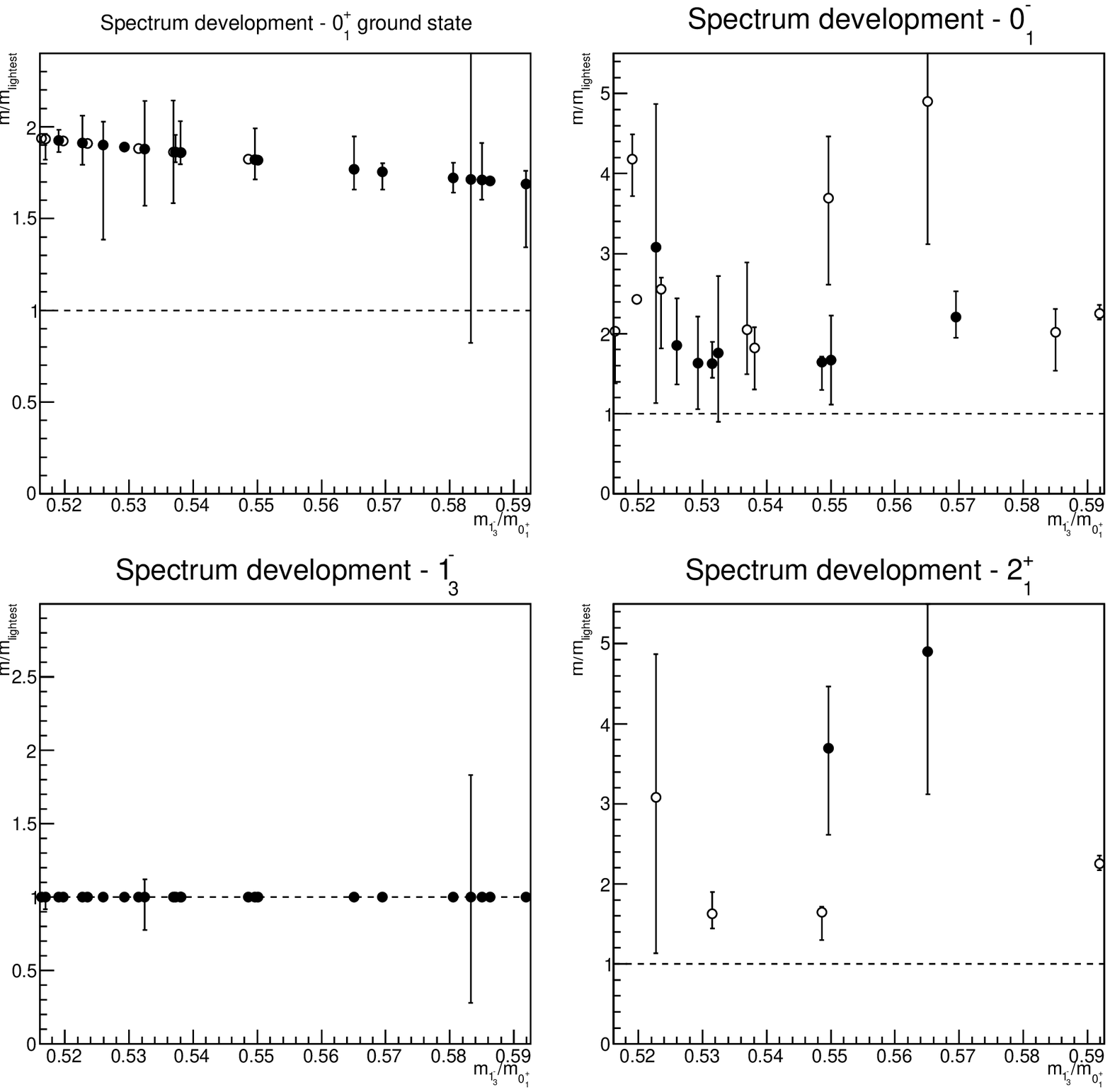}
\caption{\label{hh-gs}The ground states in the different quantum number channels. Open symbols have an energy greater than one, but below 1.5, in lattice units, while closed symbols are below 1. All levels are normalized to the lightest mass.}
\end{figure}

\begin{figure}
\centering
\textsf{The spectrum in the heavy-Higgs region, normalized to the decay channels}
\includegraphics[width=\linewidth]{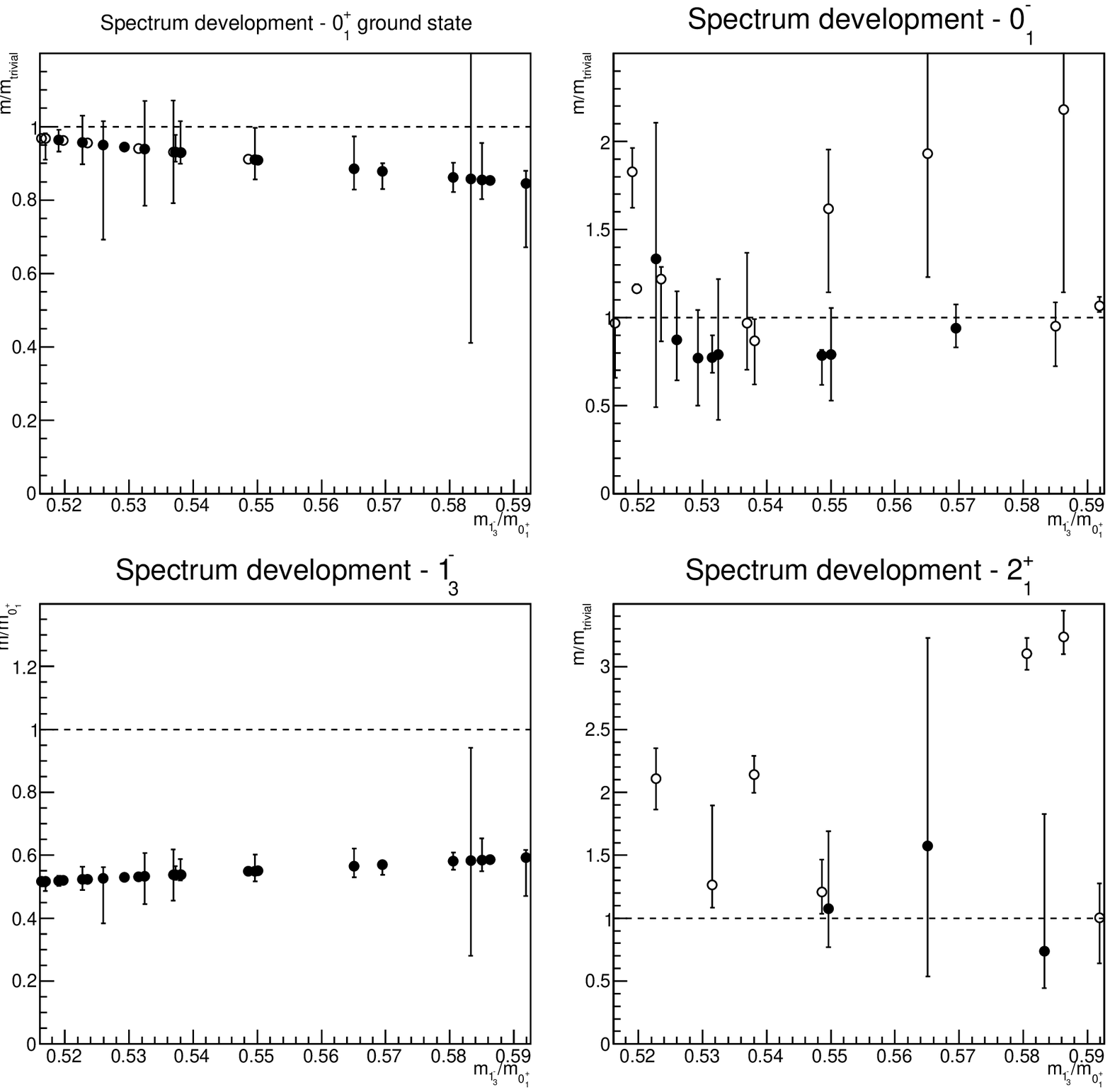}
\caption{\label{hh-gs-dc}The ground states in the different quantum number channels. Open symbols have an energy greater than one, but below 1.5, in lattice units, while closed symbols are below 1. For the normalization, see text.}
\end{figure}

The energies in the different channels are again shown in figure \ref{hh-gs} and \ref{hh-gs-dc}. The ground state in the $2^+_1$ channel, when it can be extracted, is always at or above the elastic threshold. Hence, no sign of any stable excitations does exist in this channel. The $0^-_1$ is in essentially all cases compatible with the elastic threshold, especially the heavier the $0^+_1$ is. However, in these cases the average values are most often below the threshold, and only due to the size of the errors the elastic threshold are reached. In addition, though this is not statistically reliable, there appears little dependence of the masses in the other channels on the $0^+_1$ mass.

\begin{figure}
\centering
\textsf{2$^\text{nd}$ and 3$^\text{rd}$ levels in the heavy-Higgs region}
\includegraphics[width=\linewidth]{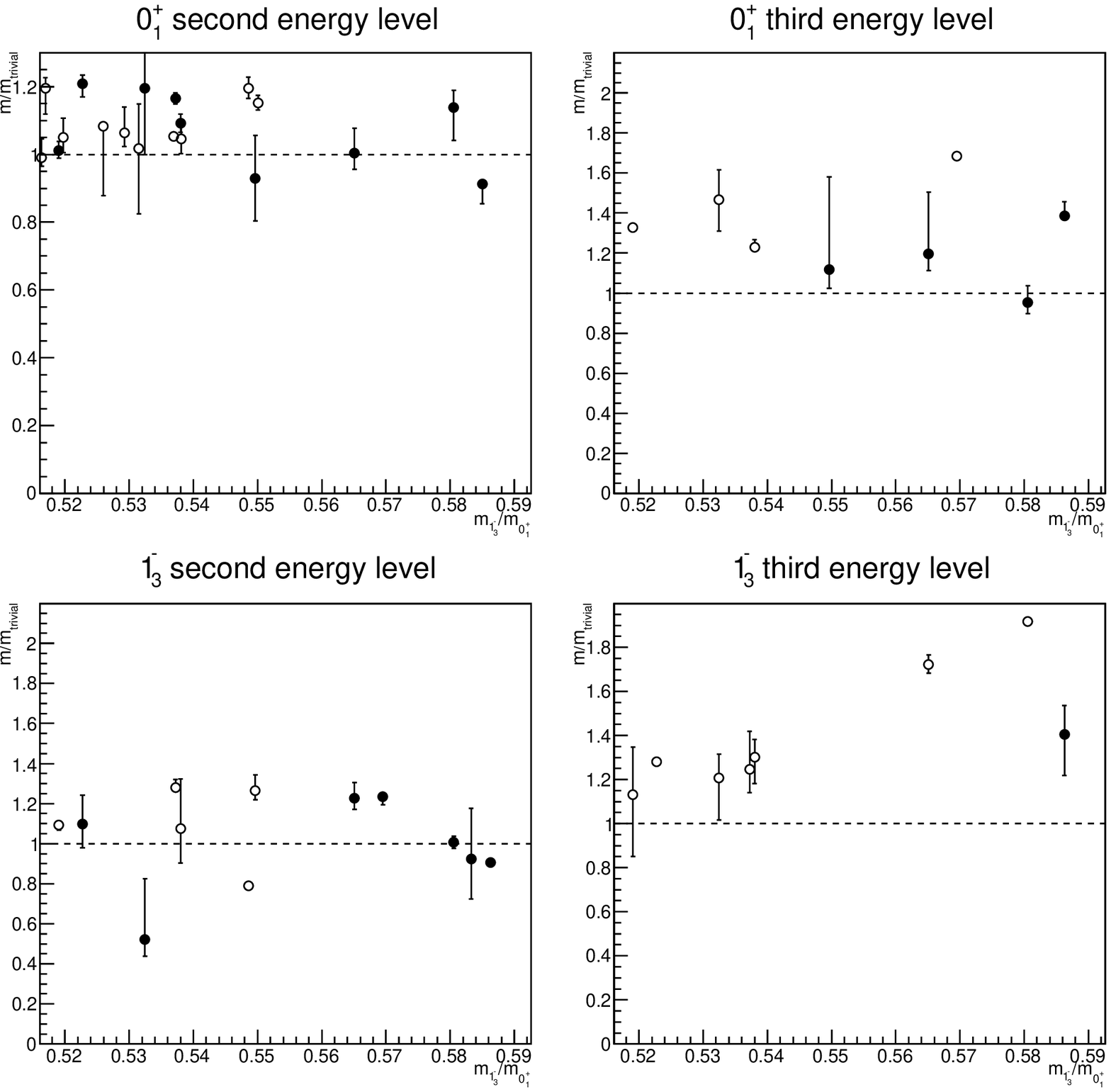}
\caption{\label{hh-higher}The two next states after the ground state in the $0^+_1$ channel (upper panels) and the $1^-_3$ channel (bottom panels). Open symbols have an energy greater than one, but below 1.5, in lattice units, while closed symbols are below 1. All levels are normalized to scattering states, as detailed in the text.}
\end{figure}

The higher states in the $0^+_1$ and $1^-_3$ channels are shown in figure \ref{hh-higher}. In all cases, the states are in agreement with scattering states, tending to be even somewhat above the scattering state level. Thus, in case of a rather heavy but stable $0^+_1$ there is no indication for any internal excitations in either channel.

\subsection{At threshold}\label{s:th}

The last reasonable case when ordering the states according to the mass ratio $m_{0^+_1}/m_{1^-_3}$ is reached when the lowest level in the $0^+_1$ channel is at the elastic threshold, and therefore there is no stable state anymore in this channel. In perturbation theory there is then still a reasonable stable resonance existing \cite{Bohm:2001yx} for quite a large mass range above. The interesting question is whether this can be confirmed in the full non-perturbative case. Due to the uncertainties in determining the mass, the threshold region here covers any ground state mass of the $0^+_1$ in the mass interval 155 GeV$<m_{0^+_1}<170$ GeV.

\begin{figure}
\centering
\textsf{The threshold region}
\includegraphics[width=0.5\linewidth]{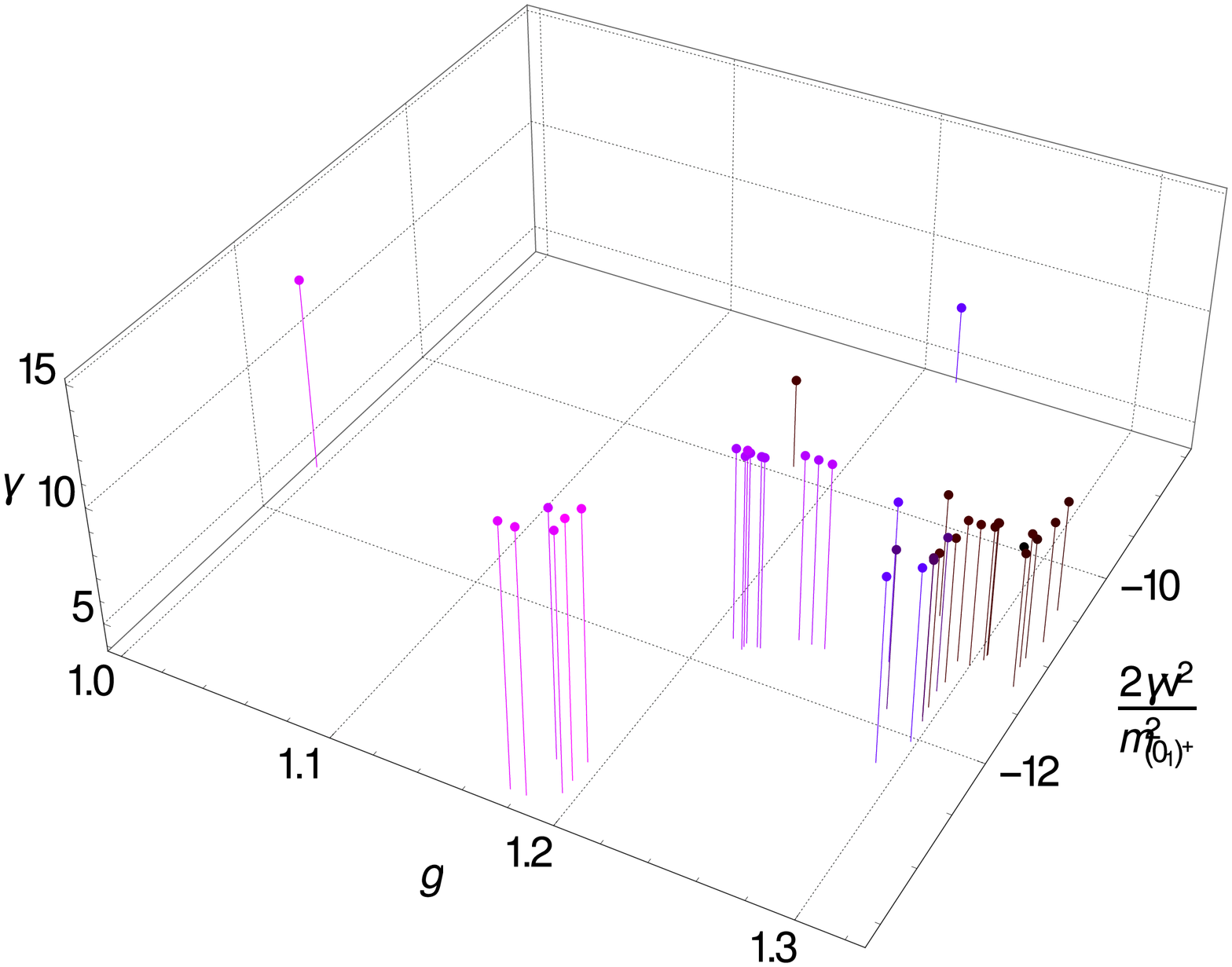}\includegraphics[width=0.5\linewidth]{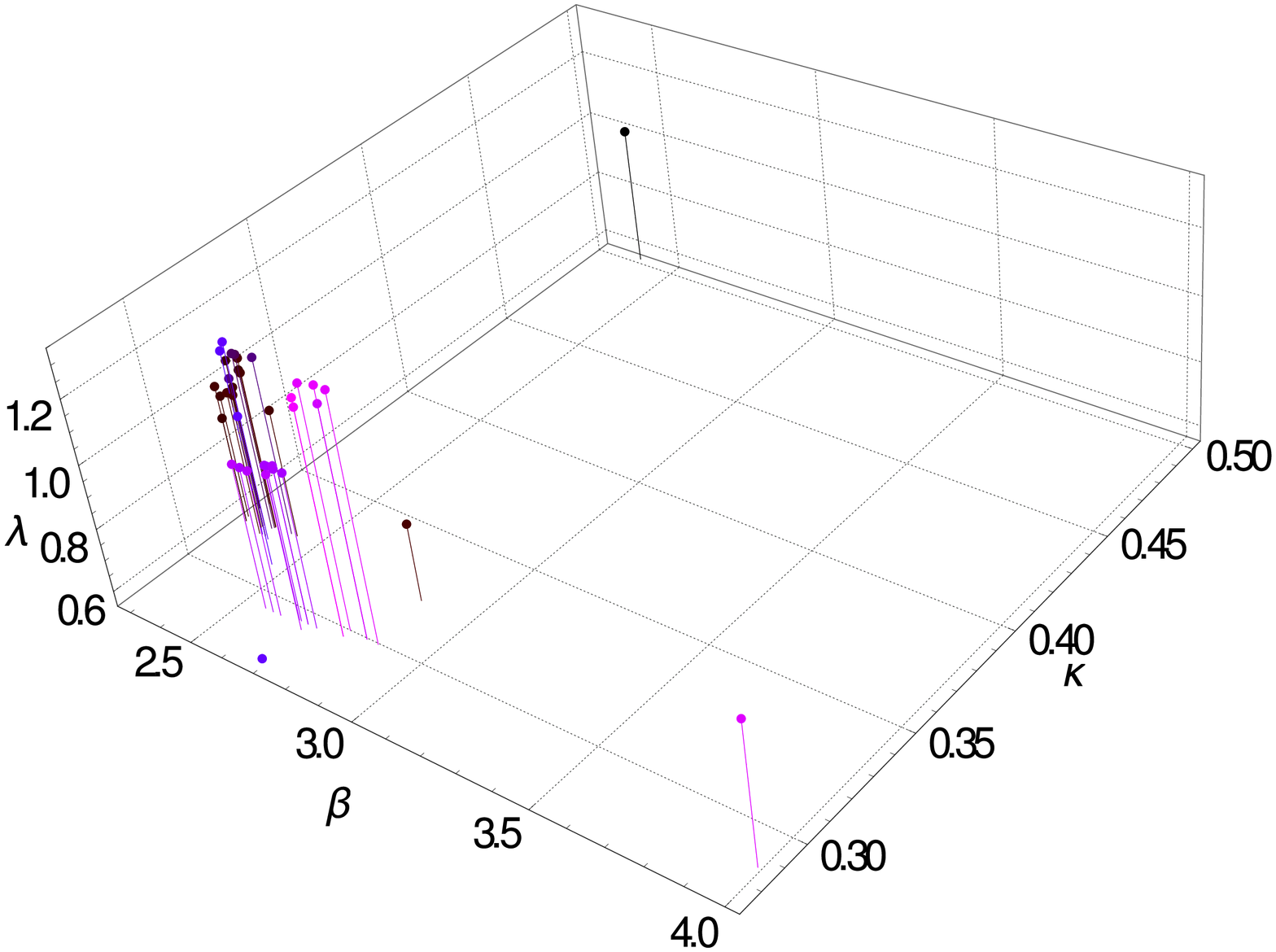}
\caption{\label{lcp-th}The left-hand plot shows the bare continuum parameters exhibiting a $0^+_1$ ground state in the HLD with the $0^+_1$ at the decay threshold into two $1^-_3$, while the right-hand plot shows the same in terms of the lattice bare parameters.}
\end{figure}

The parameter sets qualifying for this are shown in figure \ref{lcp-th}. There is a clear tendency for smaller gauge couplings to reach small lattice spacings. In continuum terms, this corresponds to rather small gauge couplings and deep classical potentials.

\begin{figure}
\centering
\textsf{The spectrum in the threshold region, normalized to the lightest mass}
\includegraphics[width=\linewidth]{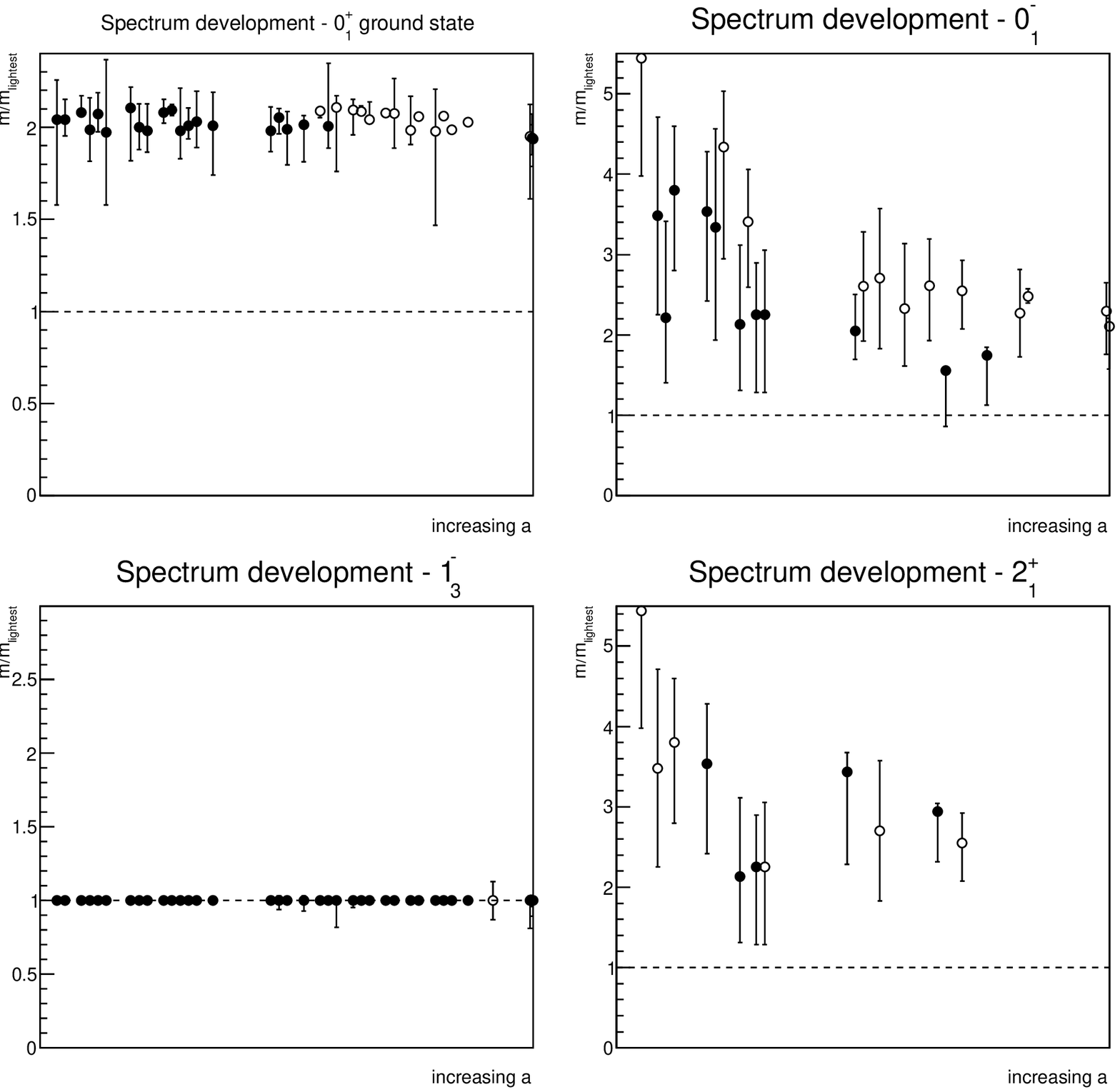}
\caption{\label{th-gs}The ground states in the different quantum number channels, lattice spacing decreasing from right to left. Open symbols have an energy greater than one, but below 1.5, in lattice units, while closed symbols are below 1. All levels are normalized to the lightest mass.}
\end{figure}

\begin{figure}
\centering
\textsf{The spectrum in the threshold region, normalized to the decay channels}
\includegraphics[width=\linewidth]{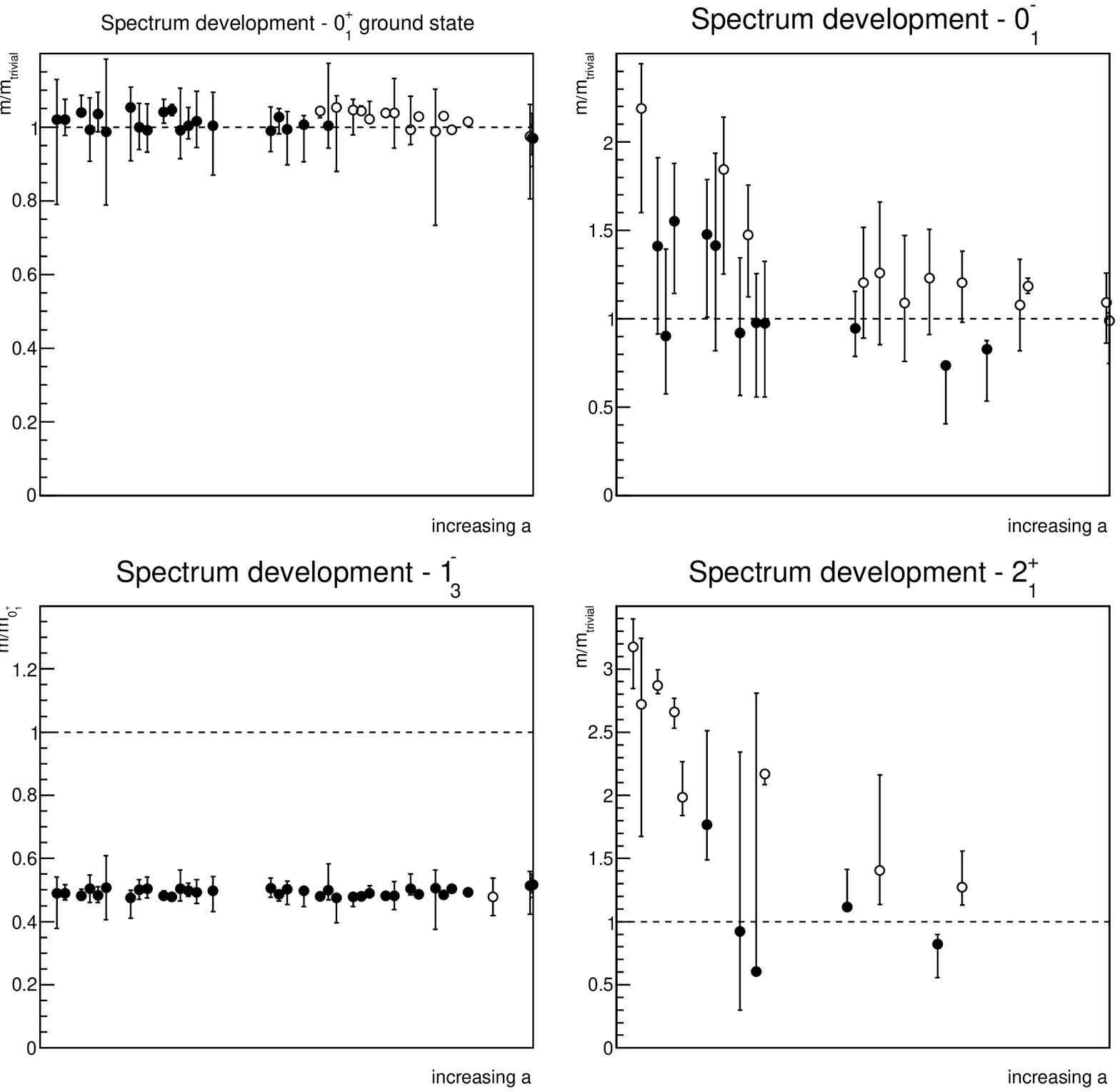}
\caption{\label{th-gs-dc}The ground states in the different quantum number channels, lattice spacing decreasing from right to left. Open symbols have an energy greater than one, but below 1.5, in lattice units, while closed symbols are below 1. For the normalization, see text.}
\end{figure}

The results for the ground states are shown in figure \ref{th-gs} and \ref{th-gs-dc}. Especially from figure \ref{th-gs-dc} it is clear that the ground state in the $0^+_1$ channel is compatible with being the scattering state at the elastic threshold, but partly with rather large errors. The behavior in the $0^-_1$ channel is essentially compatible with scattering states. The results in the $2^+_1$ channel have rather large errors, but tend to be also in agreement with scattering states. 

\begin{figure}
\centering
\textsf{2$^\text{nd}$ and 3$^\text{rd}$ levels in the threshold region}
\includegraphics[width=\linewidth]{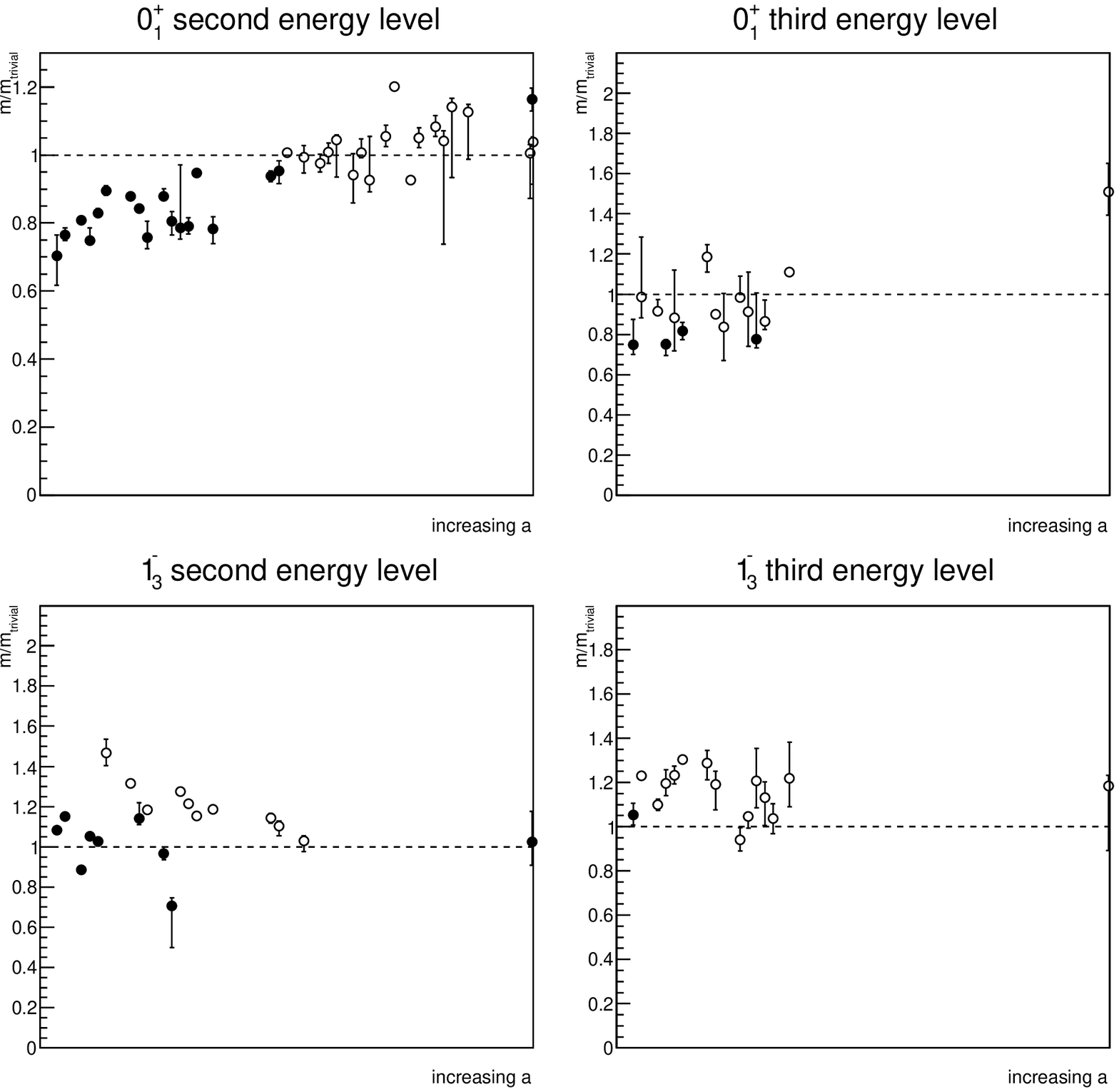}
\caption{\label{th-higher}The two next states after the ground state in the $0^+_1$ channel (upper panels) and the $1^-_3$ channel (bottom panels), lattice spacing decreasing from right to left. Open symbols have an energy greater than one, but below 1.5, in lattice units, while closed symbols are below 1. All levels are normalized to scattering states, as detailed in the text.}
\end{figure}

A similar pattern also emerges for the higher states shown in figure \ref{th-higher}. The levels in the $1^-_3$ channel are essentially compatible with  scattering states. The additional channels, most notably the second level, in the $0^+_1$ channel are, however, decreasing with decreasing lattice spacing. This systematic dependence seems to exclude the possibility that the lowest level was erroneously assigned in some cases to be a scattering state, and is, in truth, just a stable state just below threshold. Comparing to the actual numerical values in table \ref{t:th} shows, however, that the energy of the second state is at the same time just a little bit above the threshold.

There are a number of possible interpretations. The simplest would be that there is indeed a metastable excitation above threshold, in line with the perturbative expectation, which can be better and better resolved with decreasing lattice spacing. More likely is, however the following: This state will not be a state at rest, but has relative momentum. Such operators are not included in the basis here, and this can lead to a misidentification or shift of the energy levels, as has been observed in \cite{Wurtz:2013ova}. Also, mixing with tolerons is a possibility \cite{Wurtz:2013ova,Philipsen:1996af}. To distinguish these possibilities, more volumes and/or an enlarged operator basis will be necessary in a future investigation.

It should be noted however that this more likely explanation from the point of view of systematics has far-reaching consequences. If not hidden in even further different other operators or other regions of the phase diagram, this would indicate that it is not possible to have a reasonable stable $0^+_1$ above threshold. This would be in marked contradiction to the perturbative expectation, and would have rendered any search for a heavy standard-model Higgs with standard methods at the LHC possibly meaningless. It is also an interesting conceptual question what this would mean for the theories with additional heavier Higgs states or for the question of naturalness, albeit it is too easy to get lost in speculations without further facts.

\subsection{Anomalous}\label{s:a}

There is a set of parameters, for which the described procedure yielded a ground state significantly above the elastic threshold. This, in itself, is not a bad sign, but just indicates that the overlap of the chosen operator basis with the scattering state at threshold is not good enough. The interesting point is that this is not just the fault of the operator basis. The previous section \ref{s:th} has demonstrated that in a large number of cases the scattering state at the elastic threshold can be correctly identified using the employed operator basis. The question is hence whether there is a difference in the physics case, e.\ g.\ that these are the cases with a non-trivial resonance so acutely missing in section \ref{s:th}. It is therefore worthwhile to check whether there appears any regularity for these cases.

\begin{figure}
\centering
\textsf{Anomalous points}
\includegraphics[width=0.5\linewidth]{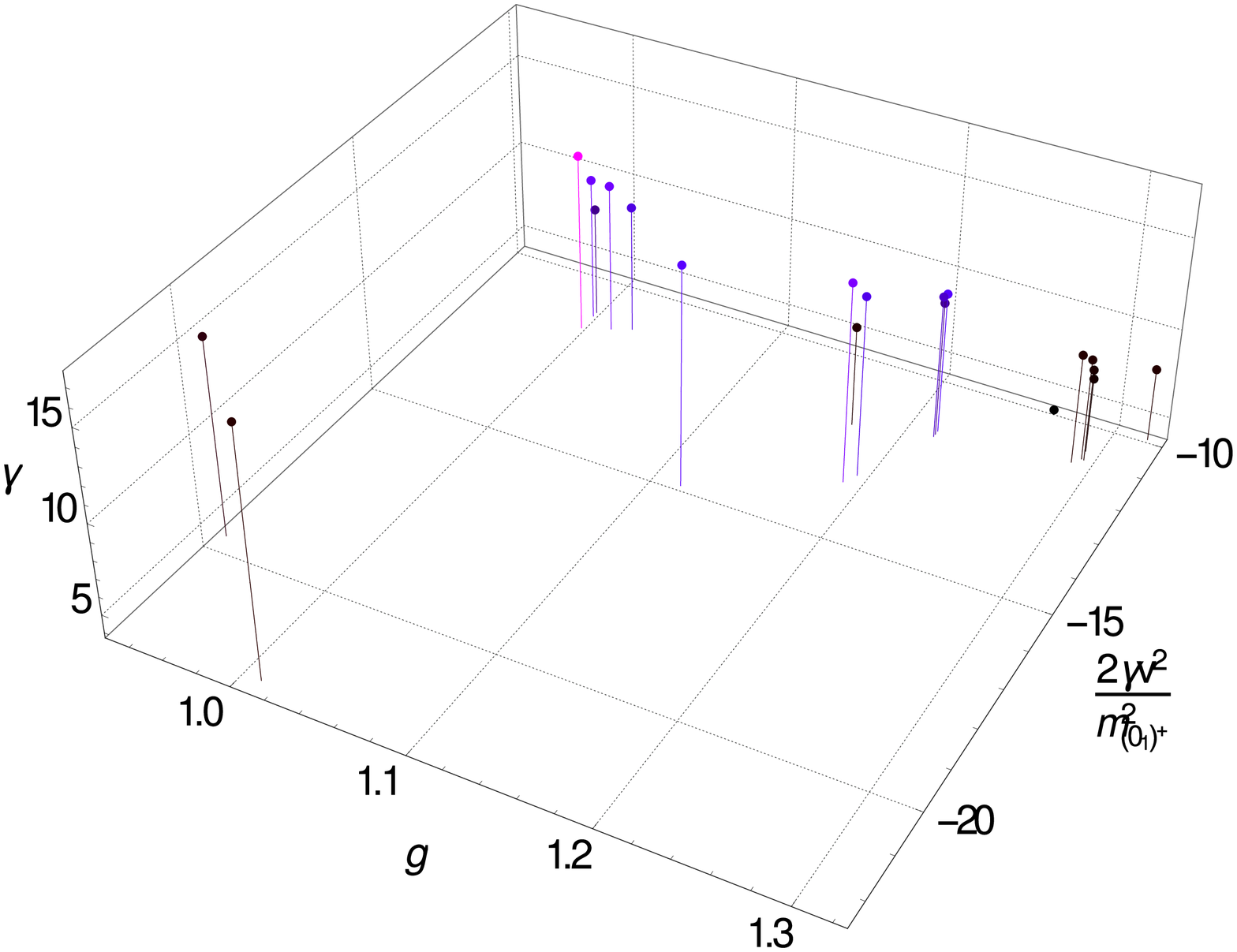}\includegraphics[width=0.5\linewidth]{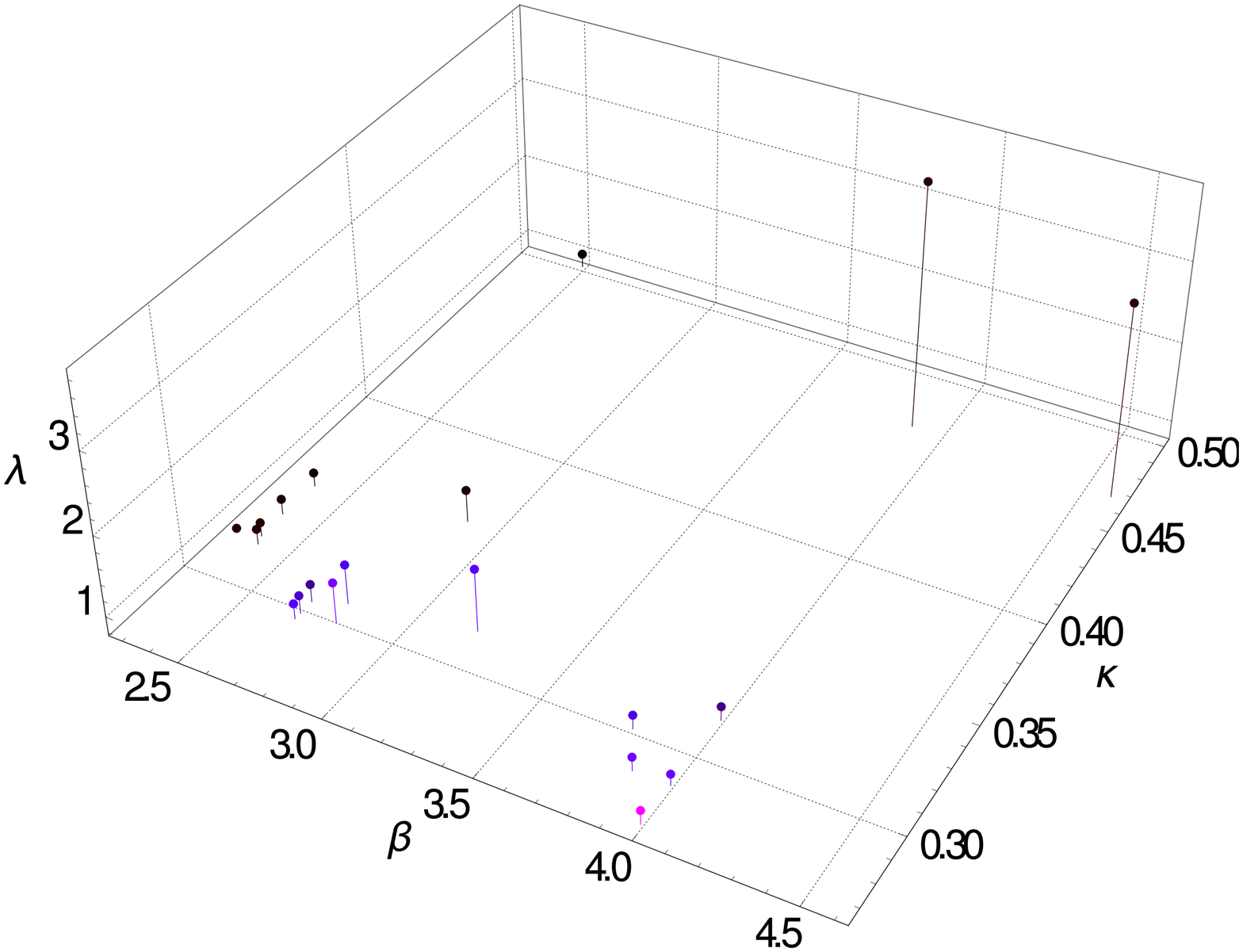}
\caption{\label{lcp-a}The left-hand plot shows the bare continuum parameters exhibiting a $0^+_1$ ground state for anomalous points, while the right-hand plot shows the same in terms of the lattice bare parameters.}
\end{figure}

The points in the phase diagram exhibiting this behavior are shown in figure \ref{lcp-a}. At first sight, no particular features stand out.

\begin{figure}
\centering
\textsf{The spectrum for the anomalous cases, normalized to the lightest mass}
\includegraphics[width=\linewidth]{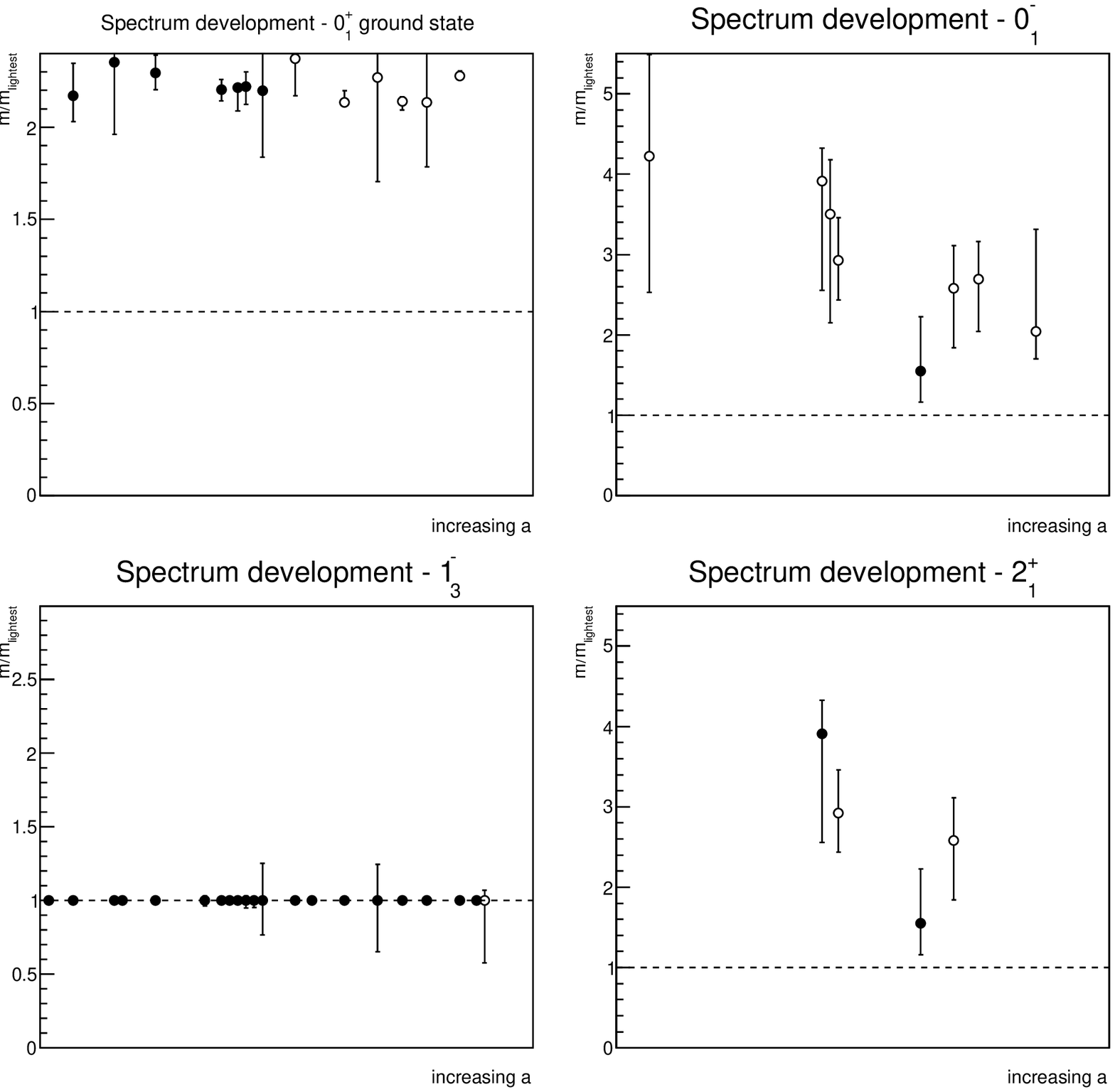}
\caption{\label{a-gs}The ground states in the different quantum number channels, lattice spacing decreasing from right to left. Open symbols have an energy greater than one, but below 1.5, in lattice units, while closed symbols are below 1. All levels are normalized to the lightest mass.}
\end{figure}

\begin{figure}
\centering
\textsf{The spectrum for the anomalous cases, normalized to the decay channels}
\includegraphics[width=\linewidth]{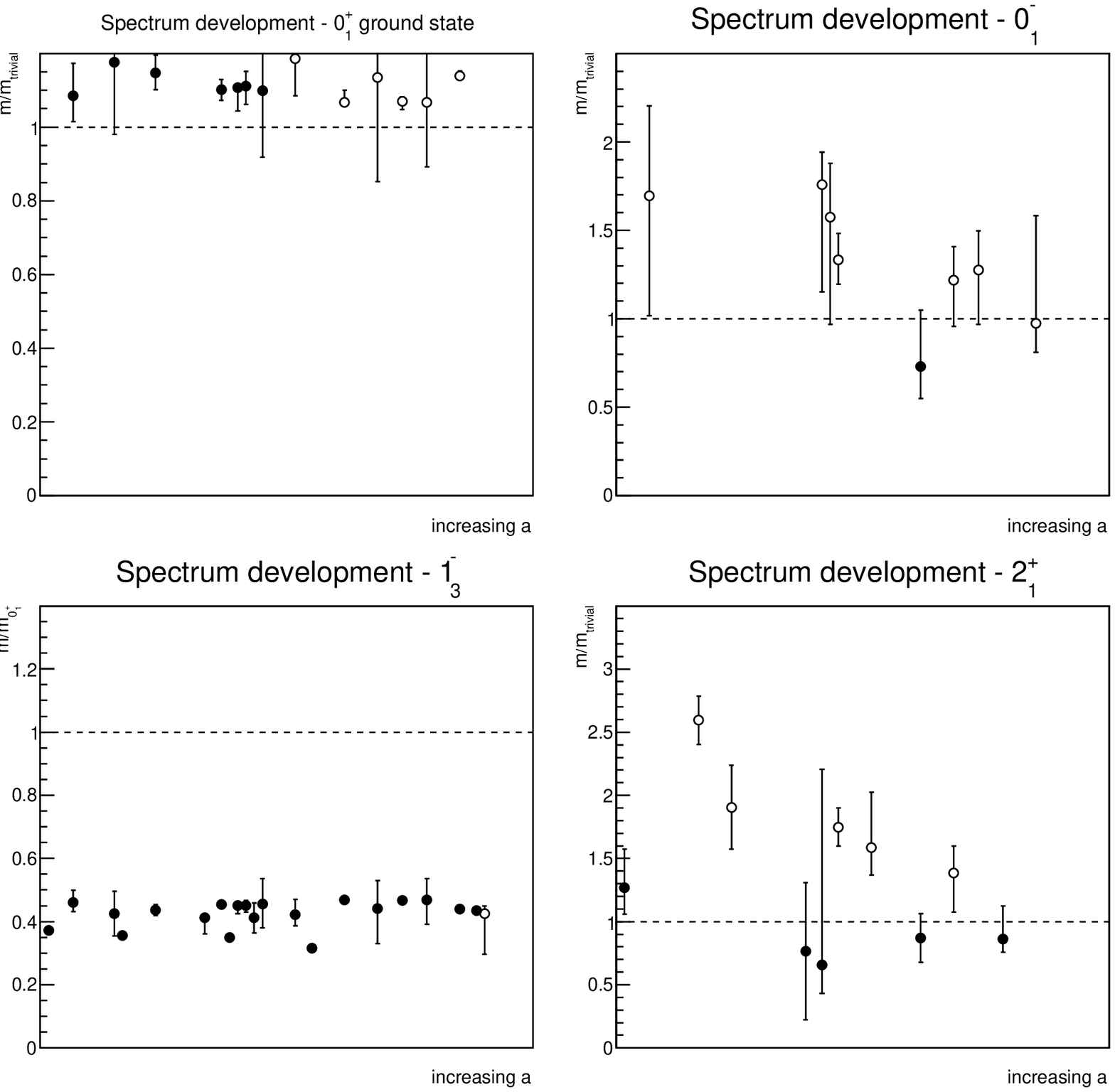}
\caption{\label{a-gs-dc}The ground states in the different quantum number channels, lattice spacing decreasing from right to left. Open symbols have an energy greater than one, but below 1.5, in lattice units, while closed symbols are below 1. For the normalization, see text.}
\end{figure}

The results for the ground states are shown in figures \ref{a-gs} and \ref{a-gs-dc}. The $0^+_1$ ground state is only slightly above the decay threshold, and in some cases within $1\sigma$ still compatible with it. Hence, there is no indication for a substantial increase above the threshold. All other channels, especially where the results are reliable, are mostly consistent with trivial states. Only the $2^+_1$ channels seems to be different, but given its absolute values, this is not a real surprise, and the channel can hardly be considered reliable.

\begin{figure}
\centering
\textsf{2$^\text{nd}$ and 3$^\text{rd}$ levels for the anomalous cases}
\includegraphics[width=\linewidth]{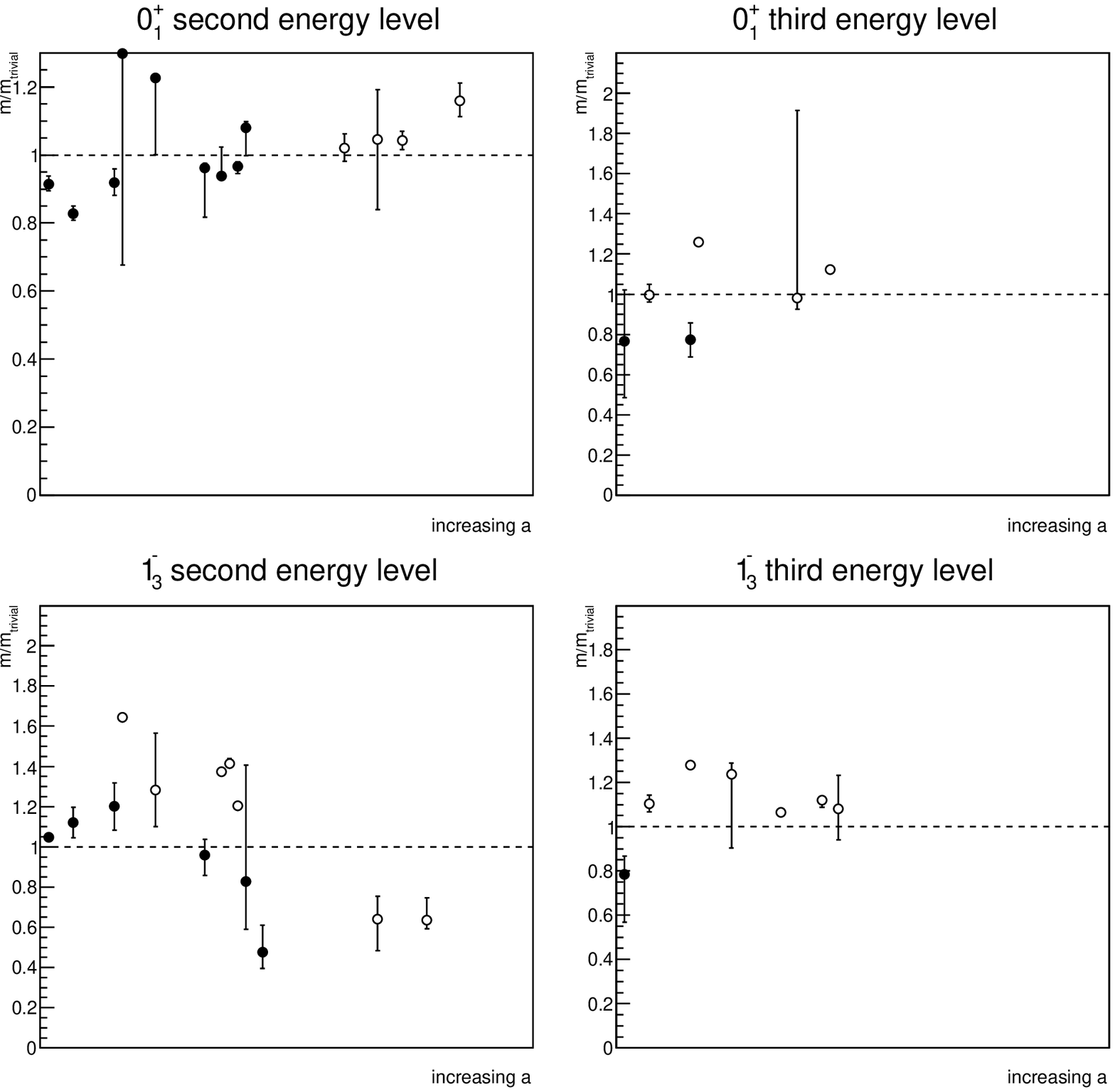}
\caption{\label{a-higher}The two next states after the ground state in the $0^+_1$ channel (upper panels) and the $1^-_3$ channel (bottom panels), lattice spacing decreasing from right to left. Open symbols have an energy greater than one, but below 1.5, in lattice units, while closed symbols are below 1. All levels are normalized to scattering states, as detailed in the text.}
\end{figure}

The results for the higher levels shown in figure \ref{a-higher} show an interesting pattern. The $0^+_1$ channel, in contrast to the previous case with the $0^+_1$ ground state being close to threshold, remains even at small lattice spacing much better consistent with a scattering state, and so does the third state. Here it should be noted that the smallest lattice spacings reached are in both cases similar, so this is not a pure lattice artifact, as can be seen from table \ref{t:a}.

Something curious happens also for the $1^-_3$ levels. For large lattice spacings, the second level is substantially below the decay threshold. However, at smaller spacings it is again compatible with a scattering state or above. The third level shows no distinct behavior, though at large lattice spacings it anyway cannot be resolved.

Still, these results suggest that, at least at sufficiently small lattice spacing, the picture is somewhat different from the situation where the ground state in the $0^+_1$ channel is located at threshold. Thus, these anomalous cases are potentially somewhat different from the cases with results compatible to threshold, especially for the second state in the $0^+_1$ channel. To better understand this, it will be necessary to analyze the properties of this channel better, e.\ g.\ using a phase-shift analysis.

\section{Summary and conclusions}\label{sconclusions}

We have presented an extensive qualitative study of the spectroscopy of Yang-Mills-Higgs theory, interpreting the results in the context of the FMS mechanism. We find, in agreement with other studies \cite{Jersak:1985nf,Langguth:1985eu,Evertz:1985fc,Langguth:1985dr,Karsch:1996aw,Philipsen:1996af,Knechtli:1998gf,Knechtli:1999qe,Iida:2007qp,Bonati:2009pf,Wurtz:2013ova}, that there are many cases, even at rather large bare gauge coupling, where the spectrum is consistent with just a Higgs-like and a $W$-like state, and no further (meta-)stable states in the spectrum. In some cases, however, we do find other states, especially in the $0^-_1$ channel, i.\ e.\ custodial singlets. Would any of them survive in the standard model, this would be genuinely new background to new physics searches \cite{Maas:2012tj}. That these are not observed at weaker coupling \cite{Wurtz:2013ova} is, however, discouraging, though it should always be kept in mind that the present theory is much weaker interacting than when embedded in the standard model \cite{Maas:2013aia}. Also, because of the triviality problem smaller gauge couplings are by no means preferred.

Outside the physical region the theory provides more unexpected results. In line with earlier results \cite{Jersak:1985nf,Langguth:1985eu,Evertz:1985fc,Langguth:1985dr}, we also find that it is not possible to keep the $0^+_1$ lighter than the $1^-_3$ without obtaining QCD-like physics. This is not expected in a perturbative setting. Nonetheless, since we again find few indications of excited states, this domain seems to be weaker interacting than expected. In turn, in the HLD we were not able to find any reliable candidate for a $0^+_1$ state substantially above the decay threshold, which is again not expected from a perturbative point of view. If these findings could be corroborated in more detailed and quantitative investigations, this would strongly suggest that the naive perturbative picture, and the FMS mechanism, can only be maintained in a very narrow parameter range, with substantial implications for beyond-the-standard-model physics \cite{Maas:2014nya,Maas:2015gma}.

Still, the present study is only at a qualitative level, with just one fixed (lattice) volume, and a somewhat small operator basis. There is always the possibility that it is just lattice artifacts which obscure the picture. Hence, dedicated investigations will be necessary to clarify the situation, allowing eventually for a more definitive conclusion to be reached.\\

\no{\bf Acknowledgments}\\

We are grateful to C.\ B.\ Lang and D.\ Mohler for helpful discussions, and to G.\ Eichmann for a critical reading of the manuscript and many helpful comments. We are especially grateful to M. Wurtz for pointing out an error in the first version of this paper. This project was supported by the DFG under grant number MA 3935/5-1. A.\ M.\ was also supported by the Heisenberg program of the DFG under grant number MA 3935/8-1 and T.\ M.\ by the DFG graduate school GRK 1523/1 and GRK 1523/2. Simulations were performed on the HPC cluster at the Universities of Jena and Graz. The authors are grateful to the HPC team for the very good performance of the clusters. The ROOT framework \cite{Brun:1997pa} has been used in this project.

\appendix

\section{Lattice artifacts}\label{a:sys}

Though a detailed study of lattice artefacts will require substantially more statistics, especially for larger volumes, and parameter sets, we provide here some preliminary estimates of the effects. Also a few more remarks on fitting and statistics are given. The result is that the qualitative conclusions of the main text remain unaltered.

\subsection{Volume}\label{a:vol}

To study the volume dependence, we repeat the investigation of the main text on additionally smaller volumes of $8^4$ and $16^4$. We also add some results on $32^4$ lattices, though these have substantially larger statistical fluctuations, and thus will require much more statistics. The volume dependence will be studied for some of the different settings in turn, depending on where sufficient statistics is already available. The different sets of parameters also vary substantially with respect to lattice spacing, so that the interplay of both effects is also visible to some extent.

\begin{figure}
\includegraphics[width=\linewidth]{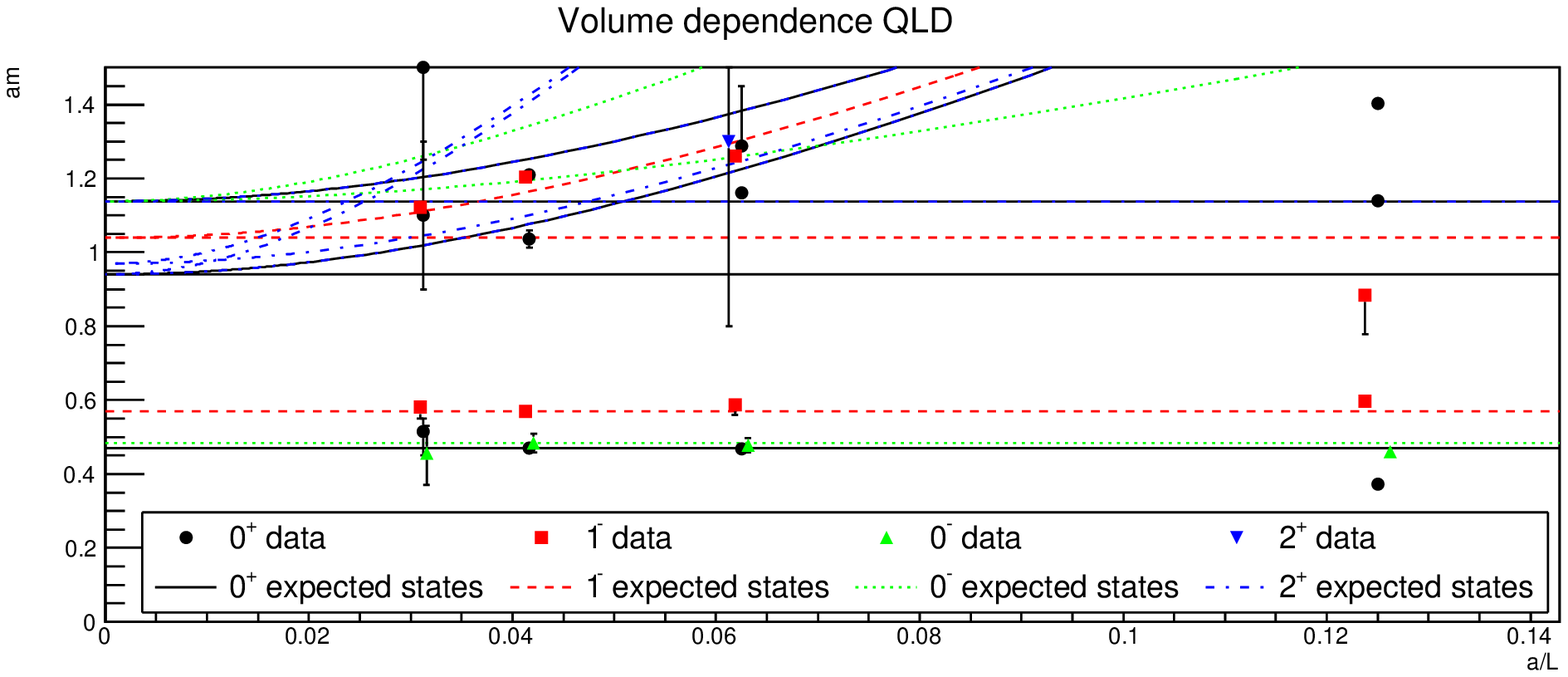}
\caption{\label{qld-vol}The volume dependence of the extracted states in the QLD. Note that here and hereafter the different channels have been slightly displaced horizontally for better visibility.}
\end{figure}

In the QLD, the parameters $\beta=2.2171$, $\kappa=0.3182$, and $\lambda=1.046$ have been used. The result is shown in figure \ref{qld-vol}. Besides the data for the four-volumes also a number of expected states are shown. One is in all channels the ground-state mass, as extracted from the 24$^4$ lattice in the main text. The others are the possible scattering states, as discussed in section \ref{s:qld}. Here, the input masses are also the masses from the $24^4$ lattices.

For this lattice setup, the results for the $2^+$ are not such that any realistic conclusion can be drawn, except that its mass is likely above the elastic decay thresholds. In the other channels, always a stable ground state exists. The mass of these are essentially volume-independent from the $16^4$ lattice onwards, and even on the $8^4$ lattice the masses are quite close to the values of the larger lattices.

The situation for the second levels is less clear. For the $0^+$, the data points first follow the second elastic decay threshold in two $1^-$, but then become better comparable with a state made up from two $0^+$ with relative momenta. The state with two $0^+$ at rest is not seen. Similarly, for the second state in the $1^-$ level, it follows the line of a $0^+$ and a $1^-$ with relative momenta, rather than the one at rest. Nonetheless, both behaviors are still compatible with the observation in the main text that these are probably scattering states. Given that no non-zero momentum states are in the operator basis, but the elastic thresholds are, it is likely that also other artifacts play a role.

\begin{figure}
\includegraphics[width=\linewidth]{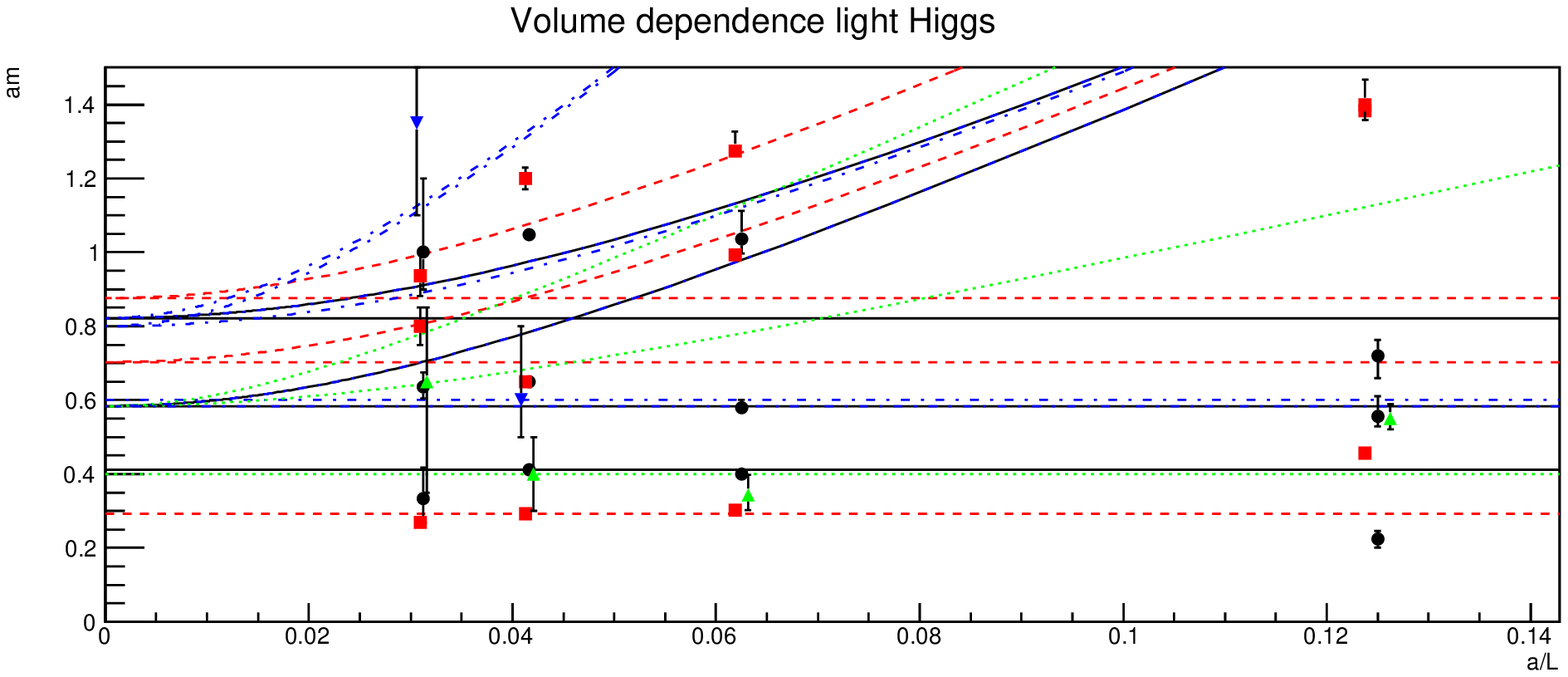}
\caption{\label{lh-vol}The volume dependence of the extracted states for a light Higgs. Symbols and lines have the same meaning as in figure \ref{qld-vol}.}
\end{figure}

For a light Higgs, the parameters $\beta=2.4728$, $\kappa=0.2939$, and $\lambda=1.036$ are used. The results are shown in figure \ref{lh-vol}.

The ground states both in the $0^+$ and $1^-$ channel show again an essentially stable behavior starting with the $16^4$ lattices. The same is also true for the $0^-$ state, though with rather large error bars, while for the $2^+$ no conclusive results are available.

The second level of the $0^+$ is, even for all volumes, stable, and essentially consistent with the threshold. This is not true for the third level, which comes out too high. The situation for the higher levels in the 1$^-$ channel is more involved. The second level is far off for any volumes smaller than $24^4$. For the largest two volumes, it clusters around the expected level, but not fully convincing yet. The third level is also consistently to high compared to the expected level. Still, in all cases none of the higher levels show a behavior which strongly suggest that any of the states is surplus compared to the expected scattering states.

\begin{figure}
\includegraphics[width=\linewidth]{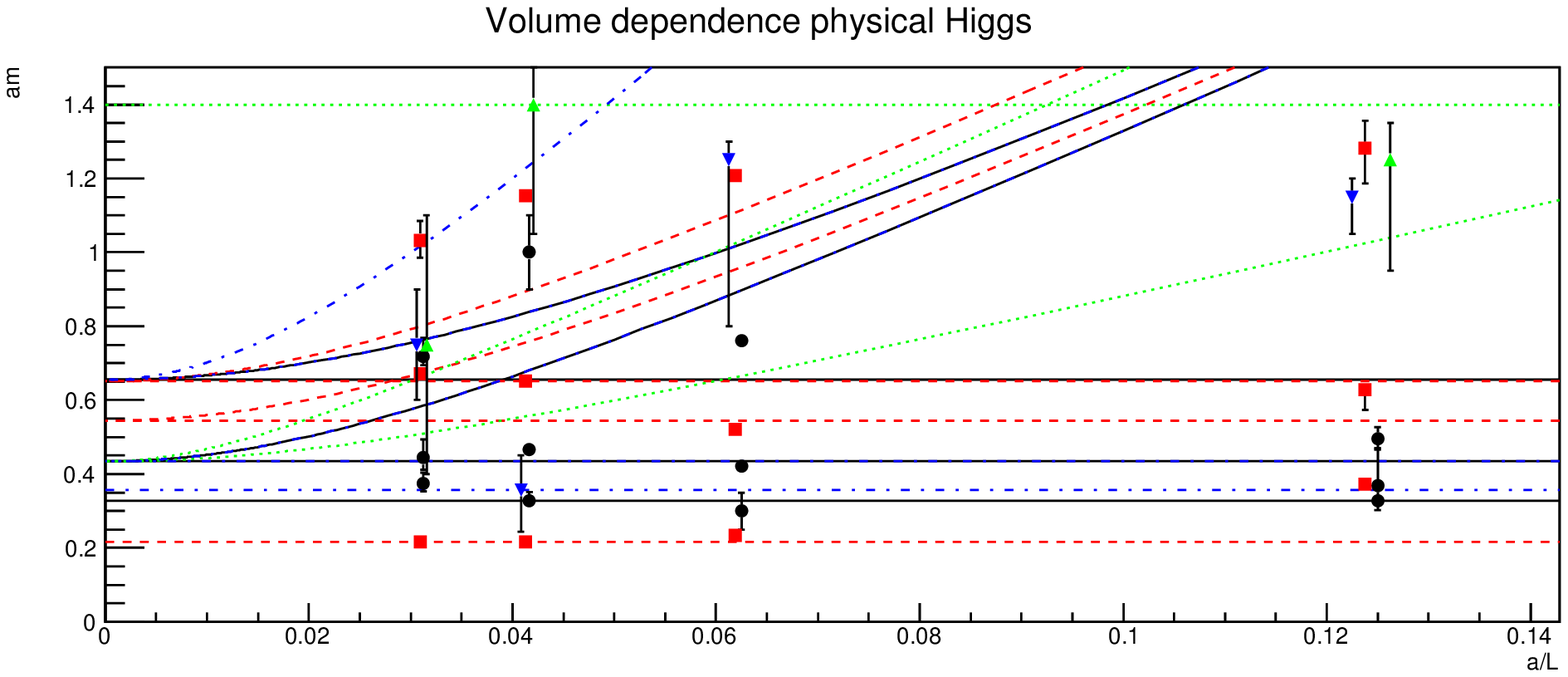}
\caption{\label{ph-vol}The volume dependence of the extracted states for a physical Higgs. Symbols and lines have the same meaning as in figure \ref{qld-vol}.}
\end{figure}

For a physical Higgs, the parameters $\beta=2.7984$, $\kappa=0.2954$, and $\lambda=1.317$ are used. The results are shown in figure \ref{ph-vol}.

The ground state in the $0^+$ channel is essentially volume independent. However, the $1^-$ state only converges to its final volume on the 24$^4$ lattice, though it is very close already at $16^4$. The results in the $0^-$ and $2^+$ channels are statistically too unreliable to give a final result, but within two standard deviations the results agree on the larger volumes.

The second state in the $0^+$ channel is for all volumes pretty good in agreement with a scattering state. The third state comes out too high for all volumes larger than $8^4$. Still, the result is consistent with the absence of further states. The second state in the $1^-$ channel shows a drift to the 3-particle scattering state for the large volumes, while it is compatible with a 2-particle scattering state for the smaller volumes. Though the operator basis includes both states, it appears that the resolution for larger volumes is difficult. The third states comes out substantially too high. In total, the result is compatible with the statement in the main text that no reasonable signal for additional states is seen.

\begin{figure}
\includegraphics[width=\linewidth]{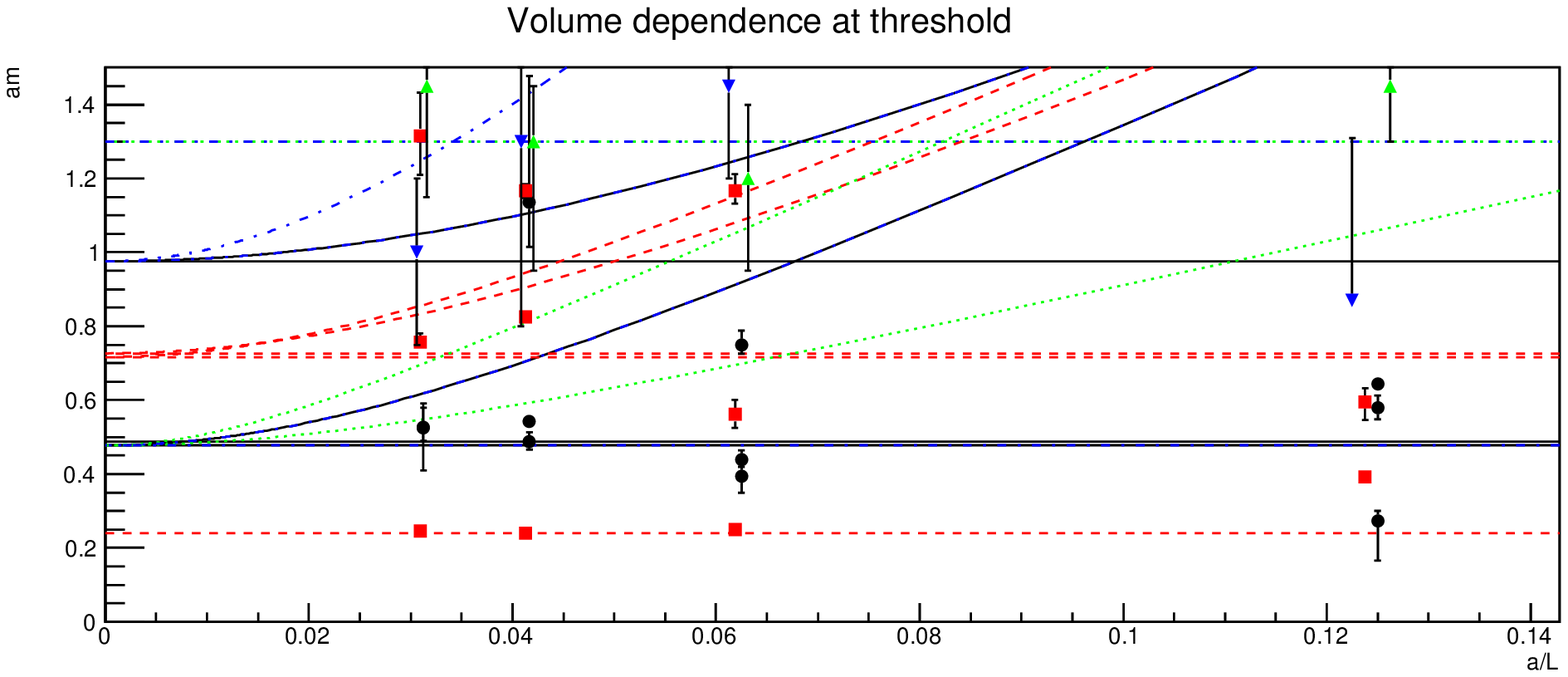}
\caption{\label{th-vol}The volume dependence of the extracted states for the threshold case. Symbols and lines have the same meaning as in figure \ref{qld-vol}.}
\end{figure}

Finally, at threshold, the parameters $\beta=2.7984$, $\kappa=0.2954$, and $\lambda=1.264$ are used. At this mass, there are potentially two scenarios. One is that there is indeed no bound state in the 0$^+$ channel. The other is that there is a bound state at threshold, in addition to a scattering state. To illustrate both options, both interpretations are included for the expected states in figure \ref{th-vol}, where also the results are shown.

Indeed, in the $0^+$ channel there are two states around threshold, only lightly dependent on volume. The third state increases substantially with volume, and may therefore correspond to the next expected state. Thus, the results in this case correspond best to a state close to threshold. Whether it is slightly above or below threshold would require statistically better results and possibly a phase shift analysis.

The lowest state in the $1^-$ channel is again essentially volume-independent at above $16^4$ volumes. The next state moves substantially around, indicating still some problems with fitting. Nonetheless, it predominantly clusters around the threshold region, winch has two very close-by independent thresholds, one based on a $0^+$-$1^-$ combination and the other a three $1^-$ combination. However, the next state is again substantially higher, showing that again the coupling to either of the scattering states is absent.

The states in the $0^-$ and $2^+$ channels are essentially consistent for the different volumes, tough with large error bars. They are also both above threshold, and therefore possibly scattering states.

In total, all cases show the same results: the ground states on the $24^4$ lattices employed in the main text are essentially already in an essentially volume-independent regime. Thus, the corresponding results appear rather reliable. Any other states in the HLD are usually above or around scattering states. However, not all scattering states are seen, even though they are present in the operator basis. On the other hand, no surplus states are seen. Thus, in agreement with the main text, the additional states appear to be all scattering states, except for the QLD. However, the assignment to which scattering states would require better statistics, and possibly a larger operator basis. Thus, the volume-dependence of the results seems, at least to some extent, to qualitatively support the main text.

\subsection{Discretization}\label{a:a}

Analyzing discretization artifacts is much more complicated. The original hope that some of the analyzed states could be sufficiently stable throughout the phase diagram could not be confirmed. Thus, only the ground state in the $0^+$ and $1^-$ channels can be used to identify LCPs. Deep inside the threshold region or the QLD, where one of the states would become also unstable, even only one. Thus, a third parameter is missing. Once a statistically reliable set of different volumes is obtained, the phase shift will provide a possibility to have further physical parameters to fix LCPs. But this is not yet the case.

To circumvent the problem, here the running gauge coupling will be used, as determined from the renormalization-scale invariant combination
\be
\alpha(p)=\alpha(\mu_0)p^6D_G^2(p)D_W(p)\nn,
\ee
\no where $\alpha(\mu_0)$ is the coupling at the lattice cutoff obtained from $\beta$, $D_G$ is the (minimal-)Landau gauge ghost propagator and $D_W$ is the $W$-propagator. Details of their determination can be found in \cite{Maas:2013aia}. Note that this running gauge coupling is also used to obtain an independent determination of $\Lambda_{\overline{\text{MS}}}$ for QCD \cite{pdg}, and therefore represents a sufficiently sound basis to analyze discretization artifacts.

To fix the other two parameters, only parameter combinations from the physical Higgs region have been included, i.\ e.\ $m_{0^+}=125\pm10$ GeV. Finally, the last condition is that $\alpha(200$ GeV$)=0.66\pm0.05$. To avoid any consequences due to violations of rotational symmetry, all momenta are evaluated on the $x-$axis, and a linear interpolation is performed between the two closest momentum values above an below 200 GeV.

\begin{figure}
\includegraphics[width=\linewidth]{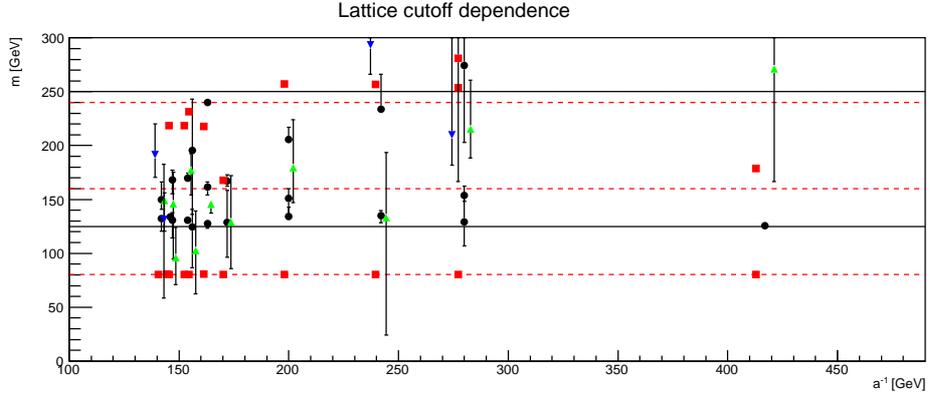}
\caption{\label{sp-a}The cutoff dependence of the extracted states for a physical Higgs. Different channels have been slightly displaced for better visibility. Symbols and lines have the same meaning as in figure \ref{qld-vol}. Especially, the black solid lines are in the $0^+$ channel made from constituents without relative momenta and the red dashed lines are in the $1^-$ channel made from constituents without relative momenta.}
\end{figure}

The dependence of the masses on the lattice cutoff for these requirements are shown in figure \ref{sp-a}. By construction, the lowest states in the $0^+$ and $1^-$ channel are independent of the lattice cutoff.

The $0^-$ states have rather large error bars. However, consistent with the interpretation that they are made up from two $1^-$ particles with relative momenta, they are, within $2\sigma$, always consistent with this result. Furthermore, they have a trend upward for finer lattices, though not a statistically relevant one, for smaller $a$, which would be expected in case of relative momentum of the constituents. The few $2^+$ state also have rather large errors, but except for one they are also in a $2\sigma$ band around the expected 2-$1^-$ structure with no relative momentum.

The second state in the $0^+$ channel is, where seen and except in one case, always consistent with the expected 2-$1^-$-particle substructure with no relative internal momenta, for all lattice cutoffs. The third state only appears relatively late, which is to be expected given the cut to include only states with energy in lattice units below 1.5, and then consistent with two $0^+$ ground states with no internal momentum, again for all lattice cutoffs.

The only state showing a trend with better discretization is the second state in the $1^-$ channel, which moves upwards, and somewhat overshots then the expected scattering threshold. What happens at the finest discretization is unclear. However, the correlator shows a rather large fluctuation at large times, so it may be possible that substantially increased statistics will change the situation. Again, as in the case of the third $0^+$ state, the third $1^-$ almost never appears, due to the cut on the lattice mass of 1.5.

Concluding, within statistical error, so far only the second $1^-$ state shows a systematic dependence on the lattice cutoff. Though better statistics may reveal more systematic behavior, at the current time it appears that the results of the main text are mostly unaffected by discretization artifacts.

\subsection{Fitting}\label{a:fit}

As can be seen in figure \ref{exp-1-3}-\ref{exp-rest}, many correlators show a pronounced behavior as a function of time, and not a single plateau. Especially, at larger times often a relaxation to a smaller mass is observed, necessitating two-states fits. On the other hand, especially for states higher up in the spectrum, a fit over the full time extent is rarely possible, with the available statistics. Hence, it may well be that especially for the higher states some intermediate state are missed. This would explain that sometimes particular scattering states in the systematic studies of sections \ref{a:vol} and \ref{a:a} are not seen: It may well be that they would appear at longer time extent. This is supported by the fact that usually expected states higher up in the spectrum are observed. Thus, the limited statistics plays here an important role. This problem is aggravated by the fact that the density of states here is, compared to e.\ g.\ QCD, very small, as no additional bound states seem to appear. Hence, statistics so far seriously limits the identification of states above the ground states. In the $0^-$ and $2^+$ channels, where there appear to be no bound ground state in the HLD, this problem becomes even more severe, as thus the lowest state is already very heavy in lattice units, and therefore the correlators often drown early in noise.

Concluding, the limited statistics plays an important role in the identification of non-bound states. More statistics will be needed to reliably exclude the existence of non-trivial states. Of course, especially for higher states non-zero momentum states will also be relevant. A possible set of further operators is given, e.\ g., in \cite{Wurtz:2013ova}.

\section{Numerical values}\label{numval}

In this appendix we provide the raw numerical values for the energy levels in lattice units. We report only values which have been used in the main text. Especially, we only give those results where the average value in lattice units are at or below 1.5. This implies that some (very few) entries are completely empty, indicating that for these parameters no states with a mass below 1.5 have been found. The errors are 1$\sigma$ statistical ones only. As discussed in section \ref{s:fitex} the systematic uncertainties in the fitting procedure for energies above 1 in lattice units alone will probably exceed these errors. Hence, caution should be exercised when comparing to these values. Furthermore, especially for states significantly above the lightest, finite-volume corrections may be large \cite{Luscher:1985dn,Luscher:1986pf}. The results shown in sections \ref{s:qld}-\ref{s:a} are given in tables \ref{t:qld}-\ref{t:a}, respectively. The ordering in the table is, as in the plots in the main text, according to the ratio of the ground states masses in the $1^-_3$ to the $0^+_1$ channel in decreasing order, as long as the $0^+_1$ channel satisfies the stability criterion discussed in the main text. After that, the entries are in order of decreasing lattice spacing.

\afterpage{\begin{landscape}
\begin{small}

\begin{longtable}{|c|c|c|c|c|c|c|c|c|c|c|}
\caption{\label{t:qld}The raw numerical values for the energy levels for the various values of $\beta$, $\kappa$, and $\lambda$ in the QCD-like domain, which have been used in section \ref{s:qld}. 1$^\text{st}$, 2$^\text{nd}$, and 3$^\text{rd}$ refer to the respective levels in the corresponding channel.}\\
 \hline
 $\beta$ & $\kappa$ & $\lambda$ & $0^+_1$ (1$^\text{st}$) & $0^+_1$ (2$^\text{nd}$) & $0^+_1$ (3$^\text{rd}$) & $1^-_3$ (1$^\text{st}$) & $1^-_3$ (2$^\text{nd}$) & $1^-_3$ (3$^\text{rd}$) & $0^-_1$ & $2^+_1$\endfirsthead
 \hline
 \multicolumn{11}{|l|}{Table \ref{t:qld} continued}\\
 \hline
 $\beta$ & $\kappa$ & $\lambda$ & $0^+_1$ (1$^\text{st}$) & $0^+_1$ (2$^\text{nd}$) & $0^+_1$ (3$^\text{rd}$) & $1^-_3$ (1$^\text{st}$) & $1^-_3$ (2$^\text{nd}$) & $1^-_3$ (3$^\text{rd}$) & $0^-_1$ & $2^+_1$ \endhead
 \hline
 \multicolumn{11}{|r|}{Continued on next page}\\
 \hline\endfoot
 \endlastfoot
 \hline
2.4882 & 0.125 & 0 & 0.88$^{+0.04}_{-0.01}$ &  &  &  &  &  & 1.1$^{+0.1}_{-0.3}$ & 1.2$^{+0.2}_{-0.6}$\cr
\hline
2.3724 & 0.125 & 0 & 1.0$^{+0.1}_{-0.3}$ &  &  &  &  &  & 1.2$^{+0.1}_{-0.3}$ & \cr
\hline
2.4492 & 0.2939 & 1.036 & 0.111$^{+0.003}_{-0.003}$ & 0.225$^{+0.003}_{-0.003}$ & 0.490$^{+0.002}_{-0.001}$ & 0.283$^{+0.003}_{-0.003}$ & 0.512$^{+0.009}_{-0.009}$ & 1.09$^{+0.02}_{-0.02}$ & 0.19$^{+0.01}_{-0.01}$ & \cr
\hline
2.7752 & 0.125 & 0 & 0.91$^{+0.05}_{-0.02}$ &  &  &  &  &  &  & 1.2$^{+0.2}_{-0.3}$\cr
\hline
2.7984 & 0.2895 & 1.317 & 0.32$^{+0.03}_{-0.02}$ & 0.82$^{+0.04}_{-0.02}$ &  & 0.74$^{+0.08}_{-0.07}$ &  &  &  & 1.41$^{+0.02}_{-0.02}$\cr
\hline
4.2230 & 0.2249 & 0.9762 & 0.9$^{+0.3}_{-0.3}$ &  &  &  &  &  & 1.2$^{+0.3}_{-0.7}$ & 0.9$^{+0.3}_{-0.1}$\cr
\hline
2.5331 & 0.2860 & 1.121 & 0.6$^{+0.1}_{-0.1}$ & 0.96$^{+0.01}_{-0.01}$ & 1.4$^{+0.1}_{-0.1}$ & 1.3$^{+0.0}_{-0.1}$ &  &  & 0.6$^{+0.2}_{-0.5}$ & \cr
\hline
2.6626 & 0.125 & 0 & 1.0$^{+0.1}_{-0.3}$ &  &  &  &  &  &  & 0.68$^{+0.06}_{-0.06}$\cr
\hline
2.2651 & 0.125 & 0 & 1.1$^{+0.1}_{-0.1}$ &  &  &  &  &  & 1.4$^{+0.0}_{-0.2}$ & 1.1$^{+0.2}_{-0.3}$\cr
\hline
2.2210 & 0.125 & 0 & 1.5$^{+0.0}_{-0.1}$ &  &  &  &  &  & 0.9$^{+0.1}_{-0.2}$ & \cr
\hline
2.2410 & 0.125 & 0 & 1.39$^{+0.02}_{-0.09}$ &  &  &  &  &  & 1.5$^{+0.0}_{-0.2}$ & \cr
\hline
2.4245 & 0.125 & 0 & 1.06$^{+0.03}_{-0.02}$ &  &  &  &  &  & 0.9$^{+0.2}_{-0.1}$ & \cr
\hline
2.4728 & 0.2939 & 1.088 & 0.247$^{+0.005}_{-0.004}$ & 0.41$^{+0.02}_{-0.02}$ & 0.87$^{+0.03}_{-0.03}$ & 0.47$^{+0.02}_{-0.02}$ &  &  & 0.58$^{+0.08}_{-0.09}$ & \cr
\hline
2.4728 & 0.2792 & 1.036 & 0.47$^{+0.02}_{-0.02}$ & 1.26$^{+0.02}_{-0.02}$ &  & 0.891$^{+0.000}_{-0.001}$ &  &  & 0.9$^{+0.3}_{-0.4}$ & \cr
\hline
2.2010 & 0.125 & 0 &  &  &  &  &  &  & 1.2$^{+0.1}_{-0.1}$ & \cr
\hline
2.3296 & 0.125 & 0 & 1.21$^{+0.03}_{-0.02}$ &  &  &  &  &  & 1.1$^{+0.3}_{-0.2}$ & \cr
\hline
2.1810 & 0.125 & 0 &  &  &  &  &  &  &  & \cr
\hline
2.3095 & 0.2668 & 0.5254 & 0.47$^{+0.02}_{-0.03}$ & 0.88$^{+0.02}_{-0.02}$ & 1.22$^{+0.02}_{-0.01}$ & 0.9$^{+0.1}_{-0.1}$ & 1.4$^{+0.5}_{-0.2}$ &  & 0.44$^{+0.06}_{-0.03}$ & \cr
\hline
2.3050 & 0.2857 & 1.455 & 1.17$^{+0.06}_{-0.07}$ &  &  &  &  &  & 1.2$^{+0.1}_{-0.4}$ & \cr
\hline
2.6000 & 0.2808 & 1.020 & 0.7$^{+0.2}_{-0.0}$ & 0.93$^{+0.02}_{-0.03}$ & 1.3$^{+0.5}_{-0.1}$ & 1.24$^{+-0.14}_{-0.02}$ &  &  &  & 1.3$^{+0.2}_{-0.6}$\cr
\hline
2.6434 & 0.2905 & 1.398 & 0.754$^{+0.006}_{-0.002}$ & 0.92$^{+0.06}_{-0.06}$ & 1.5$^{+0.5}_{-0.1}$ & 1.19$^{+0.03}_{-0.07}$ &  &  & 0.9$^{+0.4}_{-0.4}$ & 0.6$^{+0.3}_{-0.1}$\cr
\hline
2.7622 & 0.2914 & 1.353 & 0.48$^{+0.02}_{-0.01}$ & 0.88$^{+0.00}_{-0.02}$ & 1.19$^{+0.08}_{-0.04}$ & 0.74$^{+0.05}_{-0.05}$ &  &  &  & 1.1$^{+0.2}_{-0.0}$\cr
\hline
4.0000 & 0.2500 & 1.0000 & 1.0$^{+0.3}_{-0.2}$ & 1.43$^{+0.02}_{-0.01}$ &  & 1.5$^{+0.0}_{-0.7}$ &  &  &  & 1.2$^{+0.1}_{-0.1}$\cr
\hline
2.3637 & 0.2939 & 1.293 & 1.08$^{+0.02}_{-0.02}$ &  &  & 1.5$^{+0.1}_{-0.2}$ &  &  & 1.0$^{+0.2}_{-0.4}$ & \cr
\hline
2.7984 & 0.2836 & 1.317 & 0.849$^{+0.008}_{-0.005}$ & 1.034$^{+0.008}_{-0.006}$ &  & 1.2$^{+0.0}_{-0.1}$ &  &  &  & 1.16$^{+0.05}_{-0.03}$\cr
\hline
2.3346 & 0.2590 & 0.4979 & 0.835$^{+0.009}_{-0.008}$ & 1.215$^{+0.010}_{-0.008}$ &  & 1.2$^{+0.1}_{-0.1}$ &  &  & 0.87$^{+0.07}_{-0.06}$ & \cr
\hline
2.1712 & 0.2889 & 0.7224 & 1.2$^{+0.0}_{-0.3}$ &  &  &  &  &  & 0.9$^{+0.5}_{-0.2}$ & 1.4$^{+0.5}_{-0.8}$\cr
\hline
2.0000 & 0.2500 & 0 & 0.7$^{+0.1}_{-0.1}$ &  &  & 1.0$^{+1.1}_{-0.5}$ &  &  &  & \cr
\hline
4.9409 & 0.125 & 0 & 1.17$^{+0.07}_{-0.00}$ &  &  & 1.5$^{+0.1}_{-0.1}$ &  &  &  & 1.1$^{+0.1}_{-0.1}$\cr
\hline
3.0000 & 0.3150 & 5.000 & 1.147$^{+0.010}_{-0.008}$ &  &  & 1.4$^{+0.1}_{-0.1}$ &  &  &  & 1.08$^{+0.05}_{-0.03}$\cr
\hline
2.2171 & 0.3182 & 1.046 & 0.47$^{+0.01}_{-0.02}$ & 1.04$^{+0.02}_{-0.02}$ & 1.209$^{+0.001}_{-0.001}$ & 0.569$^{+0.004}_{-0.006}$ & 1.203$^{+0.001}_{-0.001}$ &  & 0.48$^{+0.02}_{-0.03}$ & \cr
\hline
4.0000 & 0.3150 & 1.050 & 0.26$^{+0.03}_{-0.00}$ & 0.91$^{+0.03}_{-0.03}$ & 1.317$^{+0.005}_{-0.002}$ & 0.314$^{+0.002}_{-0.002}$ &  &  &  & 1.1$^{+0.0}_{-0.5}$\cr
\hline
2.9147 & 0.125 & 0 & 0.6$^{+0.1}_{-0.0}$ &  &  & 0.8$^{+0.4}_{-0.3}$ &  &  &  & 1.1$^{+0.2}_{-0.2}$\cr
\hline
2.8739 & 0.2919 & 1.551 & 0.69$^{+0.04}_{-0.03}$ & 1.4$^{+0.2}_{-0.1}$ &  & 0.8$^{+0.1}_{-0.1}$ & 1.5$^{+0.2}_{-0.1}$ &  &  & 0.9$^{+0.2}_{-0.0}$\cr
\hline
4.6000 & 0.2500 & 1.0000 & 1.3$^{+0.0}_{-0.3}$ & 1.3$^{+0.2}_{-0.2}$ &  & 1.4$^{+0.1}_{-0.2}$ &  &  & 1.3$^{+0.8}_{-0.8}$ & 1.3$^{+0.0}_{-0.5}$\cr
\hline
\end{longtable}

\begin{longtable}{|c|c|c|c|c|c|c|c|c|c|c|}
\caption{\label{t:cor}The raw numerical values for the energy levels for the various values of $\beta$, $\kappa$, and $\lambda$ in the crossover region, which have been used in section \ref{s:cor}. 1$^\text{st}$, 2$^\text{nd}$, and 3$^\text{rd}$ refer to the respective levels in the corresponding channel.}\\
 \hline
 $\beta$ & $\kappa$ & $\lambda$ & $0^+_1$ (1$^\text{st}$) & $0^+_1$ (2$^\text{nd}$) & $0^+_1$ (3$^\text{rd}$) & $1^-_3$ (1$^\text{st}$) & $1^-_3$ (2$^\text{nd}$) & $1^-_3$ (3$^\text{rd}$) & $0^-_1$ & $2^+_1$\endfirsthead
 \hline
 \multicolumn{11}{|l|}{Table \ref{t:cor} continued}\\
 \hline
 $\beta$ & $\kappa$ & $\lambda$ & $0^+_1$ (1$^\text{st}$) & $0^+_1$ (2$^\text{nd}$) & $0^+_1$ (3$^\text{rd}$) & $1^-_3$ (1$^\text{st}$) & $1^-_3$ (2$^\text{nd}$) & $1^-_3$ (3$^\text{rd}$) & $0^-_1$ & $2^+_1$ \endhead
 \hline
 \multicolumn{11}{|r|}{Continued on next page}\\
 \hline\endfoot
 \endlastfoot
\hline
2.2060 & 0.3084 & 0.9983 & 0.926$^{+0.008}_{-0.008}$ &  &  & 0.931$^{+0.002}_{-0.008}$ &  &  &  & \cr
\hline
2.2943 & 0.125 & 0 & 1.29$^{+0.04}_{-0.02}$ &  &  & 1.3$^{+0.8}_{-0.5}$ &  &  & 1.2$^{+0.2}_{-0.3}$ & \cr
\hline
2.5008 & 0.2829 & 1.015 & 0.92$^{+0.02}_{-0.03}$ &  &  & 0.9$^{+0.1}_{-0.1}$ & 1.29$^{+0.01}_{-0.01}$ &  & 0.5$^{+0.5}_{-0.3}$ & 1.1$^{+0.2}_{-0.0}$\cr
\hline
2.2975 & 0.3254 & 1.371 & 0.48$^{+0.08}_{-0.02}$ & 0.7$^{+0.1}_{-0.1}$ & 1.22$^{+0.08}_{-0.00}$ & 0.471$^{+0.006}_{-0.007}$ & 1.279$^{+0.002}_{-0.002}$ &  & 0.73$^{+0.06}_{-0.09}$ & 1.4$^{+0.1}_{-0.1}$\cr
\hline
2.7665 & 0.2954 & 1.317 & 0.22$^{+0.05}_{-0.02}$ & 0.43$^{+0.01}_{-0.01}$ & 0.65$^{+0.02}_{-0.02}$ & 0.205$^{+0.002}_{-0.002}$ & 0.60$^{+0.01}_{-0.01}$ & 1.05$^{+0.02}_{-0.02}$ &  & \cr
\hline
2.7824 & 0.2954 & 1.317 & 0.23$^{+0.08}_{-0.04}$ & 0.31$^{+0.04}_{-0.01}$ & 0.52$^{+0.02}_{-0.01}$ & 0.213$^{+0.002}_{-0.002}$ & 0.703$^{+0.010}_{-0.010}$ & 1.12$^{+0.02}_{-0.02}$ & 0.8$^{+0.3}_{-0.4}$ & 1.2$^{+0.1}_{-0.4}$\cr
\hline
\end{longtable}

\begin{longtable}{|c|c|c|c|c|c|c|c|c|c|c|}
\caption{\label{t:lh}The raw numerical values for the energy levels for the various values of $\beta$, $\kappa$, and $\lambda$ for a light Higgs, which have been used in section \ref{s:lh}. 1$^\text{st}$, 2$^\text{nd}$, and 3$^\text{rd}$ refer to the respective levels in the corresponding channel.}\\
 \hline
 $\beta$ & $\kappa$ & $\lambda$ & $0^+_1$ (1$^\text{st}$) & $0^+_1$ (2$^\text{nd}$) & $0^+_1$ (3$^\text{rd}$) & $1^-_3$ (1$^\text{st}$) & $1^-_3$ (2$^\text{nd}$) & $1^-_3$ (3$^\text{rd}$) & $0^-_1$ & $2^+_1$\endfirsthead
 \hline
 \multicolumn{11}{|l|}{Table \ref{t:lh} continued}\\
 \hline
 $\beta$ & $\kappa$ & $\lambda$ & $0^+_1$ (1$^\text{st}$) & $0^+_1$ (2$^\text{nd}$) & $0^+_1$ (3$^\text{rd}$) & $1^-_3$ (1$^\text{st}$) & $1^-_3$ (2$^\text{nd}$) & $1^-_3$ (3$^\text{rd}$) & $0^-_1$ & $2^+_1$ \endhead
 \hline
 \multicolumn{11}{|r|}{Continued on next page}\\
 \hline\endfoot
 \endlastfoot
\hline
2.7987 & 0.2953 & 1.320 & 0.23$^{+0.04}_{-0.02}$ & 0.45$^{+0.02}_{-0.02}$ & 0.82$^{+0.02}_{-0.04}$ & 0.209$^{+0.002}_{-0.002}$ & 0.62$^{+0.01}_{-0.01}$ & 1.07$^{+0.02}_{-0.02}$ & 0.9$^{+0.5}_{-0.5}$ & \cr
\hline
2.7984 & 0.3013 & 1.317 & 0.31$^{+0.10}_{-0.02}$ & 0.58$^{+0.05}_{-0.04}$ & 0.740$^{+0.002}_{-0.003}$ & 0.284$^{+0.003}_{-0.003}$ & 1.020$^{+0.008}_{-0.007}$ &  & 0.9$^{+0.2}_{-0.4}$ & \cr
\hline
2.7704 & 0.2954 & 1.317 & 0.23$^{+0.02}_{-0.01}$ & 0.39$^{+0.04}_{-0.03}$ & 0.7$^{+0.2}_{-0.1}$ & 0.211$^{+0.002}_{-0.002}$ & 0.593$^{+0.008}_{-0.008}$ & 1.15$^{+0.02}_{-0.02}$ & 0.6$^{+0.3}_{-0.5}$ & 1.5$^{+0.0}_{-0.2}$\cr
\hline
3.5010 & 0.2992 & 1.145 & 0.4$^{+0.1}_{-0.1}$ & 0.74$^{+0.02}_{-0.02}$ & 1.14$^{+0.04}_{-0.01}$ & 0.307$^{+0.002}_{-0.002}$ & 1.387$^{+0.003}_{-0.001}$ &  &  & \cr
\hline
4.0000 & 0.3000 & 0.9500 & 0.4$^{+0.1}_{-0.0}$ & 0.87$^{+0.03}_{-0.02}$ & 1.43$^{+0.05}_{-0.01}$ & 0.331$^{+0.003}_{-0.004}$ &  &  &  & \cr
\hline
2.2667 & 0.3141 & 1.043 & 0.67$^{+0.01}_{-0.01}$ & 1.36$^{+0.02}_{-0.02}$ &  & 0.50$^{+0.01}_{-0.03}$ & 1.320$^{+0.002}_{-0.002}$ &  & 0.57$^{+0.10}_{-0.07}$ & 1.5$^{+0.7}_{-0.1}$\cr
\hline
2.3862 & 0.3174 & 1.169 & 0.6$^{+0.2}_{-0.2}$ & 0.921$^{+0.008}_{-0.006}$ &  & 0.45$^{+0.01}_{-0.04}$ & 1.465$^{+0.003}_{-0.001}$ &  & 1.3$^{+0.2}_{-0.2}$ & 1.2$^{+0.2}_{-0.2}$\cr
\hline
2.2847 & 0.3152 & 1.098 & 0.675$^{+0.004}_{-0.005}$ & 1.00$^{+0.02}_{-0.01}$ & 1.295$^{+0.006}_{-0.003}$ & 0.491$^{+0.003}_{-0.003}$ & 1.275$^{+0.001}_{-0.001}$ &  & 0.71$^{+0.04}_{-0.04}$ & 0.72$^{+1.21}_{-0.44}$\cr
\hline
2.4728 & 0.2939 & 1.036 & 0.411$^{+0.007}_{-0.007}$ & 0.65$^{+0.01}_{-0.01}$ & 1.048$^{+0.003}_{-0.004}$ & 0.292$^{+0.002}_{-0.002}$ & 0.650$^{+0.003}_{-0.003}$ & 1.20$^{+0.03}_{-0.02}$ & 0.4$^{+0.1}_{-0.1}$ & 0.6$^{+0.2}_{-0.1}$\cr
\hline
2.8518 & 0.2862 & 1.334 & 0.74$^{+0.05}_{-0.03}$ & 1.09$^{+0.02}_{-0.02}$ &  & 0.5$^{+0.2}_{-0.1}$ & 1.30$^{+0.06}_{-0.02}$ &  & 0.6$^{+0.2}_{-0.1}$ & 1.26$^{+0.05}_{-0.03}$\cr
\hline
4.0000 & 0.2850 & 1.030 & 0.35$^{+0.05}_{-0.05}$ & 0.66$^{+0.02}_{-0.02}$ & 0.94$^{+0.01}_{-0.01}$ & 0.245$^{+0.001}_{-0.001}$ & 1.02$^{+0.06}_{-0.05}$ & 1.237$^{+0.007}_{-0.006}$ & 0.9$^{+0.4}_{-0.3}$ & 1.3$^{+0.1}_{-0.2}$\cr
\hline 
\end{longtable}

\begin{longtable}{|c|c|c|c|c|c|c|c|c|c|c|}
\caption{\label{t:ph}The raw numerical values for the energy levels for the various values of $\beta$, $\kappa$, and $\lambda$ for a physical Higgs, which have been used in section \ref{s:ph}. 1$^\text{st}$, 2$^\text{nd}$, and 3$^\text{rd}$ refer to the respective levels in the corresponding channel.}\\
 \hline
 $\beta$ & $\kappa$ & $\lambda$ & $0^+_1$ (1$^\text{st}$) & $0^+_1$ (2$^\text{nd}$) & $0^+_1$ (3$^\text{rd}$) & $1^-_3$ (1$^\text{st}$) & $1^-_3$ (2$^\text{nd}$) & $1^-_3$ (3$^\text{rd}$) & $0^-_1$ & $2^+_1$\endfirsthead
 \hline
 \multicolumn{11}{|l|}{Table \ref{t:ph} continued}\\
 \hline
 $\beta$ & $\kappa$ & $\lambda$ & $0^+_1$ (1$^\text{st}$) & $0^+_1$ (2$^\text{nd}$) & $0^+_1$ (3$^\text{rd}$) & $1^-_3$ (1$^\text{st}$) & $1^-_3$ (2$^\text{nd}$) & $1^-_3$ (3$^\text{rd}$) & $0^-_1$ & $2^+_1$ \endhead
 \hline
 \multicolumn{11}{|r|}{Continued on next page}\\
 \hline\endfoot
 \endlastfoot
\hline
4.0000 & 0.3000 & 1.0000 & 0.46$^{+0.10}_{-0.07}$ & 0.81$^{+0.03}_{-0.03}$ & 1.40$^{+0.06}_{-0.03}$ & 0.315$^{+0.004}_{-0.005}$ & 0.7$^{+0.2}_{-0.1}$ &  &  & 0.3$^{+0.1}_{-0.2}$\cr
\hline
2.8518 & 0.2981 & 1.387 & 0.34$^{+0.07}_{-0.02}$ & 0.43$^{+0.03}_{-0.02}$ & 0.63$^{+0.03}_{-0.01}$ & 0.229$^{+0.002}_{-0.001}$ & 0.729$^{+0.007}_{-0.007}$ & 1.09$^{+0.03}_{-0.02}$ & 0.8$^{+0.5}_{-0.4}$ & 0.4$^{+0.2}_{-0.1}$\cr
\hline
2.7984 & 0.2954 & 1.330 & 0.32$^{+0.02}_{-0.02}$ & 0.59$^{+0.02}_{-0.01}$ & 0.9$^{+0.1}_{-0.3}$ & 0.213$^{+0.001}_{-0.001}$ & 0.677$^{+0.007}_{-0.007}$ & 1.18$^{+0.02}_{-0.01}$ & 1.2$^{+0.1}_{-0.4}$ & 1.5$^{+0.1}_{-0.2}$\cr
\hline
2.7984 & 0.2954 & 1.317 & 0.33$^{+0.02}_{-0.02}$ & 0.466$^{+0.006}_{-0.006}$ & 1.0$^{+0.1}_{-0.1}$ & 0.2171$^{+0.0007}_{-0.0008}$ & 0.65$^{+0.01}_{-0.01}$ & 1.153$^{+0.010}_{-0.009}$ & 1.4$^{+0.2}_{-0.4}$ & 0.4$^{+0.1}_{-0.1}$\cr
\hline
2.3000 & 0.3100 & 1.0000 & 0.72$^{+0.02}_{-0.05}$ & 0.91$^{+0.04}_{-0.06}$ & 1.377$^{+0.001}_{-0.002}$ & 0.47$^{+0.01}_{-0.01}$ & 1.271$^{+0.002}_{-0.002}$ &  & 0.90$^{+0.07}_{-0.05}$ & 1.5$^{+0.2}_{-0.3}$\cr
\hline
2.8303 & 0.2954 & 1.317 & 0.337$^{+0.002}_{-0.003}$ & 0.38$^{+0.04}_{-0.03}$ & 0.601$^{+0.006}_{-0.004}$ & 0.222$^{+0.002}_{-0.002}$ & 0.70$^{+0.02}_{-0.02}$ & 1.18$^{+0.03}_{-0.03}$ & 0.9$^{+0.5}_{-0.5}$ & 0.3$^{+0.2}_{-0.1}$\cr
\hline
2.3000 & 0.3200 & 1.100 & 0.8$^{+0.1}_{-0.2}$ & 1.3$^{+0.3}_{-0.4}$ &  & 0.516$^{+0.003}_{-0.004}$ & 1.483$^{+0.003}_{-0.002}$ &  & 0.7$^{+0.2}_{-0.3}$ & \cr
\hline
2.7000 & 0.2851 & 1.050 & 0.301$^{+0.007}_{-0.007}$ & 0.75$^{+0.05}_{-0.04}$ & 1.00$^{+0.01}_{-0.01}$ & 0.193$^{+0.001}_{-0.001}$ & 0.429$^{+0.007}_{-0.007}$ & 1.048$^{+0.008}_{-0.007}$ & 0.6$^{+0.2}_{-0.2}$ & 1.2$^{+0.2}_{-0.0}$\cr
\hline
2.7984 & 0.2954 & 1.343 & 0.32$^{+0.03}_{-0.03}$ & 0.58$^{+0.04}_{-0.02}$ & 1.196$^{+0.008}_{-0.008}$ & 0.205$^{+0.001}_{-0.001}$ & 0.585$^{+0.005}_{-0.005}$ & 1.14$^{+0.01}_{-0.01}$ & 0.8$^{+0.3}_{-0.2}$ & 1.4$^{+0.1}_{-0.2}$\cr
\hline
2.2945 & 0.3191 & 1.143 & 0.78$^{+0.01}_{-0.03}$ & 0.99$^{+0.03}_{-0.05}$ & 1.47$^{+0.01}_{-0.02}$ & 0.49$^{+0.00}_{-0.01}$ & 1.337$^{+0.001}_{-0.001}$ &  & 0.89$^{+0.04}_{-0.05}$ & \cr
\hline
2.3767 & 0.3178 & 1.154 & 0.7$^{+0.2}_{-0.2}$ & 0.97$^{+0.03}_{-0.03}$ &  & 0.468$^{+0.004}_{-0.009}$ & 0.98$^{+0.01}_{-0.01}$ &  & 0.8$^{+-0.2}_{-0.2}$ & \cr
\hline
2.7984 & 0.2954 & 1.304 & 0.34$^{+0.03}_{-0.02}$ & 0.50$^{+0.01}_{-0.01}$ & 0.9$^{+0.2}_{-0.1}$ & 0.215$^{+0.002}_{-0.002}$ & 0.73$^{+0.01}_{-0.01}$ & 1.03$^{+0.02}_{-0.02}$ &  & 1.3$^{+0.2}_{-0.5}$\cr
\hline
4.0000 & 0.4000 & 2.000 & 0.72$^{+0.03}_{-0.10}$ & 0.9$^{+0.5}_{-0.5}$ &  & 0.45$^{+0.04}_{-0.05}$ & 0.5$^{+1.0}_{-0.0}$ &  &  & 1.4$^{+0.1}_{-0.8}$\cr
\hline
2.6000 & 0.2925 & 1.061 & 0.46$^{+0.09}_{-0.08}$ & 0.55$^{+0.03}_{-0.02}$ & 1.0$^{+0.1}_{-0.3}$ & 0.287$^{+0.003}_{-0.003}$ & 0.906$^{+0.005}_{-0.005}$ & 1.0$^{+0.1}_{-0.4}$ & 0.8$^{+0.2}_{-0.1}$ & 0.8$^{+0.5}_{-0.1}$\cr
\hline
2.3579 & 0.3208 & 1.010 & 0.86$^{+0.04}_{-0.00}$ & 1.127$^{+0.006}_{-0.006}$ &  & 0.54$^{+0.01}_{-0.02}$ & 0.8$^{+0.0}_{-0.2}$ &  & 1.5$^{+0.2}_{-0.4}$ & \cr
\hline
2.2674 & 0.3157 & 0.9920 & 0.85$^{+0.01}_{-0.01}$ & 1.10$^{+0.03}_{-0.02}$ &  & 0.52$^{+0.01}_{-0.02}$ & 1.419$^{+0.006}_{-0.006}$ &  & 1.1$^{+0.1}_{-0.1}$ & \cr
\hline
2.2696 & 0.3195 & 1.034 & 0.9$^{+0.0}_{-0.1}$ & 1.14$^{+0.06}_{-0.09}$ &  & 0.547$^{+0.002}_{-0.002}$ & 1.487$^{+0.004}_{-0.003}$ &  & 0.7$^{+0.2}_{-0.2}$ & \cr
\hline
2.3634 & 0.3223 & 1.066 & 0.9$^{+0.1}_{-0.1}$ & 1.105$^{+0.004}_{-0.004}$ &  & 0.533$^{+0.002}_{-0.002}$ &  &  & 1.2$^{+0.2}_{-0.3}$ & \cr
\hline
4.5000 & 0.3000 & 1.0000 & 0.5$^{+0.2}_{-0.1}$ & 0.80$^{+0.03}_{-0.03}$ &  & 0.306$^{+0.002}_{-0.002}$ &  &  &  & 1.4$^{+0.1}_{-0.6}$\cr
\hline
2.3827 & 0.3017 & 1.018 & 0.635$^{+0.009}_{-0.008}$ & 0.73$^{+0.06}_{-0.03}$ & 1.1$^{+0.1}_{-0.0}$ & 0.39$^{+0.02}_{-0.02}$ & 1.121$^{+0.003}_{-0.003}$ &  & 0.9$^{+0.2}_{-0.2}$ & 1.1$^{+0.1}_{-0.6}$\cr
\hline
2.2500 & 0.3200 & 1.0000 & 0.93$^{+0.01}_{-0.08}$ & 1.1$^{+0.1}_{-0.1}$ &  & 0.567$^{+0.002}_{-0.002}$ &  &  & 1.0$^{+0.1}_{-0.2}$ & 1.4$^{+0.2}_{-0.1}$\cr
\hline
2.2664 & 0.3163 & 0.9745 & 0.916$^{+0.006}_{-0.005}$ &  &  & 0.552$^{+-0.003}_{-0.003}$ &  &  & 1.0$^{+0.2}_{-0.4}$ & 0.9$^{+0.4}_{-0.5}$\cr
\hline
2.4526 & 0.3161 & 1.313 & 0.67$^{+0.04}_{-0.01}$ & 0.75$^{+0.05}_{-0.10}$ & 1.03$^{+0.06}_{-0.01}$ & 0.403$^{+0.004}_{-0.005}$ & 1.285$^{+0.002}_{-0.002}$ &  & 0.9$^{+0.2}_{-0.2}$ & \cr
\hline
4.2000 & 0.3000 & 1.0000 & 0.52$^{+0.01}_{-0.01}$ & 0.97$^{+0.04}_{-0.03}$ & 1.46$^{+0.02}_{-0.02}$ & 0.309$^{+0.006}_{-0.001}$ & 0.7$^{+0.3}_{-0.1}$ &  &  & 1.4$^{+0.1}_{-0.2}$\cr
\hline
2.8144 & 0.2954 & 1.317 & 0.36$^{+0.03}_{-0.02}$ & 0.48$^{+0.01}_{-0.01}$ & 0.84$^{+0.06}_{-0.04}$ & 0.212$^{+0.002}_{-0.002}$ & 0.727$^{+0.009}_{-0.009}$ & 1.02$^{+0.02}_{-0.02}$ & 1.2$^{+0.4}_{-0.7}$ & \cr
\hline
2.4728 & 0.2939 & 0.9842 & 0.56$^{+0.02}_{-0.03}$ & 1.0$^{+0.1}_{-0.0}$ & 1.35$^{+0.01}_{-0.10}$ & 0.331$^{+0.004}_{-0.004}$ & 1.061$^{+0.002}_{-0.002}$ &  & 0.6$^{+0.2}_{-0.5}$ & 1.2$^{+0.1}_{-0.1}$\cr
\hline
\end{longtable}

\begin{longtable}{|c|c|c|c|c|c|c|c|c|c|c|}
\caption{\label{t:hh}The raw numerical values for the energy levels for the various values of $\beta$, $\kappa$, and $\lambda$ for a heavy Higgs, which have been used in section \ref{s:hh}. 1$^\text{st}$, 2$^\text{nd}$, and 3$^\text{rd}$ refer to the respective levels in the corresponding channel.}\\
 \hline
 $\beta$ & $\kappa$ & $\lambda$ & $0^+_1$ (1$^\text{st}$) & $0^+_1$ (2$^\text{nd}$) & $0^+_1$ (3$^\text{rd}$) & $1^-_3$ (1$^\text{st}$) & $1^-_3$ (2$^\text{nd}$) & $1^-_3$ (3$^\text{rd}$) & $0^-_1$ & $2^+_1$\endfirsthead
 \hline
 \multicolumn{11}{|l|}{Table \ref{t:hh} continued}\\
 \hline
 $\beta$ & $\kappa$ & $\lambda$ & $0^+_1$ (1$^\text{st}$) & $0^+_1$ (2$^\text{nd}$) & $0^+_1$ (3$^\text{rd}$) & $1^-_3$ (1$^\text{st}$) & $1^-_3$ (2$^\text{nd}$) & $1^-_3$ (3$^\text{rd}$) & $0^-_1$ & $2^+_1$ \endhead
 \hline
 \multicolumn{11}{|r|}{Continued on next page}\\
 \hline\endfoot
 \endlastfoot
\hline
2.2837 & 0.3248 & 1.120 & 0.9$^{+0.0}_{-0.2}$ &  &  & 0.548$^{+0.003}_{-0.003}$ &  &  & 1.23$^{+0.06}_{-0.04}$ & 1.1$^{+0.3}_{-0.4}$\cr
\hline
2.7984 & 0.2924 & 1.317 & 0.289$^{+0.004}_{-0.004}$ & 0.49$^{+0.03}_{-0.02}$ & 0.80$^{+0.04}_{-0.00}$ & 0.1695$^{+0.0007}_{-0.0007}$ & 0.415$^{+0.008}_{-0.008}$ & 0.71$^{+0.07}_{-0.10}$ & 1.0$^{+0.7}_{-0.5}$ & 1.10$^{+0.07}_{-0.05}$\cr
\hline
2.3107 & 0.3183 & 1.048 & 0.9$^{+0.1}_{-0.1}$ & 0.95$^{+0.00}_{-0.06}$ &  & 0.521$^{+0.002}_{-0.003}$ & 1.50$^{+0.00}_{-0.02}$ &  & 1.0$^{+0.2}_{-0.2}$ & 1.5$^{+0.1}_{-0.9}$\cr
\hline
3.8515 & 0.2771 & 2.708 & 0.6$^{+0.1}_{-0.1}$ &  &  & 0.4$^{+0.2}_{-0.2}$ & 0.9$^{+0.1}_{-0.1}$ &  &  & 0.5$^{+0.8}_{-0.1}$\cr
\hline
2.7984 & 0.2954 & 1.370 & 0.33$^{+0.02}_{-0.02}$ & 0.43$^{+0.02}_{-0.04}$ & 0.62$^{+0.05}_{-0.04}$ & 0.191$^{+0.001}_{-0.001}$ & 0.522$^{+0.006}_{-0.005}$ & 1.096$^{+0.009}_{-0.008}$ & 1.4$^{+0.2}_{-0.5}$ & 1.18$^{+0.05}_{-0.05}$\cr
\hline
2.4964 & 0.2939 & 1.036 & 0.51$^{+0.01}_{-0.03}$ & 1.030$^{+0.009}_{-0.009}$ & 1.310$^{+0.001}_{-0.001}$ & 0.288$^{+0.003}_{-0.004}$ & 0.981$^{+0.002}_{-0.002}$ &  & 0.64$^{+0.09}_{-0.07}$ & \cr
\hline
2.8264 & 0.2954 & 1.317 & 0.40$^{+0.04}_{-0.02}$ & 0.45$^{+0.03}_{-0.02}$ & 0.8$^{+0.2}_{-0.1}$ & 0.224$^{+0.002}_{-0.002}$ & 0.76$^{+0.01}_{-0.01}$ & 1.16$^{+0.03}_{-0.03}$ & 1.1$^{+0.5}_{-0.4}$ & 0.7$^{+0.7}_{-0.5}$\cr
\hline
2.3030 & 0.3168 & 0.9900 & 0.979$^{+0.005}_{-0.005}$ & 1.24$^{+0.02}_{-0.02}$ &  & 0.538$^{+0.002}_{-0.002}$ &  &  & 0.9$^{+0.3}_{-0.3}$ & \cr
\hline
2.6000 & 0.2925 & 0.9792 & 0.59$^{+0.06}_{-0.03}$ & 0.60$^{+0.08}_{-0.08}$ & 0.9$^{+0.4}_{-0.1}$ & 0.325$^{+0.005}_{-0.005}$ & 1.159$^{+0.003}_{-0.002}$ &  & 1.2$^{+0.2}_{-0.4}$ & 0.7$^{+0.4}_{-0.2}$\cr
\hline
2.2655 & 0.3220 & 0.9811 & 1.056$^{+0.008}_{-0.007}$ & 1.38$^{+0.04}_{-0.03}$ &  & 0.579$^{+0.003}_{-0.003}$ & 1.292$^{+0.009}_{-0.007}$ &  & 1.0$^{+0.0}_{-0.2}$ & 1.4$^{+0.3}_{-0.2}$\cr
\hline
2.2786 & 0.3224 & 0.9801 & 1.071$^{+0.006}_{-0.005}$ & 1.21$^{+0.04}_{-0.05}$ &  & 0.576$^{+0.002}_{-0.002}$ &  &  & 1.0$^{+0.2}_{-0.3}$ & \cr
\hline
2.8518 & 0.3100 & 1.334 & 0.63$^{+0.06}_{-0.02}$ & 0.74$^{+0.02}_{-0.01}$ & 1.05$^{+0.03}_{-0.01}$ & 0.338$^{+0.004}_{-0.005}$ & 1.0$^{+0.2}_{-0.2}$ & 1.3$^{+0.1}_{-0.1}$ &  & 1.5$^{+0.1}_{-0.1}$\cr
\hline
2.7000 & 0.2939 & 1.082 & 0.55$^{+0.03}_{-0.01}$ & 0.691$^{+0.005}_{-0.004}$ &  & 0.297$^{+0.003}_{-0.004}$ & 1.086$^{+0.003}_{-0.002}$ & 1.1$^{+0.2}_{-0.1}$ &  & \cr
\hline
2.3827 & 0.3176 & 0.9671 & 1.0$^{+0.1}_{-0.1}$ & 1.13$^{+0.01}_{-0.01}$ &  & 0.537$^{+0.003}_{-0.004}$ &  &  & 1.1$^{+0.4}_{-0.3}$ & \cr
\hline
2.6000 & 0.3042 & 1.020 & 0.69$^{+0.08}_{-0.03}$ & 0.88$^{+0.06}_{-0.03}$ & 1.3$^{+0.1}_{-0.0}$ & 0.37$^{+0.03}_{-0.06}$ & 0.6$^{+0.3}_{-0.1}$ & 1.34$^{+0.03}_{-0.01}$ & 0.6$^{+0.4}_{-0.3}$ & 1.5$^{+0.1}_{-0.2}$\cr
\hline
2.3023 & 0.3211 & 1.020 & 1.041$^{+0.005}_{-0.004}$ & 1.1$^{+0.1}_{-0.2}$ &  & 0.553$^{+0.002}_{-0.002}$ &  &  & 0.9$^{+0.1}_{-0.1}$ & 1.4$^{+0.7}_{-0.2}$\cr
\hline
2.3146 & 0.3201 & 1.071 & 0.983$^{+0.004}_{-0.004}$ & 1.11$^{+0.08}_{-0.04}$ &  & 0.520$^{+0.004}_{-0.006}$ &  &  & 0.9$^{+0.3}_{-0.3}$ & 1.5$^{+0.1}_{-0.2}$\cr
\hline
2.4072 & 0.3228 & 1.110 & 1.0$^{+0.1}_{-0.3}$ & 1.1$^{+0.0}_{-0.2}$ &  & 0.512$^{+0.003}_{-0.003}$ &  &  & 0.9$^{+0.3}_{-0.2}$ & \cr
\hline
2.2846 & 0.3224 & 0.9755 & 1.10$^{+0.00}_{-0.01}$ &  &  & 0.578$^{+0.002}_{-0.002}$ &  &  & 1.5$^{+0.1}_{-0.4}$ & \cr
\hline
2.7210 & 0.2939 & 1.050 & 0.59$^{+0.05}_{-0.04}$ & 0.74$^{+0.01}_{-0.02}$ &  & 0.308$^{+0.004}_{-0.004}$ & 1.0$^{+0.1}_{-0.1}$ & 1.184$^{+0.002}_{-0.001}$ & 0.9$^{+0.6}_{-0.6}$ & 1.3$^{+0.2}_{-0.1}$\cr
\hline
2.2630 & 0.3245 & 0.9179 & 1.181$^{+0.004}_{-0.005}$ & 1.29$^{+0.07}_{-0.06}$ &  & 0.614$^{+0.002}_{-0.002}$ &  &  & 1.49$^{+0.03}_{-0.03}$ & \cr
\hline
2.5500 & 0.2950 & 1.040 & 0.62$^{+0.02}_{-0.02}$ & 0.65$^{+0.02}_{-0.01}$ & 1.10$^{+0.01}_{-0.01}$ & 0.323$^{+0.002}_{-0.002}$ & 1.033$^{+0.007}_{-0.010}$ & 1.1$^{+0.2}_{-0.3}$ & 1.4$^{+0.1}_{-0.1}$ & \cr
\hline
2.3000 & 0.3100 & 1.0000 & 1.050$^{+0.007}_{-0.006}$ & 1.30$^{+0.03}_{-0.03}$ &  & 0.54$^{+0.01}_{-0.03}$ &  &  &  & \cr
\hline
2.2920 & 0.3269 & 1.042 & 1.118$^{+0.005}_{-0.005}$ & 1.14$^{+0.06}_{-0.03}$ &  & 0.577$^{+0.002}_{-0.002}$ &  &  & 1.2$^{+0.0}_{-0.4}$ & \cr
\hline
\end{longtable}

\begin{longtable}{|c|c|c|c|c|c|c|c|c|c|c|}
\caption{\label{t:th}The raw numerical values for the energy levels for the various values of $\beta$, $\kappa$, and $\lambda$ for a Higgs at threshold, which have been used in section \ref{s:th}. 1$^\text{st}$, 2$^\text{nd}$, and 3$^\text{rd}$ refer to the respective levels in the corresponding channel.}\\
 \hline
 $\beta$ & $\kappa$ & $\lambda$ & $0^+_1$ (1$^\text{st}$) & $0^+_1$ (2$^\text{nd}$) & $0^+_1$ (3$^\text{rd}$) & $1^-_3$ (1$^\text{st}$) & $1^-_3$ (2$^\text{nd}$) & $1^-_3$ (3$^\text{rd}$) & $0^-_1$ & $2^+_1$\endfirsthead
 \hline
 \multicolumn{11}{|l|}{Table \ref{t:th} continued}\\
 \hline
 $\beta$ & $\kappa$ & $\lambda$ & $0^+_1$ (1$^\text{st}$) & $0^+_1$ (2$^\text{nd}$) & $0^+_1$ (3$^\text{rd}$) & $1^-_3$ (1$^\text{st}$) & $1^-_3$ (2$^\text{nd}$) & $1^-_3$ (3$^\text{rd}$) & $0^-_1$ & $2^+_1$ \endhead
 \hline
 \multicolumn{11}{|r|}{Continued on next page}\\
 \hline\endfoot
 \endlastfoot
\hline
2.3744 & 0.3214 & 1.123 & 0.97$^{+0.04}_{-0.02}$ & 1.04$^{+0.01}_{-0.10}$ &  & 0.50$^{+0.01}_{-0.04}$ &  &  & 1.0$^{+-0.0}_{-0.2}$ & 1.5$^{+0.0}_{-0.2}$\cr
\hline
2.6240 & 0.2925 & 1.020 & 0.57$^{+0.04}_{-0.02}$ & 0.687$^{+0.007}_{-0.006}$ & 1.2$^{+0.1}_{-0.1}$ & 0.295$^{+0.008}_{-0.008}$ & 0.9$^{+0.1}_{-0.1}$ & 1.0$^{+-0.0}_{-0.3}$ &  & \cr
\hline
2.2916 & 0.3220 & 0.9560 & 1.1$^{+0.1}_{-0.1}$ & 1.140$^{+0.008}_{-0.006}$ &  & 0.57$^{+0.01}_{-0.08}$ &  &  & 1.3$^{+0.2}_{-0.2}$ & \cr
\hline
2.4000 & 0.5000 & 1.0000 &  &  &  & 1.1$^{+0.1}_{-0.1}$ &  &  &  & \cr
\hline
2.3000 & 0.3300 & 1.0000 & 1.225$^{+0.006}_{-0.005}$ & 1.5$^{+0.0}_{-0.2}$ &  & 0.604$^{+0.002}_{-0.002}$ &  &  & 1.5$^{+0.2}_{-0.1}$ & \cr
\hline
2.3000 & 0.3200 & 0.9000 & 1.166$^{+0.005}_{-0.005}$ & 1.5$^{+0.0}_{-0.3}$ &  & 0.587$^{+0.002}_{-0.002}$ &  &  & 1.46$^{+0.05}_{-0.05}$ & \cr
\hline
2.3498 & 0.3206 & 0.9991 & 1.134$^{+0.009}_{-0.006}$ & 1.3$^{+0.0}_{-0.4}$ &  & 0.550$^{+0.003}_{-0.003}$ &  &  & 1.2$^{+0.3}_{-0.3}$ & \cr
\hline
2.4486 & 0.3228 & 0.9699 & 1.1$^{+0.1}_{-0.3}$ & 1.32$^{+0.04}_{-0.03}$ &  & 0.550$^{+0.002}_{-0.002}$ &  &  &  & \cr
\hline
2.3504 & 0.3214 & 1.010 & 1.125$^{+0.007}_{-0.006}$ & 1.27$^{+0.04}_{-0.03}$ &  & 0.547$^{+0.003}_{-0.003}$ &  &  &  & \cr
\hline
2.3500 & 0.3200 & 1.0000 & 1.07$^{+0.10}_{-0.04}$ & 1.113$^{+0.009}_{-0.007}$ &  & 0.541$^{+0.003}_{-0.003}$ &  &  & 0.9$^{+0.1}_{-0.3}$ & \cr
\hline
2.8650 & 0.3197 & 0.8098 & 1.1$^{+0.1}_{-0.1}$ & 1.420$^{+0.006}_{-0.003}$ &  & 0.530$^{+0.004}_{-0.003}$ &  &  &  & \cr
\hline
2.3798 & 0.3224 & 1.061 & 1.101$^{+0.010}_{-0.008}$ & 1.25$^{+0.04}_{-0.03}$ &  & 0.530$^{+0.004}_{-0.004}$ &  &  & 1.4$^{+0.2}_{-0.2}$ & 1.4$^{+0.3}_{-0.1}$\cr
\hline
2.3636 & 0.3176 & 1.018 & 1.06$^{+0.05}_{-0.01}$ & 1.1$^{+0.1}_{-0.0}$ &  & 0.520$^{+0.003}_{-0.003}$ &  &  & 0.8$^{+0.0}_{-0.4}$ & \cr
\hline
2.3827 & 0.3176 & 1.018 & 1.080$^{+-0.015}_{-0.003}$ & 1.17$^{+0.05}_{-0.02}$ &  & 0.518$^{+0.002}_{-0.001}$ &  &  &  & 0.9$^{+0.1}_{-0.3}$\cr
\hline
2.3788 & 0.3227 & 1.104 & 1.08$^{+-0.03}_{-0.06}$ & 1.09$^{+0.07}_{-0.09}$ &  & 0.52$^{+0.00}_{-0.02}$ &  &  & 1.4$^{+0.3}_{-0.4}$ & \cr
\hline
2.3783 & 0.3221 & 1.118 & 1.1$^{+0.0}_{-0.1}$ & 1.18$^{+0.00}_{-0.03}$ &  & 0.50$^{+0.01}_{-0.06}$ &  &  & 1.5$^{+0.4}_{-0.6}$ & \cr
\hline
2.4305 & 0.3227 & 1.036 & 1.0$^{+0.2}_{-0.1}$ & 1.13$^{+-0.03}_{-0.03}$ &  & 0.495$^{+0.007}_{-0.009}$ &  &  & 1.1$^{+0.4}_{-0.3}$ & \cr
\hline
2.3827 & 0.3176 & 1.069 & 1.032$^{+0.006}_{-0.005}$ & 1.09$^{+0.03}_{-0.02}$ &  & 0.494$^{+0.004}_{-0.008}$ &  &  &  & \cr
\hline
2.3714 & 0.3178 & 1.148 & 0.93$^{+0.01}_{-0.08}$ & 1.05$^{+0.03}_{-0.03}$ &  & 0.46$^{+0.01}_{-0.02}$ & 1.428$^{+0.007}_{-0.007}$ &  & 1.2$^{+0.4}_{-0.4}$ & 1.3$^{+0.7}_{-0.2}$\cr
\hline
2.6957 & 0.2754 & 0.5674 & 0.88$^{+0.04}_{-0.09}$ & 1.034$^{+0.002}_{-0.004}$ &  & 0.442$^{+0.002}_{-0.003}$ &  &  & 1.1$^{+0.3}_{-0.3}$ & \cr
\hline
2.3964 & 0.3169 & 1.189 & 0.903$^{+0.006}_{-0.005}$ & 0.98$^{+0.02}_{-0.02}$ &  & 0.44$^{+0.01}_{-0.02}$ & 1.455$^{+0.003}_{-0.003}$ &  & 0.9$^{+0.2}_{-0.1}$ & \cr
\hline
2.4728 & 0.3086 & 1.036 & 0.86$^{+0.05}_{-0.05}$ & 0.955$^{+0.007}_{-0.006}$ &  & 0.437$^{+0.009}_{-0.009}$ & 1.497$^{+0.004}_{-0.006}$ &  & 1.5$^{+0.1}_{-0.5}$ & 1.0$^{+0.3}_{-0.0}$\cr
\hline
2.6790 & 0.2939 & 1.050 & 0.63$^{+0.06}_{-0.08}$ & 0.64$^{+0.03}_{-0.03}$ & 1.36$^{+0.01}_{-0.01}$ & 0.315$^{+0.003}_{-0.003}$ & 1.119$^{+0.002}_{-0.002}$ & 1.4$^{+0.2}_{-0.1}$ &  & \cr
\hline
2.7270 & 0.2939 & 1.050 & 0.63$^{+0.05}_{-0.04}$ & 0.772$^{+0.003}_{-0.003}$ &  & 0.312$^{+0.002}_{-0.002}$ & 1.081$^{+0.007}_{-0.006}$ & 1.17$^{+0.08}_{-0.08}$ &  & \cr
\hline
2.6730 & 0.2939 & 1.050 & 0.62$^{+0.03}_{-0.02}$ & 0.64$^{+0.02}_{-0.02}$ & 1.1$^{+0.1}_{-0.0}$ & 0.311$^{+0.002}_{-0.002}$ & 1.133$^{+0.002}_{-0.002}$ & 1.3$^{+0.1}_{-0.1}$ & 0.7$^{+0.2}_{-0.3}$ & 1.35$^{+0.02}_{-0.05}$\cr
\hline
2.7000 & 0.2939 & 1.050 & 0.62$^{+0.07}_{-0.05}$ & 0.6$^{+0.1}_{-0.0}$ & 0.9$^{+0.3}_{-0.1}$ & 0.311$^{+0.003}_{-0.003}$ & 1.188$^{+0.001}_{-0.002}$ & 1.4$^{+0.2}_{-0.1}$ & 0.7$^{+0.2}_{-0.3}$ & 0.4$^{+1.4}_{-0.0}$\cr
\hline
2.6000 & 0.2925 & 1.020 & 0.645$^{+0.007}_{-0.007}$ & 0.65$^{+0.02}_{-0.03}$ & 1.1$^{+0.2}_{-0.2}$ & 0.308$^{+0.003}_{-0.003}$ & 0.7$^{+0.0}_{-0.2}$ & 1.17$^{+0.01}_{-0.06}$ & 1.0$^{+0.2}_{-0.2}$ & \cr
\hline
2.7000 & 0.2939 & 1.060 & 0.63$^{+0.02}_{-0.02}$ & 0.71$^{+0.02}_{-0.01}$ & 1.2$^{+0.1}_{-0.0}$ & 0.305$^{+0.002}_{-0.002}$ & 0.88$^{+0.03}_{-0.03}$ & 1.04$^{+0.06}_{-0.06}$ & 0.6$^{+0.3}_{-0.2}$ & 0.6$^{+0.9}_{-0.4}$\cr
\hline
2.5760 & 0.2925 & 1.020 & 0.57$^{+0.04}_{-0.03}$ & 0.59$^{+0.04}_{-0.03}$ & 1.0$^{+0.2}_{-0.2}$ & 0.288$^{+0.004}_{-0.004}$ & 1.024$^{+0.002}_{-0.003}$ &  & 1.2$^{+0.2}_{-0.4}$ & \cr
\hline
2.7000 & 0.2910 & 1.050 & 0.57$^{+0.04}_{-0.03}$ & 0.652$^{+0.004}_{-0.004}$ & 1.07$^{+0.02}_{-0.02}$ & 0.285$^{+0.002}_{-0.002}$ & 0.98$^{+0.07}_{-0.03}$ & 1.3$^{+0.1}_{-0.1}$ & 0.9$^{+0.3}_{-0.4}$ & \cr
\hline
2.8518 & 0.2981 & 1.281 & 0.57$^{+0.03}_{-0.08}$ & 0.659$^{+0.007}_{-0.001}$ & 1.40$^{+0.07}_{-0.09}$ & 0.269$^{+0.002}_{-0.002}$ & 1.061$^{+0.001}_{-0.001}$ & 1.31$^{+0.06}_{-0.08}$ & 0.9$^{+0.2}_{-0.3}$ & 0.9$^{+0.4}_{-0.1}$\cr
\hline
4.0000 & 0.2850 & 1.0000 & 0.5$^{+0.1}_{-0.1}$ & 0.65$^{+0.01}_{-0.01}$ & 0.95$^{+0.05}_{-0.05}$ & 0.2534$^{+0.0007}_{-0.0008}$ & 1.12$^{+0.05}_{-0.05}$ & 1.281$^{+0.004}_{-0.005}$ &  & 1.0$^{+0.1}_{-0.1}$\cr
\hline
2.7984 & 0.2984 & 1.317 & 0.52$^{+0.03}_{-0.02}$ & 0.600$^{+0.010}_{-0.008}$ & 1.0$^{+0.3}_{-0.2}$ & 0.250$^{+0.002}_{-0.002}$ & 0.77$^{+0.01}_{-0.02}$ & 1.20$^{+0.04}_{-0.04}$ & 0.9$^{+0.2}_{-0.2}$ & 1.33$^{+0.06}_{-0.06}$\cr
\hline
2.8859 & 0.2981 & 1.334 & 0.49$^{+0.04}_{-0.04}$ & 0.54$^{+0.03}_{-0.00}$ & 0.87$^{+0.03}_{-0.06}$ & 0.249$^{+0.002}_{-0.002}$ & 0.79$^{+0.01}_{-0.01}$ & 1.16$^{+0.06}_{-0.05}$ & 0.6$^{+0.3}_{-0.2}$ & \cr
\hline
2.8518 & 0.2981 & 1.334 & 0.51$^{+0.02}_{-0.00}$ & 0.579$^{+0.009}_{-0.006}$ & 1.06$^{+0.07}_{-0.02}$ & 0.244$^{+0.001}_{-0.001}$ & 0.648$^{+0.007}_{-0.007}$ & 1.05$^{+0.02}_{-0.02}$ & 0.9$^{+0.3}_{-0.3}$ & 1.40$^{+0.06}_{-0.03}$\cr
\hline
2.7984 & 0.2954 & 1.264 & 0.49$^{+0.03}_{-0.02}$ & 0.54$^{+0.01}_{-0.01}$ & 1.1$^{+0.3}_{-0.1}$ & 0.2389$^{+0.0010}_{-0.0009}$ & 0.82$^{+0.01}_{-0.01}$ & 1.17$^{+0.02}_{-0.02}$ & 1.3$^{+0.2}_{-0.3}$ & 1.3$^{+0.2}_{-0.5}$\cr
\hline
2.7984 & 0.2954 & 1.291 & 0.5$^{+0.0}_{-0.1}$ & 0.49$^{+0.04}_{-0.06}$ & 0.9$^{+0.1}_{-0.1}$ & 0.228$^{+0.002}_{-0.002}$ & 0.742$^{+0.007}_{-0.007}$ & 0.97$^{+0.05}_{-0.04}$ &  & 1.5$^{+0.1}_{-0.2}$\cr
\hline
\end{longtable}

\begin{longtable}{|c|c|c|c|c|c|c|c|c|c|c|}
\caption{\label{t:a}The raw numerical values for the energy levels for the various values of $\beta$, $\kappa$, and $\lambda$ for the anomalous cases, which have been used in section \ref{s:a}. 1$^\text{st}$, 2$^\text{nd}$, and 3$^\text{rd}$ refer to the respective levels in the corresponding channel.}\\
 \hline
 $\beta$ & $\kappa$ & $\lambda$ & $0^+_1$ (1$^\text{st}$) & $0^+_1$ (2$^\text{nd}$) & $0^+_1$ (3$^\text{rd}$) & $1^-_3$ (1$^\text{st}$) & $1^-_3$ (2$^\text{nd}$) & $1^-_3$ (3$^\text{rd}$) & $0^-_1$ & $2^+_1$\endfirsthead
 \hline
 \multicolumn{11}{|l|}{Table \ref{t:a} continued}\\
 \hline
 $\beta$ & $\kappa$ & $\lambda$ & $0^+_1$ (1$^\text{st}$) & $0^+_1$ (2$^\text{nd}$) & $0^+_1$ (3$^\text{rd}$) & $1^-_3$ (1$^\text{st}$) & $1^-_3$ (2$^\text{nd}$) & $1^-_3$ (3$^\text{rd}$) & $0^-_1$ & $2^+_1$ \endhead
 \hline
 \multicolumn{11}{|r|}{Continued on next page}\\
 \hline\endfoot
 \endlastfoot
\hline
2.5000 & 0.5000 & 1.0000 &  &  &  & 1.0$^{+0.0}_{-0.3}$ &  &  &  & \cr
\hline
2.3900 & 0.3500 & 1.0000 &  &  &  & 0.672$^{+0.004}_{-0.004}$ &  &  &  & \cr
\hline
2.3827 & 0.3335 & 1.018 & 1.34$^{+0.01}_{-0.01}$ & 1.49$^{+0.07}_{-0.06}$ &  & 0.588$^{+0.004}_{-0.005}$ &  &  & 1.2$^{+0.8}_{-0.2}$ & \cr
\hline
2.9026 & 0.3578 & 1.226 & 1.2$^{+0.2}_{-0.2}$ &  &  & 0.548$^{+0.003}_{-0.002}$ & 1.0$^{+0.2}_{-0.1}$ &  &  & 0.9$^{+0.3}_{-0.1}$\cr
\hline
2.3881 & 0.3214 & 0.9974 & 1.15$^{+0.01}_{-0.01}$ & 1.25$^{+0.03}_{-0.03}$ &  & 0.54$^{+0.00}_{-0.01}$ &  &  & 1.5$^{+0.2}_{-0.4}$ & \cr
\hline
2.2989 & 0.3208 & 0.8284 & 1.2$^{+0.1}_{-0.1}$ & 1.22$^{+0.03}_{-0.03}$ &  & 0.5$^{+0.1}_{-0.1}$ & 1.00$^{+0.02}_{-0.02}$ &  & 1.4$^{+0.1}_{-0.2}$ & 1.5$^{+0.2}_{-0.3}$\cr
\hline
2.4018 & 0.3176 & 1.018 & 1.102$^{+-0.033}_{-0.008}$ & 1.18$^{+-0.05}_{-0.04}$ &  & 0.516$^{+0.003}_{-0.003}$ &  &  & 0.8$^{+0.3}_{-0.2}$ & 0.9$^{+0.2}_{-0.2}$\cr
\hline
3.8980 & 0.4689 & 3.859 &  &  &  & 0.490$^{+0.006}_{-0.006}$ &  &  &  & \cr
\hline
4.5915 & 0.4647 & 3.219 & 1.1$^{+0.1}_{-0.1}$ &  &  & 0.457$^{+0.005}_{-0.004}$ &  &  &  & 1.5$^{+0.4}_{-0.2}$\cr
\hline
2.3100 & 0.3130 & 1.0000 & 0.881$^{+0.006}_{-0.005}$ &  &  & 0.40$^{+0.07}_{-0.07}$ & 0.6$^{+0.1}_{-0.0}$ & 1.467$^{+0.004}_{-0.004}$ & 1.17$^{+0.04}_{-0.04}$ & 1.4$^{+0.1}_{-0.1}$\cr
\hline
4.0000 & 0.3150 & 1.0000 & 0.9$^{+0.1}_{-0.1}$ & 1.39$^{+0.07}_{-0.07}$ & 1.44$^{+0.03}_{-0.01}$ & 0.37$^{+0.00}_{-0.01}$ &  &  & 1.3$^{+0.2}_{-0.5}$ & \cr
\hline
2.7000 & 0.3027 & 1.050 & 0.82$^{+0.02}_{-0.02}$ & 0.98$^{+0.01}_{-0.07}$ &  & 0.37$^{+0.01}_{-0.01}$ & 0.9$^{+0.6}_{-0.3}$ & 1.432$^{+0.005}_{-0.005}$ & 1.5$^{+0.1}_{-0.5}$ & 0.5$^{+1.1}_{-0.2}$\cr
\hline
2.7000 & 0.2968 & 1.050 & 0.75$^{+0.00}_{-0.04}$ & 0.83$^{+0.01}_{-0.02}$ &  & 0.339$^{+0.003}_{-0.004}$ & 1.226$^{+0.009}_{-0.006}$ &  &  & \cr
\hline
3.8000 & 0.3000 & 1.0000 & 0.94$^{+0.03}_{-0.02}$ & 1.29$^{+0.05}_{-0.03}$ &  & 0.328$^{+0.002}_{-0.002}$ & 1.39$^{+0.02}_{-0.02}$ &  &  & 0.5$^{+0.4}_{-0.4}$\cr
\hline
2.7984 & 0.3072 & 1.317 & 0.72$^{+0.02}_{-0.02}$ & 0.78$^{+0.07}_{-0.01}$ & 1.21$^{+1.15}_{-0.07}$ & 0.324$^{+0.004}_{-0.004}$ & 1.336$^{+0.003}_{-0.003}$ &  &  & \cr
\hline
2.7000 & 0.2939 & 1.019 & 0.76$^{+0.01}_{-0.09}$ & 0.8$^{+0.0}_{-0.1}$ &  & 0.315$^{+0.007}_{-0.008}$ & 0.91$^{+0.07}_{-0.09}$ & 1.207$^{+0.006}_{-0.006}$ & 1.5$^{+0.1}_{-0.1}$ & \cr
\hline
3.2000 & 0.3150 & 1.600 & 0.69$^{+0.03}_{-0.03}$ & 1.0$^{+0.0}_{-0.2}$ &  & 0.302$^{+0.004}_{-0.004}$ & 1.2$^{+0.3}_{-0.2}$ & 1.4$^{+0.1}_{-0.4}$ &  & 1.1$^{+0.2}_{-0.2}$\cr
\hline
4.0000 & 0.2850 & 0.9700 & 0.73$^{+0.03}_{-0.03}$ & 1.0$^{+0.0}_{-0.5}$ & 1.473$^{+0.004}_{-0.002}$ & 0.260$^{+0.002}_{-0.003}$ & 1.283$^{+0.008}_{-0.007}$ &  &  & 1.4$^{+0.1}_{-0.1}$\cr
\hline
3.8800 & 0.2850 & 1.0000 & 0.6$^{+0.1}_{-0.1}$ & 0.67$^{+0.03}_{-0.03}$ & 0.9$^{+0.1}_{-0.1}$ & 0.255$^{+0.002}_{-0.002}$ & 0.92$^{+0.09}_{-0.09}$ & 1.26$^{+0.01}_{-0.01}$ &  & \cr
\hline
2.8177 & 0.2981 & 1.334 & 0.51$^{+0.04}_{-0.03}$ & 0.58$^{+0.02}_{-0.01}$ & 1.14$^{+0.06}_{-0.04}$ & 0.237$^{+0.002}_{-0.002}$ & 0.80$^{+0.05}_{-0.05}$ & 1.04$^{+0.04}_{-0.03}$ & 1.0$^{+0.3}_{-0.4}$ & \cr
\hline
4.0000 & 0.2679 & 1.0000 & 0.355$^{+0.008}_{-0.007}$ & 0.54$^{+0.01}_{-0.01}$ & 0.8$^{+-0.3}_{-0.3}$ & 0.1319$^{+0.0007}_{-0.0007}$ & 0.414$^{+0.006}_{-0.005}$ & 0.6$^{+0.1}_{-0.2}$ &  & 0.33$^{+0.08}_{-0.05}$\cr
\hline
\end{longtable}

\end{small}
\end{landscape}}

\bibliographystyle{bibstyle}
\bibliography{bib}
%%%%%%%%%%%%%%%%%%%%%%%%%%%%%%%%%%%%%%%%%%%%%%%%%%%%%%%%%%%%%%%%%%%%%%%%%%%%%%%%%%%%%%%%%%%%%%%%%%

\end{document}